\documentclass[12pt,a4paper,oneside,titlepage,onecolumn,openright]{book}

\usepackage[T1]{fontenc}
\usepackage[utf8]{inputenc}
\usepackage{cite}

\usepackage{fancyhdr}
\usepackage{multirow}
\usepackage{geometry}
\fancyheadoffset[H]{0.1cm}
\usepackage{graphicx}
\usepackage{bm} 
\usepackage{subfigure}
\usepackage[colorlinks=true,linktocpage=true,linkcolor=blue,citecolor=blue]{hyperref}
\usepackage[usenames,dvipsnames]{color}

\usepackage{mathtools}

\usepackage{multicol}

\newcommand{\s}{{\rm s}}	
\newcommand{\Qp}{{Q^+}}		
\newcommand{\Qm}{{Q^-}}		
\newcommand{\G}{{G}}		
\newcommand{\Qpm}{{Q^\pm}}		
\newcommand{\Q}{{Q}}		
\newcommand{\eq}{{\rm eq}}  
\newcommand{\an}{{\rm a}}  
\newcommand{\LRF}{{\rm LRF}}  
\newcommand{\teq}{\tau_\eq}

\newcommand{\bpT}{{\boldsymbol p}_T} 
 
\newcommand{\I}{{\cal I}}
\newcommand{\hI}{\hat{{\cal I}}}
\newcommand{\lp}{\left(}
\newcommand{\rp}{\right)}
\def\lb{\left(}
\def\rb{\right)}
\newcommand{\lsb}{\left[}
\newcommand{\rsb}{\right]}

\newcommand{\bea}{\begin{eqnarray}}
\newcommand{\eea}{\end{eqnarray}}
\newcommand{\beal}[1]{\begin{eqnarray}\label{#1}}
\newcommand{\eeal}{\end{eqnarray}}
\newcommand{\bel}[1]{\begin{eqnarray}\label{#1}}
\newcommand{\eel}{\end{eqnarray}}
\newcommand{\nn}{\nonumber}
\newcommand{\f}[2]{\frac{#1}{#2}}
\newcommand{\VP}{\vphantom{\frac{}{}}\!}

\newcommand{\p}{\partial}
\newcommand{\twpt}{(\tau,w,\pT)}
\newcommand{\tiwpt}{(\tau_0,w,\pT)}
\newcommand{\EQ}[1]{Eq.~(\ref{#1})}
\newcommand{\rf}[1]{Eq.~(\ref{#1})}
\newcommand{\EQS}[1]{Eqs.~(\ref{#1})}

\newcommand{\EQSTWO}[2]{Eqs.~(\ref{#1})~and~(\ref{#2})}
\newcommand{\EQSM}[2]{Eqs.~(\ref{#1})--(\ref{#2})}
\newcommand{\EQB}[1]{(\ref{#1})}
\newcommand{\rfn}[1]{(\ref{#1})}
\newcommand{\SEC}[1]{Sec.~\ref{#1}}

\newcommand{\APP}[1]{App.~\ref{#1}}
 
\def\mT{m_{T}} 				
\def\pT{p_{T}} 				
\newcommand{\pt}{p_{T}}		
\def\pL{p_{L}} 				
\newcommand{\TmnU}[2]{T^{\mu\nu \, {\rm #2}}_{\rm #1}} 
\def\Lq{\Lambda_Q} 			
\def\Lg{\Lambda_G} 			
\def\xq{\xi_Q} 				
\def\xg{\xi_G} 				
\def\ed{{\cal E}}			
\def\peq{{\cal P}_{\rm eq}} 
\def\sd{{\cal S}}			
\begin{document}
\newgeometry{tmargin=2.5cm,bmargin=2.5cm,lmargin=2.5cm,rmargin=2.5cm}
%
\begin{titlepage}
\begin{center}
\begin{table}
\begin{tabular}{ll}
\multirow{2}{*}{\includegraphics[scale=0.4]{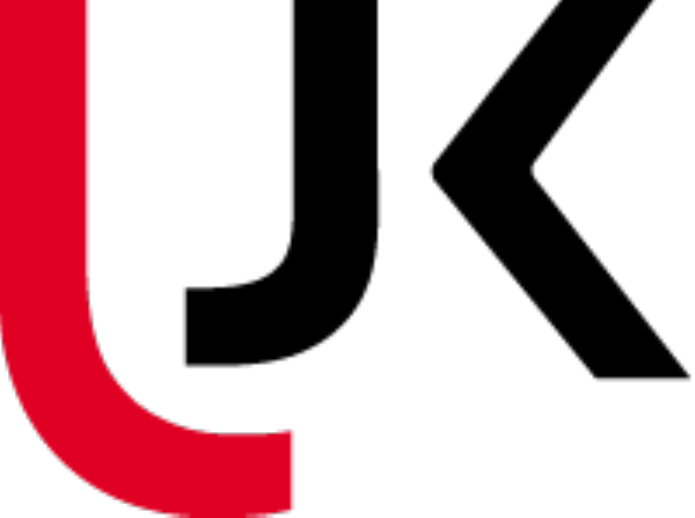}} & \\
& \textsc{\Large Jan Kochanowski University in Kielce}
\\[0.4cm]
&\textsc{\large Faculty of mathematics and natural sciences}
\end{tabular}
\end{table}

\textsc{\Large}\\[2cm]
\textmd{\Large Doctoral Thesis}\\[1cm]
\textbf{\huge EXACT SOLUTIONS}\\[0.4cm]
\textbf{\huge OF THE RELATIVISTIC }\\[0.4cm]
\textbf{\huge BOLTZMANN EQUATION}\\[0.4cm]
\textbf{\huge IN THE RELAXATION TIME}\\[0.4cm]
\textbf{\huge  APPROXIMATION}\\[3.5cm]
\textsc{\huge Ewa Maksymiuk}\\[3cm]
\textmd{\Large The thesis prepared in the Institute of Physics}\\[0.4cm]
\textmd{\Large under supervision by prof. dr hab. Wojciech Florkowski}\\[0.4cm]
\textmd{\Large and auxiliary supervision by dr hab. Rados\l aw Ryblewski}\\[2.5cm]
\textsc{\Large Kielce 2018}

\end{center}
\end{titlepage}

\setlength{\oddsidemargin}{0.46cm}   
\setlength{\evensidemargin}{-0.54cm} 
\setlength{\headheight}{27.5pt}

\newpage

\newpage
\cleardoublepage
\thispagestyle{plain}

\begin{center}
\section*{Abstract}
\end{center}

\bigskip
This Thesis concentrates on the analysis of coupled RTA (relaxation time approximation) kinetic equations for bosons and fermions. Bosons are treated as massless particles, while fermions have a finite mass. Using analytic and numerical methods we find exact solutions for such a mixture in the case of one-dimensional, boost-invariant expansion (for systems which are transversally homogeneous). In this way, we generalize several earlier results obtained for one-component systems and classical particles. 

\medskip
Throughout the text, we refer to fermions and bosons as to quarks
 and gluons, respectively. There are, however, important differences between our particles and real quarks and gluons appearing in the QCD field-theoretic calculations. In particular, we assume that the relaxation times used in the kinetic equations are the same for bosons and fermions. Moreover, the common relaxation time used in the numerical calculations is independent of the momenta of colliding particles and taken to be a constant. Due to such simplifications, our approach cannot be treated as a truly realistic model for QGP. However, it includes most of the important aspects connected with the hydrodynamization and equilibration processes, which are the main topics studied in this work. 

\medskip
Our numerical results illustrate how a non-equilibrium mixture approaches hydrodynamic regime described by the Navier-Stokes equations. We determine appropriate forms of the viscous kinetic coefficients. The shear viscosity of a mixture turns out to be the sum of the shear viscosities of boson and fermion components, while the bulk viscosity is given by the formula derived earlier for a gas of fermions, however, with the thermodynamic coefficients characterizing the whole boson-fermion mixture. Consequently, we find that massless bosons do contribute in a non-trivial way to the bulk viscosity of a mixture (provided quarks are massive). 

\medskip
We further observe the hydrodynamization effect which takes place earlier in the shear sector than in the bulk one --- the equalization of the longitudinal and transverse pressures takes place earlier than the equalization of the average and equilibrium pressures. The numerical studies of the ratio of the longitudinal and transverse pressures show that, to a good approximation, it depends on the ratio of the relaxation and evolution times only. We find that this behavior is connected with the existence of an attractor for conformal systems. 

\medskip
Our system of kinetic equations is subsequently used to construct a corresponding scheme of anisotropic hydrodynamics, which can be treated as an effective description of the underlying microscopic dynamics defined by the kinetic model. The comparisons between predictions of anisotropic hydrodynamics and exact solutions of the kinetic equations are used to validate our hydrodynamic approach.


\cleardoublepage
\thispagestyle{plain}

\begin{center}
\section*{Streszczenie}
\end{center}
Niniejsza praca koncentruje się na analizie sprzężonych równań kinetycznych w tzw. przybliżeniu czasu relaksacji, opisujących mieszaninę bozonów i~fermionów, przy czym bozony traktowane są jako cząstki bezmasowe, zaś fermiony posiadają skończoną masę. Użycie metod analitycznych i~numerycznych umożliwia znalezienie dokładnych rozwiązań w~przypadku takiej mieszaniny, podlegającej jednowymiarowej i~boost-niezmienniczej ekspansji. Tym samym uogólnione zostały poprzednie wyniki, uzyskane dla układów składających się z~cząstek jednego rodzaju, rządzonych statystyką klasyczną.
\newline

W całej pracy przez fermiony rozumie się kwarkami (oznaczone symbolem $Q$), zaś bozony nazywamy gluonami (oznaczone symbolem $G$). Należy jednak zwrócić uwagę na istotną różnicę pomiędzy cząstkami dyskutowanymi w pracy, a rzeczywistymi kwarkami i gluonami, pojawiającymi się w~teoretycznych rachunkach QCD. Niniejsze rozwiązania opierają się na założeniu, że czasy relaksacji użyte w równaniu kinetycznym są takie same dla bozonów i fermionów. Co więcej, czas relaksacji używany w rachunkach numerycznych jest niezależny od pędu zderzających się cząstek i~został przyjęty jako stały. W~związku z~tymi 
uproszczeniami, 
prezentowane podejście nie może być traktowane jako bardzo \linebreak
 realistyczny model plazmy kwarkowo-gluonowej. Nie zmienia to jednak faktu, że niniejszy opis zawiera  wiekszość ważnych aspektów związanych z procesami hydrodynamizacji i~osią-gania 
równowagi, bądącymi najważniejszymi tematami poruszanymi w pracy.
\newline

Numeryczne rezultaty ilustrują w jaki sposób nierównowagowa mieszanina 
\linebreak
cząstek osiąga hydrodynamiczny obszar opisywany przez równania Naviera-Stokesa. Wyznaczone zostały odpowiednie formy współczynników kinetycznych. Lepkość dynamiczna  
\linebreak
mieszaniny (ang. shear viscosity) okazała się być sumą lepkości dynamicznych bozonów i~fermionów. Z kolei w przypadku lepkości objętościowej (ang. bulk viscosity), danej przez formułę wyprowadzoną wcześniej dla gazu fermionów okazało się, że należy uwzględniać współczynniki termodynamiczne, charakteryzujące całą mieszaninę bozonów i fermionów. Tym samym bezmasowe bozony przyczyniają się w nietrywialny sposób do lepkości objętościowej.
\newline

W dalszej kolejności zaobserwowaliśmy efekty hydrodynamizacji, które następują 
\linebreak
wcześniej w sektorze dynamicznym niż objętościowym --- wyrównanie się ciśnień 
podłuż-nego
i poprzecznego ma miejsce wcześniej niż wyrównanie się ciśnienia średniego i 
równowa-gowego. 
Analiza numeryczna stosunku ciśniena podłużnego do poprzecznego pokazuje z~bardzo dobrą dokładnością zależność jedynie od stosunku czasu relaksacji i czasu własnego ewolucji. Zachowanie to jest związane z istnieniem atraktora dla układów konforemnych.
\newline

Dyskutowany układ równań kinetycznych jest następnie użyty do konstrukcji odpowiedniego układu równań hydrodynamiki anizotropowej, która może być traktowana jako efektywny opis zasadniczej dynamiki mikroskopowej, definiowanej przez model 
kinetycz-ny. 
Porównania pomiędzy przewidywaniami hydrodynamiki anizotropowej i dokładnych rozwiązań równań kinetycznych są użyte do pozytywnego zweryfikowania naszego podejścia hydrodynamicznego.

\cleardoublepage
\thispagestyle{plain}
\begin{center}
\section*{Acknowledgments}
\end{center}

I would like to thank and express the deepest gratitude and appreciation to my supervisor, Wojciech Florkowski, and to my co-supervisor, Rados\l aw Ryblewski, for introducing me to the extremely interesting and beautiful branch of heavy ion physics, and for giving me an~opportunity to work in very friendly atmosphere during my research.

My final thanks are directed for my beloved Micha\l, my sister Monika, my Parents, and all of my friends for their huge support.

\tableofcontents

\newpage
\begin{table}
\begin{center}
  \begin{tabular}{c c}
    \hline
    \\
     RHIC & Relativistic Heavy-Ion Collider at Brookhaven National Laboratory \\
     LHC & Large Hadron Collider at CERN \\ 
     \\
     QCD & quantum chromodynamics  \\
     QGP & quark-gluon plasma \\
     wQGP & weakly coupled QGP \\
     sQGP & strongly coupled QGP \\
     \\
     IS & Israel-Stewart (hydrodynamic equations) \\
     NS & Navier-Stokes (hydrodynamic equations) \\
     KT & kinetic theory \\
     aHydro & anisotropic hydrodynamics \\ \\
     RTA & relaxation time approximation \\
     RS & Romatschke-Strickland  (ansatz for the distribution function)\\
     \\
     EOS & equation of state \\
     eq & label specifying local thermal equilibrium  \\
   \\
    \hline
  \end{tabular}
\end{center}
\caption{\small List of acronyms}
\label{tab:acronyms}
\end{table}

\begin{table}
\begin{center}
  \begin{tabular}{c c}
    \hline
    \\
    $\TmnU{}{}$ & energy-momentum tensor \\
    $T$  & effective temperature \\
    $\mu$  & effective chemical potential \\
    $U^\mu$ & flow vector defining the Landau hydrodynamic frame \\
    $\Delta^{\mu\nu}$ & operator projecting on the space orthogonal to $U^\mu$ \\
    $\Pi$ & bulk pressure \\
    $\ed$ & energy density \\
    $\sd$ & entropy density \\
    ${\cal B}$ & baryon number density \\
    $\peq(\ed)$ & equilibrium pressure corresponding to the energy density   $\ed$ \\
    &  (functional form $\peq(\ed)$ follows from the equation of state) \\
    $\eta$ & shear viscosity \\
    $\zeta$ & bulk viscosity \\
    ${\bar \eta} = \eta/\sd$  &  ratio of shear viscosity and entropy density  \\
   \\
    \hline
  \end{tabular}
\end{center}
\caption{\small Symbols denoting physical concepts and variables }
\label{tab:symbols}
\end{table}

\begin{table}
\begin{center}
  \begin{tabular}{c c}
    \hline
    \\
    \textit{spacetime variables} & \\
    &\\
    $g_{\mu\nu}={\rm diag}(1,-1,-1,-1)$ & metric tensor\\
    $$ & \\
    $x^\mu=(x^0,x^1,x^2,x^3)=(t,x,y,z)$  & space-time coordinates \\
    $r=\sqrt{x^2+y^2}$ & distance from the collision axis \\
    $\phi=\arctan(y/x)$ & azimuthal angle \\
    $\theta=\arctan(r/z)$ & polar angle \\
    $\tau=\sqrt{t^2-z^2}$ & proper (longitudinal) time\\
    $\eta=\frac{1}{2}\ln\frac{t+z}{t-z}$ & spacetime rapidity \\
    $$ & \\
    $U^\mu, X^\mu, Y^\mu, Z^\mu$ & basis four-vectors \\
   \\
    \hline
    &\\
    \textit{kinematical variables} &  \\
    \textit{describing a single particle} & \\
    &\\
    $p^\mu=(p^0,p^1,p^2,p^3)=(E_p,p_x,p_y,p_z)$  & four-momentum \\
    $p^0=p_0=E_p$& energy\\
    $E_p=\sqrt{m^2+p^2}$&mass-shell energy \\
    $p^3=-p_3=p_z=p_L$& longitudinal momentum of a particle\\
    $p_T=\sqrt{p_x^2+p_y^2}$&transverse momentum of a particle\\
    $\phi_p=\arctan(p_y/p_x)$ & momentum azimuthal angle\\
    $$&\\
   \\
    \hline
  \end{tabular}
\end{center}
\caption{\small Symbols of the physical quantities related to space, time, 
and momenta of particles.}
\label{tab:stp}
\end{table}

\begin{table}
\begin{center}
  \begin{tabular}{c c}
    \hline
    \\
       \textit{anisotropic} &  \\
    \textit{energy-momentum tensors} & \\
    \\
    $T^{\mu\nu}_{\rm a}$ & leading-order energy-momentum tensor for aHydro \\
      ${\cal P}_T^{\rm a}$ & transverse pressure \\
      ${\cal P}_L^{\rm a}$ & longitudinal pressure \\
               &   \\
    &\\
       \hline \\
    \textit{parameters of anisotropic} &  \\
    \textit{ distribution functions} & \\
    &\\
  $\xi_Q$ & anisotropy parameter for quarks \\
  $\xi_G$ & anisotropy parameter for gluons \\
  $\Lambda_Q$ & transverse-momentum scale for quarks\\
  $\Lambda_G$ & transverse-momentum scale for gluons \\
  $\lambda$ & the non-equilibrium chemical potential for quarks \\
   \\
    \hline
  \end{tabular}
\end{center}
\caption{\small Symbols and concepts used in aHydro}
\label{tab:ahydro}
\end{table}
%

\chapter{Introduction}
\label{chap:Intro}
%
\section[Ultra-relativistic heavy-ion collisions and quark-gluon plasma]{Ultra-relativistic heavy-ion collisions \\ and quark-gluon plasma}   
\label{sect:aim}
%
The topics discussed in this Thesis belong to a broad physics field of ultra-relativistic heavy-ion collisions~\cite{Stoecker:1986ci,Hwa1990,Wong1994,Muller:1994rb,Hwa1995,Mrowczynski:1998jr,letessier_rafelski_2002,Hwa2004,Yagi:2005yb,Vogt:2007zz,Bartke2008,Florkowski:2010zz,Hwa2010,Chaudhuri:2012yt,Wang2016,Satz:2018oiz,Bhalerao:2014owa,Busza:2018rrf}. This type of research combines the methods of high-energy physics of elementary particles with nuclear physics. However, a characteristic feature of heavy-ion collisions is that systems consisting of a large number of particles are produced in such processes, hence, theoretical tools based on the use of thermodynamics, statistical physics, hydrodynamics and/or kinetic theory concepts are commonly used to interpret the experimental results.

One of the conclusions of the experimental programs performed at RHIC~\!\footnote{The explanations of the acronyms used in this work are given in Table~\ref{tab:acronyms}.} and the LHC is that relatively simple thermodynamic and/or hydrodynamic models are indeed quite successful in a description of the data~\cite{ Jaiswal:2016hex,Yan:2017ivm, Florkowski:2017olj,Alqahtani:2017mhy,Romatschke:2017ejr}. This, in turn, leads to new questions about the applicability range of such methods. One of the persistent problems discussed in the context of heavy-ion collisions is why the particle abundances and spectra have thermal features. Is there enough time for such microscopic processes so that they can lead to a thermalized state? Another problem is applicability of hydrodynamics. Are the gradients of local hydrodynamic variables sufficiently small that hydrodynamics becomes a suitable framework for the determination of space-time evolution of the produced matter?

In the last years, at least partial answers to the questions mentioned above have been found. It seems that the timescales involved in heavy-ion collisions are sufficiently long so that the produced matter may emerge locally equilibrated at freeze-out. On the other hand, there is a growing evidence that hydrodynamics (at least in its modern formulations) may be applied in the situations where systems are far away from local equilibrium. In this way, relativistic hydrodynamics becomes a theoretical tool connecting early, non-equilibrium stages with the late, locally equilibrated system configurations. 

One way of introducing the equations of relativistic hydrodynamics is that one starts from the underlying kinetic theory and derives the hydrodynamic equations by taking a specific set of the moments of the kinetic equation. In this case, the efficacy of the hydrodynamic approach can be verified by the comparisons of hydrodynamic predictions with the kinetic-theory results. Such  methodology has been successfully used in the past~\cite{Florkowski:2013lza,Florkowski:2013lya}, and our present work contributes to this type of studies:  herein we report in detail on the new exact solutions of the Boltzmann equation and compare our kinetic-theory results with the outcome of anisotropic hydrodynamics \cite{Florkowski:2010cf,Martinez:2010sc} which represents one of the new hydrodynamic frameworks. 
%
\subsection{Searches for new phases of strongly interacting matter}
%
The ultimate aim of the physics of ultra-relativistic heavy-ion collisions is the observation of new states of matter predicted by quantum chromodynamics (QCD) and by model calculations inspired by QCD. In fact, the experimental results collected at RHIC are treated now as the evidence for production of a strongly coupled quark-gluon plasma (sQGP) in the analyzed collisions. One source of this evidence is a very successful use of the viscous hydrodynamics to describe the data. Relativistic viscous hydrodynamics forms the basic ingredient of the models describing heavy-ion collisions at the top RHIC energies. It turns out that a small ratio of the shear viscosity $\eta$ to the entropy density $\sd$ is required to reproduce well the data.~\footnote{Recent comparisons between model calculations and the data lead to the range $1 \leq \eta/\sd \leq 2.5$ in units of $\hbar/(4\pi k_B)$ \cite{Song:2010mg}.} The smallness of the shear viscosity signals the presence of a strongly coupled fluid. Another signal of a strongly coupled system is the phenomenon of jet quenching. 

We note that in heavy-ion collisions we deal with systems of quarks and gluons that always interact strongly, i.e., their fundamental interaction theory is QCD. However, because of the running of the QCD coupling constant, the interaction strength can have a~different magnitude. By a strongly coupled QGP, we understand the quark-gluon system with an effective large coupling constant. As the matter of fact, one sometimes argues that the very concept of particles may be not adequate for such a system. On the other hand, by a weakly coupled quark-gluon plasma (wQGP) we mean an asymptotic state where the coupling constant is small. In the next two sections, we describe how the concepts of the quark-gluon plasma evolved in time. 
%
\subsection{Perturbative concept of quark-gluon plasma}
%
Originally, the concepts of the quark-gluon plasma referred only to a weakly coupled plasma. The idea of formation of such a state in heavy-ion collisions was motivated by the phenomenon of the asymptotic freedom in QCD --- densely packed quarks and gluons should interact weakly and form a kind of gas of particles with color charges. The name ``quark-gluon plasma'' was introduced by Shuryak in 1978~\cite{Shuryak:1978ij,Shuryak:1980tp}. 

The weakly coupled QGP  has (in the leading order of the coupling) simple thermodynamic properties which follow directly from expressions for the Bose-Einstein (gluons) and Fermi-Dirac (quarks) gases. It has more effective degrees of freedom than a hadron gas and one expects the occurrence of a phase transition between the hadron gas and QGP as the system's temperature and/or density increases. 

The first heavy-ion experiments with relativistic beams were aiming at the creation of such an asymptotic QGP state. Several so-called QGP plasma signatures were proposed as the manifestation of the plasma formation (they included, for example, the phenomena of  enhanced strangeness production~\cite{Rafelski:1982pu} and $J/\psi$ suppression~\cite{Matsui:1986dk}). Although the physics interpretation of those signatures was not always quite straightforward, taken together, they motivated the physicists from CERN to announce in the year 2000 that a new state of matter was found\cite{PRESSCUT-2000-210}.
%
\subsection{Strongly coupled quark-gluon plasma}
%
The experimental data coming from the RHIC experiments (which started, by the way, just after the CERN statement of 2000), together with their theoretical explanations, changed the paradigm of a weakly coupled QGP. This was possible due to an enormous progress made in both the experimental techniques (which delivered new high-quality data) and theoretical tools and models (which allowed for a quantitative description of the measurements). As we have mentioned above, the smallness of shear viscosity and the phenomenon of jet quenching observed at RHIC pointed out to a strongly-coupled character of QGP. Recent results from the LHC seem to confirm these findings suggesting that the beam energies available at the moment in the colliders are insufficient to reach the regime of a weakly-coupled plasma. 

The fact that QGP behaves as a strongly coupled fluid has made possible application of the theoretical techniques designed to deal with such systems. In particular, several features of QGP have been explained in the framework of the so-called AdS/CFT correspondence~\cite{CasalderreySolana:2011us,Heller:2016gbp}. This approach explains, for example, the small value of the ratio of the shear viscosity to the entropy density, ${\bar \eta} = \eta/\sd = 1/(4\pi)$ \cite{Kovtun:2004de}~\footnote{This result follows also from the kinetic theory if quantum effects are included~\cite{Danielewicz:1984ww}.}.

Although the AdS/CFT correspondence is suitable for the description of conformal systems rather than of QCD (where conformal symmetry is broken by the renormalization scale and the finite quark masses), its appealing feature is that it delivers exact solutions describing the time dynamics of strongly-coupled non-equilibrium systems. This allows for precise studies of physical systems that approach the hydrodynamic regime.

The use of the AdS/CFT correspondence is frequently confronted with the applications of the kinetic theory since the latter is regarded as an appropriate framework to describe weakly coupled systems. Nevertheless, both AdS/CFT and kinetic theory offer a possibility of analyzing the exact non-equilibrium dynamics. This makes the two approaches equally attractive to study the hydrodynamization and equilibration processes.

\section{Hydrodynamic description of heavy-ion collisions}

%
\subsection{Standard model of heavy-ion collisions}
%
Nowadays, one speaks very often about the standard model of heavy-ion collisions (for a recent review see, for example, Ref.~\cite{Florkowski:2017ixx}). According to this model, the space-time evolution of matter created in heavy-ion collisions can be separated into three different stages. The first stage is highly out of equilibrium and includes the hydrodynamization process which is understood as the approach to the physical regime where dynamics is well described by a viscous relativistic hydrodynamics. The second stage is fully described by hydrodynamics and includes the phase transition from the quark-gluon plasma to a~hadron gas. This stage describes local equilibration of the system and ends when the interactions between hadrons become sufficiently weak, so the system enters the (third) stage called freeze-out \cite{Ryblewski:2017sgj}. The name ``freeze-out'' denotes freezing of the hadron momenta --- noninteracting hadrons have momenta that no longer change in time. The process of freeze-out is commonly divided into at least two subprocesses: the chemical and thermal freeze-outs. The thermal or kinetic freeze-out corresponds to a genuine freeze-out process defined above. On the other hand, the chemical freeze-out corresponds to a transition where inelastic collisions between hadrons cease. The chemical freeze-out (when inelastic processes stop) precedes the thermal freeze-out (when both inelastic and elastic processes stop)~\footnote{Although the concept of two freeze-outs is commonly used, many data can be explained in the ``single freeze-out model'' assuming that the chemical and thermal freeze-outs coincide~\cite{Broniowski:2001we}.}.

It is worth mentioning that the standard model defined in this way has a block structure with elements that can be separately exchanged and/or modified. For example, the middle hydrodynamic stage can be described with different hydrodynamics frameworks. Earlier approaches based on the use of the perfect-fluid hydrodynamics are now replaced by sophisticated viscous-hydrodynamics codes. 

\medskip
In this work, we concentrate on the first stage of heavy-ion collisions where the produced matter is out of local thermodynamic equilibrium. Using the framework of the kinetic theory, we can analyze how the analyzed system approaches the regime described by viscous hydrodynamic equations. Certainly, our approach is very simplistic and cannot be treated as a realistic modeling of the early stages. Nevertheless, it contains all important features of non-equilibrium dynamics, allows for finding exact solutions, and can be used to study the phenomenon of hydrodynamization. 

\subsection{From perfect-fluid to the viscous hydrodynamic framework}
\label{sec:pfvf}

\subsubsection{Perfect-fluid hydrodynamics}

Relativistic  perfect  fluid  dynamics provides us with the simplest relativistic fluid dynamical equations\cite{LLfluid,Misner:1974qy,deGroot:1980,rezzolla2013relativistic,Stoecker:1986ci,Florkowski:2010zz,Rischke:1998fq,Kolb:2003dz,Huovinen:2006jp,Romatschke:2009im,Gale:2013da}. 
The concept of viscous hydrodynamics is usually connected with an idea of gradient (hydrodynamic) expansion of the energy-momentum tensor~\cite{Florkowski:2017olj}. The leading order of hydrodynamic expansion is identified with the perfect-fluid framework, where the gradients are absent and the whole theory follows from the perfect-fluid form of the energy-momentum tensor valid in local equilibrium,
\begin{eqnarray}
T^{\mu\nu}_{\rm eq} = \left(\ed + \peq \right) U^\mu U^\nu
-g^{\mu\nu} \peq,
\label{eq:Tmunu0}
\end{eqnarray}
and the conservation laws for energy and momentum,
\begin{eqnarray}
\p_\mu T^{\mu\nu}_{\rm eq} = 0.
\label{eq:TmunuC}
\end{eqnarray}
Here $\ed$ is the energy density and $\peq$ is the equilibrium pressure corresponding to the energy density $\ed_{\rm eq} = \ed$~\footnote{The physics symbols are defined in Table~\ref{tab:symbols}.}. For systems with finite baryon number, Eqs.~\rfn{eq:Tmunu0} and \rfn{eq:TmunuC} should be supplemented by the conservation of the baryon number current
\begin{eqnarray}
\p_\mu ({\cal B} \, U^\mu) = 0.
\end{eqnarray}
At this level, the main dynamic variables (hydrodynamic fields) are the local temperature $T(x)$ and three independent components of the flow four-vector $U^\mu(x)$, as well as the chemical potential $\mu(x)$ for systems containing a conserved charge such as the baryon number.

\subsubsection{Navier-Stokes hydrodynamics}

The inclusion of dissipative terms leads to a modification of the energy-momentum tensor \rfn{eq:Tmunu0} that can be written in the form 
\begin{eqnarray}
T^{\mu\nu} = T^{\mu\nu}_{\rm eq} + \pi^{\mu\nu} + \Pi \Delta^{\mu\nu}.
\label{eq:Tmunu1}
\end{eqnarray}
Here $\pi^{\mu\nu}$ is the {\it shear stress tensor}, $\Pi$ is the bulk pressure, and $\Delta^{\mu\nu} = g^{\mu\nu} - U^\mu U^\nu$ is the operator projecting vectors on the space perpendicular to $U$. Choosing the Landau hydrodynamic frame, where
\begin{eqnarray}
T^{\mu\nu} U_\nu = \ed U^\mu,
\label{eq:LF}
\end{eqnarray}
we demand that $\pi^{\mu\nu} U_\nu = 0$. Moreover, the shear stress tensor should be symmetric and traceless (as the effects of the trace are included in the isotropic part related to the bulk pressure). This leaves five independent components in $\pi^{\mu\nu}$. These together with a one degree of freedom represented by $\Pi$ define six additional components of the symmetric energy-momentum tensor (in addition to $T$ and three independent components of $U^\mu$).

By including terms linear in gradients of $T$ and $U^\mu $, we obtain the Navier-Stokes theory where
\begin{eqnarray}
\pi^{\mu\nu} = 2\eta \sigma^{\mu\nu}, \qquad
\Pi = - \zeta \p_\mu U^\mu.
\label{eq:etazeta}
\end{eqnarray}
Here $\eta$ and $\zeta$ are the shear and bulk viscosity coefficients, respectively, and $\sigma^{\mu\nu}$ is the {\it shear flow tensor}. The latter is defined through the expression
\begin{eqnarray}
\sigma^{\mu\nu} =  \Delta^{\mu\nu}_{\alpha \beta} \p^\alpha U^\beta,
\end{eqnarray}
where the projection operator $ \Delta^{\mu\nu}_{\alpha \beta} $ has the form
\begin{eqnarray}
 \Delta^{\mu\nu}_{\alpha \beta}  = \f{1}{2} \left(
 \Delta^\mu_{\,\,\,\alpha}  \Delta^\nu_{\,\,\,\beta} +  \Delta^\mu_{\,\,\,\beta}  \Delta^\nu_{\,\,\,\alpha}  \right) 
 - \f{1}{3}  \Delta^{\mu\nu} \Delta_{\alpha \beta}.
 \label{eq:Delta4}
\end{eqnarray}
We note that the shear flow tensor is symmetric, orthogonal to $U^\mu$, and traceless. Hence, in the Navier-Stokes theory, these properties become immediately the properties of the shear stress tensor, due to the first relation in \rf{eq:etazeta}.

\subsubsection{Israel-Stewart hydrodynamics}

It turns out that the use of the form \rfn{eq:Tmunu1} with \rfn{eq:etazeta} directly in the conservation law \rfn{eq:TmunuC} leads to instabilities and acausal behavior~\cite{Hiscock:1985zz}. To remedy this situation, Israel and Stewart modified the hydrodynamic approach \cite{Israel:1976tn,Israel:1979wp} by upgrading both $\pi^{\mu\nu}$ and $\Pi$ to new independent hydrodynamic variables that satisfy the following equations
\begin{eqnarray}
 \dot{\pi}^{\langle\mu\nu\rangle} +
\frac{\pi^{\mu\nu}}{\tau_{\pi}} &=& 
\beta_{\pi} \sigma^{\mu\nu}, \qquad  \tau_{\pi} \beta_{\pi} = 2\eta,
\label{eq:ISeta} \\
\dot{\Pi} + \frac{\Pi}{\tau_{\Pi}} &=& -\beta_{\Pi} \p_\mu U^\mu ,   \qquad  \tau_{\Pi} \beta_{\Pi} = \zeta . 
\label{eq:ISzeta}
\end{eqnarray}
The dot denotes here the convective derivative, $U^\mu \p_\mu$, while the angular brackets denote contraction with the projector \rfn{eq:Delta4} (after the calculation of the derivative). The parameters $\tau_{\pi}$ and $\tau_{\Pi}$ are known as the relaxation times for viscous shear stress and bulk pressure, respectively. The coefficients $\beta_{\pi}$ and $\beta_{\Pi}$ are kinetic coefficients that relate the relaxation times with the viscosity coefficients, as described by the right-hand sides of  \rfn{eq:ISeta} and \rfn{eq:ISzeta}.

Equations \rfn{eq:ISeta} and \rfn{eq:ISzeta} are the simplest version of the hydrodynamic equations of the Israel-Stewart type. Nevertheless, they describe the main idea of treating dissipative terms as additional hydrodynamic variables. In practical calculations, one includes more terms on the right-hand sides of \rfn{eq:ISeta} and \rfn{eq:ISzeta}. Their appearance  is controlled essentially by the powers of gradients. The shear stress tensor and the bulk pressure are treated, due to \rf{eq:etazeta}, as first-order terms. Hence, their derivatives are of the second order in gradients. Consequently, if we want to construct a second-order hydrodynamic theory, the right-hand sides of \rfn{eq:ISeta} and \rfn{eq:ISzeta} should contain additional terms that are properly constructed products of $\pi^{\mu\nu}$ or $\Pi$ with $\sigma^{\mu\nu}$ or $\p_\mu U^\mu$ \cite{Jaiswal:2014isa,Denicol:2014mca,Chattopadhyay:2014lya,Florkowski:2015lra,Tinti:2016bav,Florkowski:2016kjj}. 

The hydrodynamic codes based on the Israel-Stewart approach are the most popular tools to model the expansion of matter produced in heavy-ion collisions. Our estimates of the magnitude of the shear and bulk viscosity follow mainly from such type of calculations. 

\subsubsection{New developments}

One expects usually that $\pi^{\mu\nu}$ and $\Pi$ are small compared to the leading order (perfect-fluid) expressions. However, it turns out that the gradients present at the early stages of heavy-ion collisions are very large and, consequently, the gradient corrections become quite substantial. This observation has produced doubts if the standard viscous hydrodynamics is indeed a reliable tool for the description of heavy-ion collisions. This, in turn, initiated broad analyses of the hydrodynamic concepts, which brought very intriguing answers. One of the new points in this kind of studies is the observation that the early time dynamics of systems is determined by the existence of attractors which ``lead the system'' towards a hydrodynamic, Navier-Stokes regime~\cite{Heller:2015dha}. This would correspond to the hydrodynamization process. On the other hand, a subsequent viscous evolution leads the system towards local equilibrium, thus, it is responsible for the process commonly known as thermalization.

%
\subsection{Anisotropic hydrodynamics}
%
The problem of ``large gradient corrections'' motivated also the construction of new hydrodynamic schemes that do not refer to the concept of gradient expansion. In 2010 two groups formulated independently the idea of anisotropic hydrodynamics (aHydro) \cite{Florkowski:2010cf,Martinez:2010sc}, what was the inspiration for the next important papers~\cite{Ryblewski:2010bs,Ryblewski:2011aq,Ryblewski:2010ch,Florkowski:2011jg,Martinez:2012tu,Ryblewski:2012rr,Ryblewski:2013jsa,Bazow:2013ifa,Tinti:2013vba,Florkowski:2014bba,Nopoush:2014pfa,Strickland:2014pga,Nopoush:2015yga,Tinti:2015xwa,Bazow:2015cha,Strickland:2015utc,Alqahtani:2015qja,Bluhm:2015raa,Bluhm:2015bzi,Molnar:2016vvu,Molnar:2016gwq,Alqahtani:2016rth,Alqahtani:2017jwl,Alqahtani:2017tnq}. This approach makes use of the kinetic-theory concepts and assumes that the phase-space distribution functions are highly anisotropic. Using simple parametrizations of such functions (in terms of the original or generalized Romatschke-Strickland form~\cite{Romatschke:2003ms}) in the RTA kinetic equation, and taking appropriate moments of this equation, one can derive a hydrodynamic framework allowing for the description of systems far away from local equilibrium.  The parameters defining anisotropic distributions play a role of hydrodynamic variables. For systems being close to local equilibrium, they can be connected with the shear stress tensor and the bulk pressure.

In constructing the framework of anisotropic hydrodynamics it is not completely obvious which moments of the kinetic equation should be taken into account (except for those leading directly to the conservation laws for energy, momentum, and baryon number). One way to solve this problem is to make comparisons of anisotropic-hydrodynamics predictions with the kinetic-theory results. This approach turned out to be very successful in the past for simple systems~\cite{Florkowski:2013lza,Florkowski:2013lya,Tinti:2015xra,Florkowski:2016zsi}. In this work, we use this method to introduce and validate aHydro equations for mixtures. We emphasize that to do such comparisons we have to know exact solutions of the kinetic equation since this allows us for making precise comparisons.

%
\section{Kinetic theory}
%

Kinetic (or transport) theory is based on the Boltzmann kinetic equation that is usually very difficult to deal with, as it contains a complicated collision term. The latter accounts for the collisions between particles and has the form of a multidimensional integral. The standard treatment of the Boltzmann equation is that one makes involved numerical simulations describing the free motion of particles as well as their collisions. 

Because of this technical difficulties, one quite often uses a simplified version of the kinetic equation with a simplified form of the collision term. This is known in the literature as the relaxation-time approximation (RTA)~\cite{Bhatnagar:1954zz,Anderson:1974a,Anderson:1974b,Czyz:1986mr}. The details of this approach will be given below. Here, we only state that the present Thesis is entirely based on this approximation, as it allows for finding exact solutions of the kinetic equation.  Similarly to earlier works we also restrict ourselves to boost invariant systems~\cite{Bjorken:1982qr,Baym:1984sr}, as this is another important assumption that allows for the exact treatment of the dynamics. 

For the sake of simplicity, we also assume that the relaxation time is constant and the same for different particles forming the interacting system. The use of a temperature dependent relaxation time, which is a natural choice for conformal theories, increases substantially the computational time but, otherwise, does not introduce any important restrictions for the performed calculations. 

We usually consider below the systems consisting of two types of particles: massive fermions and massless bosons. We continue to call them quarks and gluons but it should be emphasized that due to the simplistic character of our kinetic model they differ in many respects from real (perturbative) quarks and gluons.  

\newpage
\section{Specific aim of this work}   
\label{sect:spaims}

\bigskip
A specific aim of this Thesis is to collect, summarize, and update several previous results on the exact solutions of the Boltzmann RTA equation and to clarify their relations to anisotropic hydrodynamics. These results were obtained by the author of this Thesis and collaborators, and have been already published in the following publications:

\begin{itemize}

\item[{\bf 1.}] W. Florkowski, E. Maksymiuk, R. Ryblewski, and M. Strickland, {\it Exact solution of the (0+1)-dimensional Boltzmann equation for a massive gas}, Phys. Rev. {\bf C89} (2014) 054908 \cite{Florkowski:2014sfa}.

\item[{\bf 2.}] W. Florkowski and E. Maksymiuk, {\it Exact solution of the (0+1)-dimensional Boltzmann equation for massive Bose-Einstein and Fermi-Dirac gases}, J. Phys. {\bf G42} (2015) 045106 \cite{Florkowski:2014sda}.

\item[{\bf 3.}] W. Florkowski, E. Maksymiuk, R. Ryblewski, and L. Tinti, {\it Anisotropic hydrodynamics for a mixture of quark and gluon fluids}, Phys. Rev. {\bf C92} (2015) 054912 \cite{Florkowski:2015cba}.

\item[{\bf 4.}] W. Florkowski, E. Maksymiuk, and R. Ryblewski, {\it Coupled kinetic equations for fermions and bosons in the relaxation-time approximation}, Phys. Rev. {\bf C97} (2018) 024915 \cite{Florkowski:2017jnz}.

\item[{\bf 5.}] W. Florkowski, E. Maksymiuk, and R. Ryblewski, {\it Anisotropic-hydro\-dynamics approach to a quark-gluon fluid mixture}, Phys. Rev. {\bf C97} (2018) 014904 \cite{Florkowski:2017ovw}.

\item[{\bf 6.}] E.~Maksymiuk, \textit{Exact solutions of the (0+1)-dimensional kinetic equation in the relaxation time approximation}, J.Phys.Conf.Ser. \textbf{612} (2015) no.1, 012054.
\newline
Conference: Hot Quarks 2014, Las Negras, Spain.

\item[{\bf 7.}] E.~Maksymiuk, \textit{Mixture of quark and gluon fluids described in terms of anisotropic hydrodynamics}, Acta Phys.Polon.Supp. \textbf{10} (2017) 1171 (2017).
\newline
Conference: 9th International Winter Workshop "Excited QCD" 2017, Sintra, Portugal.

\item[{\bf 8.}] E.~Maksymiuk, \textit{Kinetic equations and anisotropic hydrodynamics for quark and gluon fluids}, EPJ Web Conf. \textbf{18} (2018) 20207.
\newline
Conference: 6th International Conference on New Frontiers in Physics (ICNFP 2017), Crete, Greece.

\end{itemize}

The papers {\bf 1}, {\bf 2} and {\bf 4} generalize the results obtained in Refs.~\cite{Florkowski:2013lza,Florkowski:2013lya} to the case of massive particles that obey quantum statistics. The article {\bf 5} extends the previous results to the case of a mixture of massive fermions and massless bosons. The papers {\bf 3} and {\bf 5} construct a framework of anisotropic hydrodynamics for a mixture. The present work is dominantly based on {\bf 4} and {\bf 5}.

The papers \textbf{6}, \textbf{7} and \textbf{8} are conference proceedings.

\newpage
\section{Summary of the main results}

\begin{itemize}

\item[1.] Using analytic and numerical methods we have found exact solutions of the coupled RTA kinetic equations for massless gluons and massive quarks. In this way we have generalized several previous results obtained for simple (one-component) systems, where particles usually obeyed classical statistics.

\item[2.] The shear and bulk viscosities of a quark-gluon mixture have been found. It turns out that the shear viscosity $\eta$ is simply a sum of the quark and gluon shear viscosities, $\eta = \eta_\Q + \eta_\G$. On the other hand,  the bulk viscosity of a mixture is given by the formula known for a massive quark gas, $\zeta$. Nevertheless, we find that $\zeta$ depends on thermodynamic coefficients characterizing the whole mixture rather than quarks alone, which means that massless gluons contribute in a non-trivial way to the bulk viscosity (provided the quarks are massive).

\item[3.] We find that the hydrodynamization effect takes place earlier in the shear sector than in the bulk one --- the equalization of the longitudinal, ${\cal P}_L$, and transverse, ${\cal P}_T$, pressures takes place earlier than the equalization of the average and equilibrium pressures.

\item[4.] Our studies of the time evolution of the ratio of the longitudinal and transverse pressures indicate that, to a very good approximation, it depends on the ratio of the relaxation and evolution times only. This behavior is related to the presence of an attractor which was found and discussed earlier for conformal systems~\cite{Heller:2015dha,Romatschke:2016hle,Spalinski:2016fnj,Romatschke:2017vte,Spalinski:2017mel,Strickland:2017kux}. 
 
\item[5.] The results collected in this work confirm that aHydro is a very good approximation for the kinetic-theory results. We find that starting from the different initial condition aHydro
reproduces very well the time dependence of several physical quantities, in particular, of the ratio of the longitudinal and transverse pressures. This brings further arguments in favor of using the framework of aHydro for phenomenological modeling of heavy-ion collisions.

\end{itemize}

\newpage
\section{Notation and units}

In this section, we define our symbols for four-momenta of particles, space-time coordinates and units, see also Table \ref{tab:stp}.

\subsection{Four-momentum parametrization}
\label{sect:cc}
Due to ultra-relativistic speeds along the $z$-axis, the particle four-momentum is most often parametrized in the following way
\begin{eqnarray}
p^\mu &=& \lb p^0, p^1, p^2, p^3 \rb = \lb E_p, p_x, p_y, p_z\equiv\pL \rb \nn\\
&=& \lb \mT\cosh y,\,\, \pt \!\cos \phi_p,\,\, \pt \!\sin \phi_p,\,\, \mT\sinh y \rb,
\label{p0}
\end{eqnarray}
where
\begin{equation}
\mT = \sqrt{m^2 + \pT^2} 
\label{trmass}
\end{equation}
  is the transverse mass, $\pT=\sqrt{p_x^2 + p_y^2}$ is the transverse momentum, 
\begin{equation}
y = \f{1}{2} \ln \f{E_p+\pL}{E_p-\pL} 
\label{rap}
\end{equation} 
 is the longitudinal rapidity, and 
\begin{equation}
\phi_p = \tan^{-1} \lb\f{p_y}{p_x} \rb
\label{phip}
\end{equation}
is the azimuthal angle in the transverse plane. Due to the on-shell condition, $p^2=p^\mu p_\mu = m^2$, only three momentum variables, say, $p_x$, $p_y$ and $p_z$ (or, equivalently, $\pT$, $\phi_p$ and $y$) are treated as independent which for the momentum covariant integration measure implies
\begin{equation}
 \int dP (\dots) = 2 \int d^4 p \, \Theta(p^0) \delta(p^2-m^2) (\dots) = \int \frac{d^3p}{E_p} (\dots),
\label{eq:dP} 
\end{equation}
where $\Theta$ is the Heaviside step function. 

\subsection{Spacetime parametrization}
\label{sect:zcp}
In an analogous way, we parameterize the space-time coordinates~\footnote{Note that we use the same notation for the spacetime coordinate $x^2$ and the longitudinal rapidity, and the meaning of the $y$ variable should be inferred each time from the context.}
\begin{eqnarray}
x^\mu &=& \lb x^0, x^1, x^2, x^3 \rb  = \lb t, x, y, z \rb  \nn\\
&=&  \lb\tau \cosh \eta, r \cos \phi, r \sin \phi, \tau \sinh \eta \rb ,
\label{x0}
\end{eqnarray}
where $\tau$ is the invariant time (often called the longitudinal proper time)
\begin{equation}
\tau = \sqrt{t^2 - z^2},
\label{tau}
\end{equation}
and $\eta$ is the spacetime rapidity
\begin{equation}
\eta = \f{1}{2} \ln \f{t+z}{t-z}.
\label{sprap}
\end{equation} 
We further define the distance in the transverse plane $r$ and the angle $\phi$ through the equations
\begin{equation}
r = \sqrt{x^2 + y^2}, \quad \phi = \tan^{-1} \lb\f{y}{x} \rb .
\label{rphi}
\end{equation}

For convenience, we use the following notation for the scalar product: $a \cdot b = a_\mu b^\mu = g_{\mu \nu} a^\mu b^\nu$ with the flat-space metric tensor 
\begin{equation}
 g^{\mu \nu} = \lb
\begin{array}{cccc}
1 & 0 & 0 & 0 \\
0 & -1 & 0 & 0 \\
0 & 0 & -1 & 0 \\
0 & 0 & 0 & -1
\end{array} \rb  ,
\label{gmn}
\end{equation}
satisfying  $g^{\mu \nu} g_{\mu \nu} = 4$.
\subsection{Units}
\label{sect:uj}
Throughout the text we use natural units with $c = \hbar = k_B = 1$.

\section{Text organization}

The Thesis is organized as follows: In Chapt.~\ref{sec:ke} we introduce the system of kinetic equations and study their moments. This leads to two Landau matching conditions related to the baryon number and energy-momentum conservations. The form of the exact solutions of the kinetic equations is discussed in Chapt.~\ref{sect:e}. Numerical results for various physics observables, obtained with the exact solutions, are presented in Chapt.~\ref{sect:numresexact}. The concept of anisotropic hydrodynamics is introduced in Chapt.~\ref{chap:ahydromix}.  Numerical comparisons of the results obtained with aHydro and kinetic theory are presented in Chapt.~\ref{sec:numresahyd}. We conclude and summarize in Chapt.~\ref{sec:sumcon}.

\chapter{Kinetic equations}
\label{sec:ke}

\section{Relaxation time approximation}
\label{sec:RTA}

In this Thesis, we analyze three coupled relativistic Boltzmann kinetic equations for quark ($\Qp$), antiquark ($\Qm$), and gluon ($\G$) phase-space distribution functions~$ f_{\s}(x,p)$ ~\cite{Florkowski:2012as,Florkowski:2013uqa,Florkowski:2014txa,Florkowski:2015cba},
\beal{kineq}
p^\mu \p_\mu  \, f_{\s}(x,p) &=&  {\cal C}\lsb  f_{\s}(x,p)\rsb,\quad \s=\Qp, \Qm, \G. 
\eeal
The left-hand sides of Eqs.~\rfn{kineq} describe free motion of particles (they are often dubbed the  {\it drift} or {\it free-streaming terms}), while the right-hand sides contain the collision terms ${\cal C}[f_{\s}(x,p)]$, which account for interactions in the system. In this work,  the latter are included in the relaxation time approximation (RTA)~\cite{Bhatnagar:1954zz,Anderson:1974a,Anderson:1974b}, namely, we use the form
\beal{colker}
{\cal C}\lsb f_\s(x,p)\rsb &=&  p\cdot U \,\, \frac{f_{\s, \eq}(x,p)-f_\s(x,p)}{\teq},
\eeal
where we introduce the notation with ``dot'' for scalar product, $p \cdot U = p_\mu U^\mu = g_{\mu \nu} p^\mu U^\nu$.
Here $\teq$ is the relaxation time and the four-vector $U^\mu(x)$ is the hydrodynamic flow of matter. We assume that $\teq$ is independent of momentum and the same for all particle species. Moreover, in numerical calculations we take $\teq$ to be a constant, which makes all the computations more straightforward and significantly faster. 

The form of the collision term \rfn{colker} has simple physical interpretation: the effect of the collisions on the actual distribution function $f_\s(x,p)$ is to bring it closer to the equilibrium distribution $f_{\s, \eq}(x,p)$. The rate at which this process occurs is governed by the value of the relaxation time. The equilibrium distribution functions are defined through the values of the effective temperature and baryon chemical potential, which are chosen in such a~way that the equilibrium and actual distribution functions yield the same densities of the conserved currents. This concept, essential for our whole framework, will be discussed in more detail below. 

In general, kinetic equations of the form \rfn{kineq} may contain additional terms which describe the effects of interaction of particles with various mean fields (for example, electromagnetic or color fields). Such formulations of the kinetic theory are used, for example, in the context of the color-flux-tube model~\cite{Casher:1978wy,Bialas:1984wv,Czyz:1986mr,Bialas:1987en,Banerjee:1989by,Florkowski:2012ax,Ryblewski:2013eja,Ryblewski:2015psh}. In this work, the mean-field terms are not taken into account. 

We note that a constant value of $\teq$ explicitly breaks the conformal symmetry of the system. The other source of breaking of  conformality is a finite quark mass $m$. 
\linebreak
In the case of vanishing quark and gluon masses, and for zero baryon chemical potential, the conformal symmetry requires that the relaxation time scales inversely with the temperature, $\teq \sim 1/T$. This has been a common choice in other works that dealt with simpler, one-component systems at zero baryon density. 

The form of the flow four-velocity $U^\mu(x)$, required also to characterize the equilibrium distributions is defined by choosing the Landau hydrodynamic frame, see \rf{eq:LF}. We note, however,  that for one-dimensional boost-invariant systems studied here the structure of $U^\mu(x)$ follows directly from the symmetry arguments. This is discussed in the following section.

\section{Boost invariance and four-vector basis}
\label{sec:bitb}

Following many studies originating from the seminal work by Bjorken~\cite{Bjorken:1982qr}, we investigate here the systems which are boost-invariant with respect to the longitudinal ($z$) direction (that agrees with the beam direction) and transversally homogeneous. In this case, the expansion is described by the following form of the hydrodynamic flow
\begin{eqnarray}
 U^\mu &=& (t/\tau,0,0,z/\tau),
 \label{Ubinv} 
\end{eqnarray}
where $\tau$ is the (longitudinal) proper time
\begin{eqnarray}
\tau = \sqrt{t^2-z^2}.
\end{eqnarray}
One can easily check that $U \cdot U = 1$, thus $U$ is timelike (note the mostly-minus convention of the metric). With the help of the space-time rapidity
\bel{eta}
\eta = \f{1}{2} \ln \f{t+z}{t-z}
\eel
the four-vector $U^\mu$ can be written in the form
\begin{eqnarray}
U^\mu = (\cosh\eta, 0, 0, \sinh\eta).
 \label{Ubinv1} 
\end{eqnarray}
By making an active Lorentz boost along the $z$-axis with arbitrary rapidity $\Delta y$, the four-velocity changes to
\bel{UbinvY}
U^\mu = (\cosh(\eta+\Delta y), 0, 0, \sinh(\eta+\Delta y)) =  (\cosh\eta', 0, 0, \sinh\eta'), 
\eel
where $\eta' = \eta+\Delta y$. We thus see that the form of $U^\mu$ in the boosted (primed) frame is the same as in the original (unprimed) one. This property reflects the boost invariance of the expression~\rfn{Ubinv}.

For mathematical convenience, following
Refs.~\cite{Florkowski:2011jg,Tinti:2013vba}, in addition to $U^\mu$ we introduce three additional four-vectors ($X^\mu, Y^\mu$ and $Z^\mu$) defined by the equations:
\begin{eqnarray}
 X^\mu &=& (0,1,0,0),   \label{Xbinv} \\
 Y^\mu &=& (0,0,1,0),  \label{Ybinv}  \\
 Z^\mu &=& (z/\tau,0,0,t/\tau) =  (\sinh\eta, 0, 0, \cosh\eta). \label{Zbinv}
\end{eqnarray}
The four-vectors $X^\mu, Y^\mu$ and $Z^\mu$ are space-like and orthogonal to $U^\mu$ 
\begin{eqnarray}
X \cdot X &=&  Y \cdot Y \,\,=\,\, Z \cdot Z \,\,=\,\, -1, \\ \label{XXYYZZ}
X \cdot U &=& Y \cdot U \,\,=\,\, Z \cdot U \,\,=\,\, 0,  \\ \label{XYZU} 
X \cdot Y &=&  Y \cdot Z \,\,=\,\, Z \cdot X \,\,=\,\, 0.  \label{XYYZZX}
\end{eqnarray}
We note that at each spacetime point $x$ the four-vectors $U^\mu, X^\mu, Y^\mu$, and $Z^\mu$ form a vector basis.

\section{Equilibrium distributions}
\label{sec:eqdistr}

In \EQS{colker} the functions $f_{\s, \eq}(x,p)$ are equilibrium distribution functions, which take the Fermi-Dirac and Bose-Einstein forms for (anti)quarks and gluons, respectively,~\footnote{Unless we take the classical limit of quantum distributions and consider the Boltzmann statistics. This is formally achieved by the substitution $h^{\pm}_\eq(a) \to \exp(-a)$.}
\bea
f_{\Qpm, \eq}(x,p) &=& h^{+}_\eq\lp\f{  p\cdot U  \mp \mu }{T }\rp ,
\label{Qeq} \\
f_{\G, \eq}(x,p) &=& h^{-}_\eq\lp\f{p\cdot U}{T }\rp.
\label{Geq}
\eea
Here $T(x)$ is the effective temperature, $\mu(x)$ is the effective chemical potential of quarks, and the functions $h^{\pm}_\eq$ are defined by the formula 
\beal{feq}
h^{\pm}_\eq(a) &=&   \lsb\VP\exp(a) \pm 1 \rsb^{-1}.
\eeal
The same value of the temperature $T(x)$ appearing in Eqs.~(\ref{Qeq}) and (\ref{Geq}), as well as the same value of the chemical potential $\mu(x)$ appearing in the quark and antiquark distributions in Eq.~(\ref{Qeq}) introduce interactions (coupling) between quarks, antiquarks and gluons -- all particles evolve toward the same local equilibrium defined by $T(x)$ and $\mu(x)$.  Besides this effect, there is no other coupling between quarks, antiquarks and gluons. Since the baryon number of quarks is 1/3, we can use the relation 
\bel{muB}
\mu= \f{\mu_B}{3} ,
\eel
with $\mu_B$ being the baryon chemical potential.

\section{Anisotropic distributions}
\label{sec:adistr}
%
In the context of anisotropic hydrodynamics discussed below, it is useful to introduce the anisotropic Romatschke-Strickland (RS) phase-space distributions~\cite{Romatschke:2003ms}. Their covariant forms appropriate for the Bjorken expansion read~\cite{Florkowski:2012as}:
\begin{eqnarray}
f_{\Qpm, \an}(x,p) \!\!\!&=&\!\!\! h^{+}_\eq\lp\f{\sqrt{\lp p \cdot U\rp^2+\xi_\Q  \lp p \cdot Z\rp^2} \mp \lambda }{\Lambda_\Q}\rp,
\,\,\,\,  \label{Qa} \\
f_{\G, \an}(x,p) \!\!\!&=&\!\!\! h^{-}_\eq\lp\f{\sqrt{\lp p \cdot U\rp^2+\xi_\G  \lp p \cdot Z\rp^2}}{\Lambda_G}\rp.
\label{Ga}
\end{eqnarray}
Here $\xi_\Q(x)=\xi_\Qp(x)=\xi_\Qm(x)$ is the quark anisotropy parameter, $\Lambda_\Q(x)=\Lambda_\Qp(x)=\Lambda_\Qm(x)$ is the quark transverse-momentum scale, and $\lambda(x)$ is the non-equilibrium chemical potential of quarks. Similarly,
$\xi_G(x)$ is the gluon anisotropy parameter and $\Lambda_G(x)$ is the gluon transverse-momentum scale. We note that the equilibrium distributions may be considered as the limiting case of the RS distributions for $\xi \to 0$, where also $\lambda \to \mu$, and $\Lambda \to T$.

\begin{figure}[!t]
\begin{center}
\includegraphics[angle=0,width=0.475\textwidth]{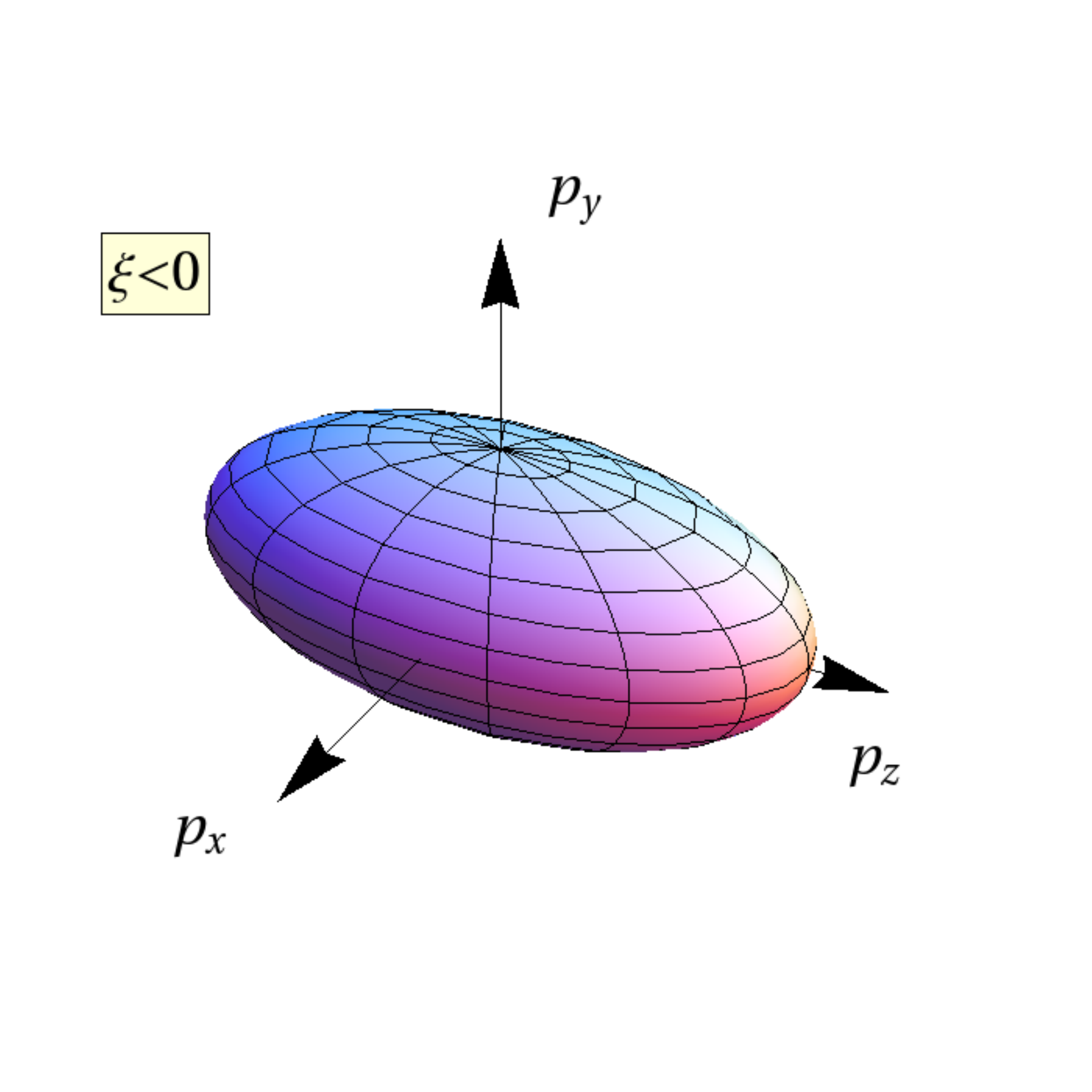} 
\includegraphics[angle=0,width=0.475\textwidth]{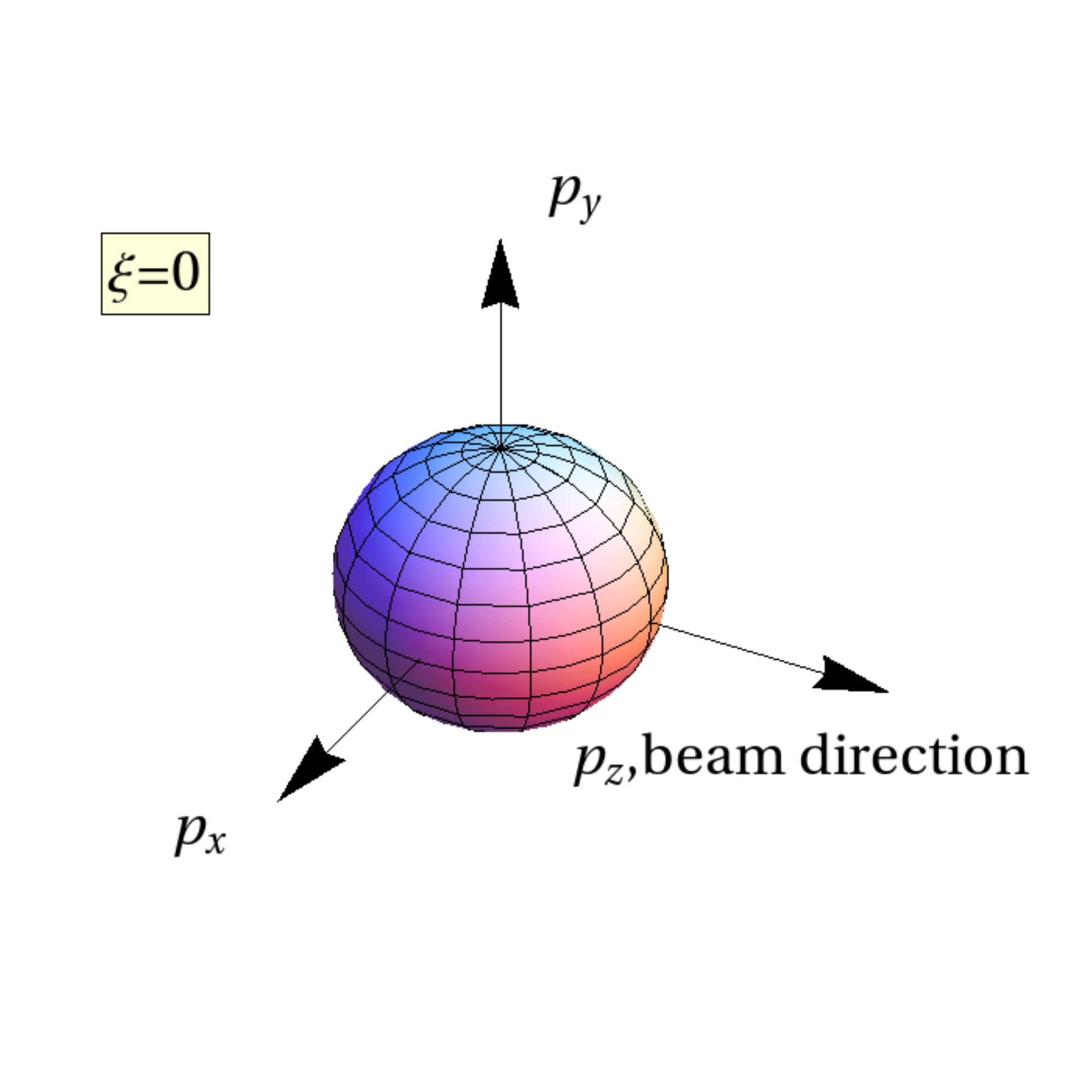} 
\includegraphics[angle=0,width=0.475\textwidth]{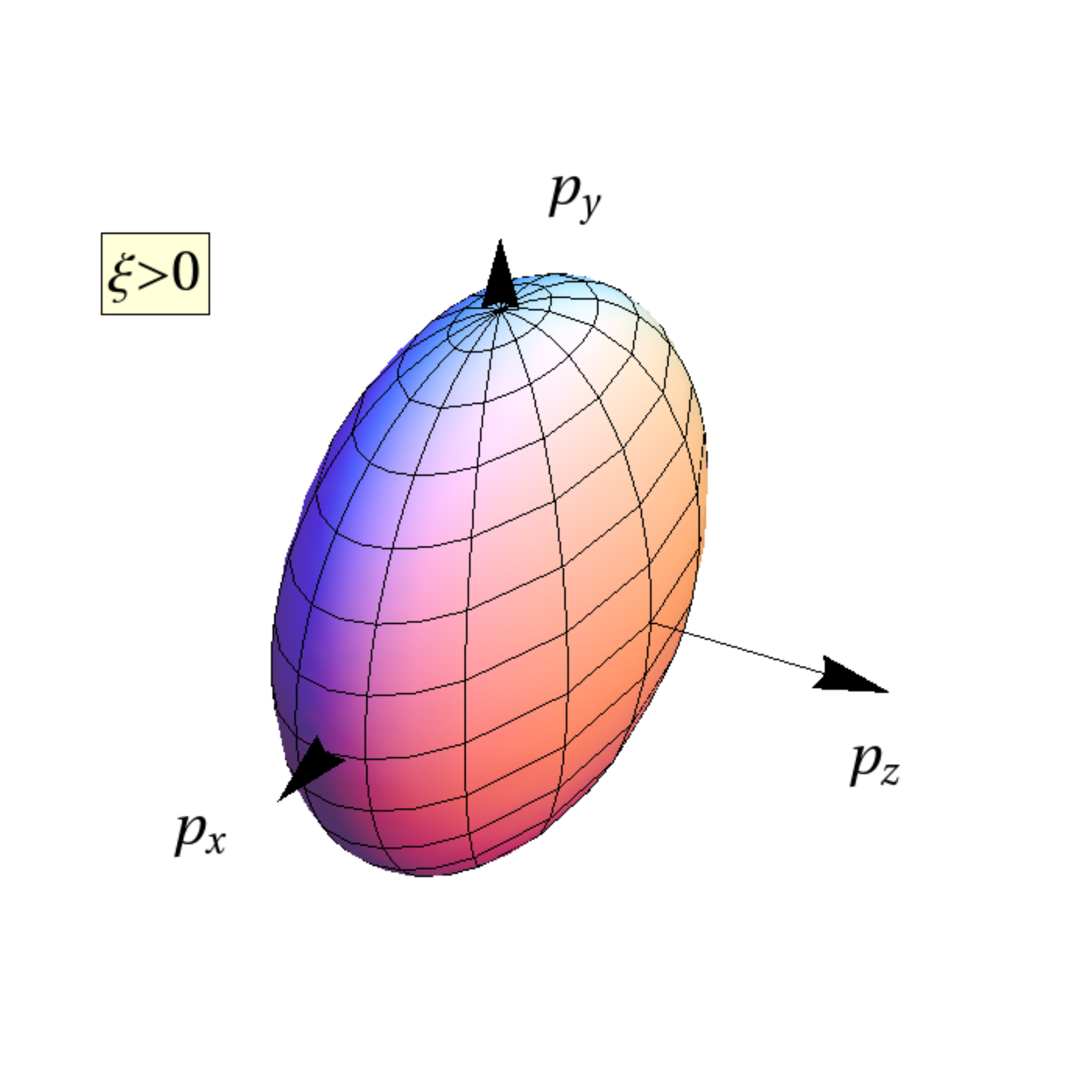} 
\end{center}
\caption{\small (Color online) Prolate (upper left part), isotropic (upper right part), and oblate (lower part) configurations of the distribution functions in momentum space.}
\label{fig:xi}
\end{figure}

The anisotropy parameters $\xi_{\rm s}$ vary in the range $-1 < \xi_{\rm s} <  \infty$, with the cases \mbox{$\xi_{\rm s} < 0$}, $\xi_{\rm s}=0$ and  $0 < \xi_{\rm s}$  corresponding to the prolate, isotropic and oblate momentum distribution, respectively, see Fig.~\ref{fig:xi}. We shall use the RS distributions also to define initial conditions for the kinetic equations that are solved exactly. The results of microscopic calculations suggest that the systems produced initially in heavy-ion collisions have transverse pressure much larger than the longitudinal one \cite{Baier:2000sb,Mrowczynski:2005ki,Lappi:2006fp,Heller:2011ju,vanderSchee:2013pia,Gelis:2013rba,Berges:2013eia}, which means that the initial distributions are most likely oblate.

%

\section{Moments of the kinetic equations}
\label{sect:motke}

For both the exact treatment of the kinetic equations \rfn{kineq} and construction of their approximation in the form of aHydro equations, it is necessary to consider moments of the kinetic equations in the momentum space. In this section, we clarify how such moments are defined. In the following section, we consider separately the zeroth and the first moments, as they are related to the baryon number and energy-momentum conservations. 

In our approach, all particles are assumed to be on the mass shell, \mbox{$p^2 =p \cdot p=m^2$}, hence we can use the momentum integration measure defined by Eq.~\rfn{eq:dP}. Gluons are always treated as massless particles, while quarks have a finite constant mass $m$.

We define the $n$-th moment  operator  (average) in the momentum space by the expression
\beal{mom}
\hI^{\mu_1\cdots\mu_n} (\dots) \equiv  \int \!dP\, p^{\mu_1}p^{\mu_2}\cdots p^{\mu_n} (\dots),  
\eeal
with the zeroth moment operator defined by
\beal{mom0}
\hI (\dots) \equiv  \int \!dP\,(\dots).
\eeal
Acting with $\hI^{\mu_1\cdots\mu_n}$ on the distribution functions $f_\s(x,p)$ and multiplying them by the degeneracy factors $k_\s$, one obtains the $n$-th moments of the distribution functions
\beal{momf}
\I_\s ^{\mu_1\cdots\mu_n} &\equiv& k_\s\, \hI^{\mu_1\cdots\mu_n} f_\s(x,p).
\eeal
Here $k_\s\equiv g_\s/(2\pi)^3$, with $g_{\Q}=g_{\Qpm}=3\times2\times N_f$ and $g_{G}=8\times2$ being the internal degeneracy factors for (anti)quarks  and gluons, respectively. In our calculations we assume that we deal with two (\emph{up} and \emph{down}) quark flavors  with equal masses, which reflects the SU(2) isospin symmetry. 

Using these definitions, the first, second, and third moments of the distribution functions read 
\bea 
N_\s^\mu(x)&\equiv&\I_\s ^{\mu}  =k_\s   \int \!dP\, p^{\mu} f_\s(x,p),
\label{ncurrs} \\
T_\s^{\mu\nu}(x)&\equiv&\I_\s ^{\mu\nu}  =   k_\s \int \!dP\, p^{\mu}p^{\nu} f_\s(x,p),
\label{emtensors}  \\
\Theta_\s^{\lambda\mu\nu}(x)  &\equiv&\I_\s ^{\lambda \mu\nu} =   k_\s \int \!dP\, p^{\lambda} p^{\mu}p^{\nu} f_\s(x,p).
\label{thetatensors} 
\eea
Equation \rfn{ncurrs} defines the \emph{particle number current}, while \rf{emtensors} defines the \emph{energy-momentum tensor} of the species ``$\s$'', respectively~\footnote{As we shall see below, the components of $\Theta_\s^{\lambda\mu\nu}$ have no direct physical interpretation but they are very useful for construction of hydrodynamic equations.}. In addition, we define the \emph{baryon number current}
\beal{bcurr}
B^\mu(x)&\equiv& \sum_\s q_\s\, N_\s^\mu(x) = \frac{k_{\Q}}{3}\int \!dP\, p^{\mu} \lsb\VP f_\Qp(x,p)-f_\Qm(x,p)\rsb, 
\eeal
where $q_\s=\left\{1/3,-1/3,0\right\}$ is the baryon number for quarks, antiquarks, and gluons, respectively. The total particle number current and total energy-momentum tensor read
\bea 
N^\mu(x)&=&\sum_\s N_\s^\mu(x),
\label{totncurr} \\
T^{\mu\nu}(x)&=& \sum_\s T_\s^{\mu\nu}(x).
\label{totemtensor} 
\eea

\medskip
We now consider the $n$-th moments of the kinetic equations (\ref{kineq}), which are obtained by acting with the operator $\hI^{\mu_1\cdots\mu_n}$ given by (\ref{mom}) on their left- and right-hand sides and multiplying them by the degeneracy factors $k_\s$. The zeroth and first moments have the form
\bea
k_\s\, \hI  p^\mu \p_\mu  f_\s(x,p) &=& k_\s\, \hI p^\mu U_\mu\frac{f_{\s, \eq}(x,p)-f_\s(x,p)}{\teq},
\label{KEzerothmom} \\
k_\s\, \hI^{\nu} p^\mu \p_\mu  f_\s(x,p) &=& k_\s\, \hI^{\nu} p^\mu U_\mu\frac{f_{\s,\eq}(x,p)-f_\s(x,p)}{\teq},\,\,\,\,\,\,
\label{KEfirstmom} 
\eea
which, with the help of Eqs.~(\ref{momf})--(\ref{emtensors}), may be rewritten as
\bea
\p_\mu {{N}}^\mu_\s  &=& U_\mu   \frac{{  N} ^\mu_{\s,\eq} -{  N} ^\mu_{\s }}{\teq},
\label{KEzerothmom2} \\
\p_\mu  {{T}}^{\mu\nu}_\s  &=& U_\mu   \frac{{ T}^{\mu\nu}_{\s,\eq} -{ T}^{\mu\nu}_{\s }}{\teq}.
\label{KEfirstmom2}  
\eea

\section{Landau matching conditions}
\label{sect:LMC}

Taking difference between $\s=\Qp$ and $\s=\Qm$ components of Eqs.~(\ref{KEzerothmom2}) we obtain  the baryon current evolution equation
\beal{KEzerothmom3a}
\p_\mu {{B}}^\mu   &=& U_\mu   \frac{{B}^\mu_\eq - {B}^\mu }{\teq} .
\eeal
Similarly, by taking the sum over $\s$ components of Eqs.~(\ref{KEfirstmom2}) and using Eq.~(\ref{totemtensor}) one gets the total energy and momentum conservation equation
\beal{KEfirsthmom3a}
\p_\mu  {{T}}^{\mu\nu}  &=& U_\mu   \frac{{T}^{\mu\nu}_\eq -{T}^{\mu\nu}_{ }}{\teq}.
\eeal

In order to have the baryon number conserved it is required that the left-hand side of Eq.~(\ref{KEzerothmom3a}) vanishes, $\p_\mu {{  B}}^\mu =0$. The latter implies vanishing of the right-hand side of Eq.~(\ref{KEzerothmom3a}), which leads to the {\it Landau matching condition for baryon current}
\beal{baryonLMC}
 U_\mu  {B}^\mu_\eq =  U_\mu { B} ^\mu  .
\eeal
Analogously, the energy and momentum conservation means that the left-hand side of Eq.~(\ref{KEfirsthmom3a}) vanishes, $\p_\mu  {{ T}}^{\mu\nu} = 0$. This condition results in vanishing of the right-hand side of Eq.~(\ref{KEfirsthmom3a}), which leads to the {\it Landau matching condition for energy and momentum} 
\beal{emLMC}
U_\mu  {T}^{\mu\nu}_\eq =U_\mu {T}^{\mu\nu}.
\eeal

Equations \rfn{baryonLMC} and \rfn{emLMC} have very natural physics interpretation. They tell us that the baryon number density ${\cal B}_{\rm eq}$ and the energy density ${\cal E}_{\rm eq}$ obtained with the reference equilibrium distribution functions $f_{\s,\eq}(x,p)$ should be the same as ${\cal B}$ and ${\cal E}$ obtained with the actual nonequilibrium distributions $f_{\s}(x,p)$ \footnote{In order to see how this comes out see Eqs.~(\ref{BIbaryonLMC}) and (\ref{energyLMC}) and the respective discussion.}. In a certain way, the Landau matching conditions define ``the closest'' equilibrium distributions to the actual ones --- the former should render the same densities of the conserved currents as the latter.  Using Eqs.~\rfn{baryonLMC} and \rfn{emLMC}, at any spacetime point $x$ we can determine effective temperature $T(x)$ and effective chemical potential $\mu(x)$, and use them to define the values of the equilibrium distribution functions. In this way, the kinetic equations \rfn{kineq} become a well-defined initial-value problem, which we aim to tackle in the following sections.

\section{Tensor decomposition}
\label{sect:tb}

We close this chapter with a section discussing algebraic structures of tensors used in our approach. Of special importance are the structures obtained for equilibrium and anisotropic RS distributions. 

\subsection{Basis vectors}
\label{sect:basis}
%
For further use, it is convenient to treat the basis introduced in Sec.~\ref{sec:bitb} in a more general way. To do so, we introduce the notation where  $A$, $B$, $C$ $\in  \lsb U,X,Y,Z \rsb$.
In the local rest frame (LRF) we have
\beal{eq:rfbasis}
  U^\mu_\LRF = (1,0,0,0), \nonumber \\
  X^\mu_\LRF = (0,1,0,0), \nonumber \\
 Y^\mu_\LRF = (0,0,1,0), \nonumber \\
 Z^\mu_\LRF = (0,0,0,1) \, .
\eeal
Using \EQS{eq:rfbasis} one may express the metric tensor as  \cite{Martinez:2012tu}
\begin{equation}
g^{\mu \nu}=U^\mu  U^\nu  - \sum_{A\neq U} A^\mu  A^\nu  \, .
\label{eq:gbasis}
\end{equation}
The projector on the space orthogonal to the four-velocity, $\Delta^{\mu\nu}\equiv g^{\mu\nu} -U^\mu U^\nu$, then takes the form
\begin{equation}
\Delta^{\mu \nu} =  - \sum_{A\neq U} A^\mu  A^\nu \, ,
\label{eq:transproj}
\end{equation}
and satisfies the conditions $U_{\mu} \Delta^{\mu\nu} = 0$, $\Delta^{\mu}_{\,\,\,\alpha}\Delta^{\alpha\nu}=\Delta^{\mu\nu}$ and $\Delta^{\mu}_{\,\,\,\mu}=3$.   
The basis (\ref{eq:rfbasis}) is a unit one in the sense that 
\begin{eqnarray}
A \cdot B =
\begin{cases}
\hphantom{-}0   & \hbox{for   } A\neq B, \\
\hphantom{-}1   & \hbox{for   } A= B=U,    \\
-1  & \hbox{for   } A= B\neq U, 
\end{cases}
\label{eq:orthonormal}
\end{eqnarray}
and complete so that any four-vector may be decomposed in the basis $\{A\}$. In particular, one may express the particle number flux as follows
\beal{eq:nfdecomp}
N_\s^\mu(x) &=& \sum_A  n_{A}^{\s} A_{ }^\mu, 
\eeal
where the coefficients $n_{A}^{\s}$, due to Eqs.~(\ref{eq:orthonormal}), are given by the projections
\beal{eq:nfdecompcoef}
n_{A}^{\s} &=&   A_{\mu}^{\,}  N_\s^\mu(x)  \, A^2, 
\eeal
with $A^2 = A \cdot A$ (note that $A^2=-1$ for space-like four-vectors of the basis (\ref{eq:rfbasis})). The tensorial basis for the rank-two tensors is constructed using tensor products of the basis four-vectors $\{A \otimes B\}$. Thus the decomposition of the energy-momentum tensor takes the form
\beal{eq:emtdecomp}
T_\s^{\mu\nu}(x) &=& \sum_{A,B} t_{AB}^{\s} A^\mu B^\nu,
\eeal
with the components of $T_\s^{\mu\nu}(x)$ defined as
\beal{eq:emtdecompcoef}
 t_{AB}^{\s} &=& A_{\mu}^{\,} B_{\nu}^{\,} T_\s^{\mu\nu}(x) \, A^2  B^2.
\eeal
Following similar methodology one can construct the tensorial basis for the rank-three tensors allowing to decompose $\Theta_\s^{\mu\nu\lambda}$ tensor as follows
\beal{eq:thetadecomp}
\Theta_\s^{\mu\nu\lambda}(x) &=& \sum_{A,B,C} c_{ABC}^{\s}   A^\mu B^\nu C^\lambda,
\eeal
where the coefficients $c_{ABC}^{\s}$ are defined through the expression 
\beal{eq:thetadecompcoef}
 c_{ABC}^{\s} &=&  A_{\mu}^{\,} B_{\nu}^{\,} C_{\lambda}^{\,} \Theta_\s^{\mu\nu\lambda}(x) \, A^2  B^2 C^2.
\eeal
Using Eqs.~(\ref{ncurrs})--(\ref{thetatensors})  in Eqs.~(\ref{eq:nfdecompcoef}), (\ref{eq:emtdecompcoef}) and (\ref{eq:thetadecompcoef}) one gets
\bea 
n_{A}^{\s} &=&  k_\s   \int \!dP\, \lp p \cdot A  \rp A^2 f_\s(x,p)   ,
\label{eq:nfdecompcoefgen}\\
t_{AB}^{\s} &=& k_\s\int \!dP\,  \lp p \cdot A \rp \lp p \cdot B \rp A^2 B^2 f_\s(x,p) ,
\label{eq:emtdecompcoefgen} \\
c_{ABC}^{\s} &=& k_\s\int \!dP\,  \lp p \cdot A \rp \lp p \cdot B \rp \lp p \cdot C \rp A^2 B^2 C^2 f_\s(x,p).
\label{eq:thetadecompcoefgen}
\eeal
%

\subsection{Equilibrium densities}
\label{sect:eqden}
%
In the case of equilibrium distribution functions $f_{\s, \eq}(x,p) =  f_{\s, \eq}(p\cdot U(x))$, as defined by Eqs.~(\ref{Qeq}) and (\ref{Geq}), which, due to momentum isotropy, are invariant with respect to $SO(3)$ rotations  in the three-momentum space, by the symmetry  of the integrands  in Eqs.~(\ref{eq:nfdecompcoefgen})--(\ref{eq:emtdecompcoefgen}) one has
\begin{eqnarray}
 n_{A}^{\s, \eq} & = &  k_\s   \int \!dP\, \lp p\cdot A \rp A^2 f_{\s, \eq}   =0 \quad \hbox{if     } A\neq U,
\label{isonfcoefgen}\\
t_{AB}^{\s, \eq} & = & k_\s\int \!dP\, \lp p\cdot A \rp \lp p\cdot B \rp A^2 B^2 f_{\s, \eq} =0  \quad \hbox{if     } A\neq B.
\label{isoemtcoefgen}
\end{eqnarray}
Hence for the momentum-isotropic state Eqs.~(\ref{eq:nfdecomp}) and (\ref{eq:emtdecomp}) have the following structure
\begin{eqnarray} 
N_{\s, \eq}^\mu(x)&=&  {\cal N}^{\s, \eq} U^\mu, \label{ncurreq} \\
T_{\s, \eq}^{\mu\nu}(x)&=& {\cal E}^{\s, \eq} U^\mu U^\nu -{\cal P}^{\s, \eq} \Delta^{\mu\nu},
\label{emtensoreq} 
\end{eqnarray} 
with 
\begin{equation}
{\cal N}^{\s, \eq} = n_{U}^{\s, \eq}, \quad {\cal E}^{\s, \eq} = t_{UU}^{\s, \eq}, 
\quad {\cal P}^{\s, \eq} = t_{XX}^{\s, \eq} = t_{YY}^{\s, \eq} = t_{ZZ}^{\s, \eq},
\label{eqvars}
\end{equation}
being the particle density, energy density, and pressure in equilibrium, respectively. For a~certain particle species ,,s'', Eqs.~\rfn{ncurreq}--\rfn{emtensoreq} define the energy-momentum tensor and the particle number flux of the  perfect fluid, see Sec.~\ref{sec:pfvf}.  Explicit forms of these expressions are given in \APP{ss:ie} that uses results of \APP{ss:ae}. Analogous calculation for the second moment \rfn{thetatensors}, with the equilibrium distributions~\rfn{Qeq} and \rfn{Geq}, gives 
\begin{eqnarray}  
\Theta_{\s, \eq}^{\mu\nu}(x)&=& \vartheta_U^{\s, \eq}\,  U^{\lambda} U^{\mu} U^{\nu}     \,-\,   \vartheta ^{\s, \eq}\, \left( U^{\lambda} \Delta^{\mu\nu} + U^{\mu} \Delta^{\lambda\nu}+    U^{\nu}\Delta ^{\lambda \mu}\right),
 \label{thetatensoreq}
\end{eqnarray} 
where $\vartheta_U^{\s, \eq} = c_{UUU}^{\s, \eq}$, and $
\vartheta^{\s, \eq} = c_{UXX}^{\s, \eq}=c_{UYY}^{\s, \eq} = c_{UZZ}^{\s, \eq}$. The explicit expressions for variables $\vartheta$ are presented in App.~\ref{s:sm}.

\subsection{Anisotropic densities}
\label{sec:aden}

The distributions defined by Eqs.~(\ref{Qa}) and (\ref{Ga}) are invariant only with respect to $SO(2)$ rotations around the $z$ direction in the three-momentum space. In this case one still has 
\begin{eqnarray}
 n_{A}^{\s, \an} &=&  0 \quad \hbox{if     } A\neq U,
\label{anisonfcoefgen}\\
t_{AB}^{\s, \an} &=& 0  \quad \hbox{if     } A\neq B,
\label{anisoemtcoefgen}
\end{eqnarray}
however, in this case, Eqs.~(\ref{ncurrs}) and (\ref{emtensors}) have the following structure~\cite{Florkowski:2008ag}
\begin{eqnarray} 
N_{\s,\an}^\mu(x)&=&  {\cal N}^{\s,\an} U^\mu, \label{ncurran} \\
T_{\s,\an}^{\mu\nu}(x)&=& {\cal E}^{\s,\an} U^\mu U^\nu -{\cal P}^{\s,\an}_T \Delta_T^{\mu\nu} +{\cal P}^{\s,\an}_L Z^\mu Z^\nu,
\label{emtensoran} 
\end{eqnarray} 
with 
\beal{anisovars}
{\cal N}^{\s,\an} &=& n_{U}^{\s, \an}, \quad {\cal E}^{\s,\an} =  t_{UU}^{\s, \an},  
\quad {\cal P}^{\s,\an}_T = t_{XX}^{\s, \an}= t_{YY}^{\s, \an}, \quad {\cal P}^{\s,\an}_L = t_{ZZ}^{\s, \an} .
\eeal
Here $\Delta_T^{\mu\nu} = -\left(X^\mu X^\nu + Y^\mu Y^\nu \right)$ is the projection operator on the direction orthogonal to $U$ and $Z$. Explicit forms of \EQS{anisovars} are given in \APP{ss:ae}. For the anisotropic distributions~\rfn{Qa} and \rfn{Ga} we also find that
\bea
\Theta_{\s, \an}^{\lambda\mu\nu} &=& \vartheta_U^{\s, \an}\,  U^{\lambda} U^{\mu} U^{\nu} \nonumber   \\ \,&-&\,   \vartheta_T^{\s, \an}\, \left( U^{\lambda} \Delta^{\mu\nu}_T + U^{\mu} \Delta^{\lambda\nu}_T+    U^{\nu}\Delta ^{\lambda \mu}_T\right)\nonumber\\
\,&+&\,  \vartheta_L^{\s, \an}\, \left( U^{\lambda}Z^{\mu} Z^{\nu} + U^{\mu}Z^{\lambda} Z^{\nu}+U^{\nu} Z^{\lambda} Z^{\mu} \right),
 \label{thetatensoran}
\eea 
where $\vartheta_U^{\s, \an} = c_{UUU}^{\s, \an}$, $
\vartheta_T^{\s, \an} = c_{UXX}^{\s, \an}=c_{UYY}^{\s, \an}$, and $ 
\vartheta_L^{\s, \an} = c_{UZZ}^{\s, \an}$.
For the explicit expressions for variables $\vartheta$ see App.~\ref{s:sm}.
%
%
%
%
%
%
\chapter{Exact solutions of kinetic equations}
\label{sect:e}
%
In order to solve the Landau matching conditions~(\ref{baryonLMC}) and (\ref{emLMC}) we need to know the form of the distribution functions $f_{\s}(x,p)$ that solve the kinetic equations~(\ref{kineq}). In~general, such solutions are difficult to find, and (\ref{kineq}) may be at best solved numerically. However, it is possible to find \textit{formal} analytic solutions of Eqs.~(\ref{kineq}) in the case where the system is boost invariant and transversally homogeneous. Below, we discuss this case in more detail.

\section{Białas-Czyż boost-invariant variables \texorpdfstring{$w$}{w} and \texorpdfstring{$v$}{v}}
\label{sect:wv}
%
In the case of the one-dimensional system exhibiting symmetries discussed above it is convenient to use the variables $w$ and $v$ which are defined as follows  \cite{Bialas:1984wv,Bialas:1987en}
\bea 
w & =& t \pL - z E_p =  - \, \tau \, p \cdot Z, \label{w}\\
v  &=& t  E_p - z \pL = \tau \, p \cdot U.  \label{v}
\eea 
We note that for $z=0$ the variable $w$ reduces to the longitudinal momentum multiplied by the time coordinate, whereas the $v$ variable is reduced to the energy multiplied by $t$.

Since particles are on the mass shell, the variables $w$ and $v$ are related by the formula
\beal{v1}
v\twpt &=&   \sqrt{w^2+\lp m^2+\pT^{\,2}\rp  \tau^2}.  
\eeal
Equations (\ref{w}) and (\ref{v}) can be inverted to express the energy and longitudinal momentum of a particle in terms of $w$ and $v$, namely
\begin{equation}   
E_p= \f{vt+wz}{\tau^2},\quad \pL=\f{wt+vz}{\tau^2}.  
\label{p0p3}
\end{equation}
The  Lorentz invariant momentum-integration measure can be then written  as
\begin{eqnarray}
dP = \f{d^3p}{E_p} = \f{dw\,d^2\pT}{v} .
\label{dP}
\end{eqnarray}
For boost-invariant systems, all scalar functions of space and time, such as the effective temperature $T$ and quark chemical potential $\mu$ may depend only on $\tau$. In addition, one can check that the phase-space distribution functions, which are Lorentz scalars, may depend only on the variables $\tau$, $w$, and $\bpT$. We use these properties in the next section.
%
\section{Formal solutions of the kinetic equations}
\label{sect:boost-inv-eq}
%
With the help of the boost-invariant variables $w$, $v$ and $\bpT$ we can rewrite Eqs.~(\ref{kineq})  in a~simple
form~\cite{Florkowski:2013lza,Florkowski:2013lya,Baym:1984np,Baym:1985tna}
\beal{BIke}
\frac{\p f_\s \left(\tau, w,  \bpT\right)}{\p\tau}   &=& \f{f_{\s,\eq}\left(\tau, w, \pT\right)-f_\s \left(\tau, w,  \bpT\right)}{\teq},  
\eeal 
where the boost-invariant versions of the equilibrium distribution functions are straightforward to find using  (\ref{w}) and (\ref{v}), namely 
\begin{eqnarray}
f_{\Qpm, \eq} \twpt &=&h^{+}_\eq \lp\f{\sqrt{\lp\f{w}{\tau}\rp^2+ \pT^2 + m^2  }\mp \mu}{T } \,  \rp  , 
\label{BIQeq0} \\
 f_{\G, \eq} \twpt &=&
 h^{-}_\eq\lp\f{\sqrt{\lp\f{w}{\tau}\rp^2+\pT^2}}{T } \,  \rp . 
\label{BIGeq0}
\end{eqnarray}
Below we assume that distribution functions $f_\s\left(\tau, w,  \bpT\right)$ are even functions of $w$, and depend only on the magnitude of $\bpT$, \footnote{In our analysis we restrict ourselves to the   initial distributions in the RS form, which are  SO(2) invariant in transverse momentum space and depend only on the magnitude of $\bpT$. }
\begin{eqnarray}
f_\s\twpt = f_\s(\tau,-w,\pT)  .
\label{symoff}
\end{eqnarray}
The formal solutions of \EQS{BIke} have the following form~\cite{Florkowski:2013lza,Florkowski:2013lya,Baym:1984np,Baym:1985tna}
\begin{eqnarray}
f_\s\twpt &=& D(\tau,\tau_0) f_\s^0(w,\pT) + \int\limits_{\tau_0}^{\tau} \f{d \tau'}{\teq^\prime}\  D(\tau,\tau') 
f_{\s,\eq}(\tau^\prime,w,p_T), \label{formsolQ}  
\end{eqnarray}
where $f_\s^0(w,p_T)\equiv f_\s (\tau_0,w,p_T)$ is the initial distribution function --- we have introduced here the notation
$\teq^\prime = \teq(\tau^\prime)$ for the general case where the equilibration time may depend on the proper time.

Perhaps at this stage, it is worth to clarify what we mean by the exact solutions of the kinetic equations. The \emph{exact solutions} are high-accuracy numerical solutions of \rf{formsolQ}. They are usually obtained by the iterative method described in more detail below. Hence, the exact solutions are obtained with the help of combined analytic and numerical methods. 

\subsection{Damping function}
%
In \EQ{formsolQ} we have introduced the damping function
\begin{equation}
D(\tau_2,\tau_1)=\mathrm{exp}\Bigg[ -\int\limits_{\tau_1}^{\tau_2} \f{d\tau''}{\teq(\tau'')}\Bigg].
\label{Damp1}
\end{equation}
The function $D(\tau_2,\tau_1)$ satisfies the two differential relations:
\begin{eqnarray}
\f{\p D(\tau_2,\tau_1)}{\p \tau_2}  = - \f{D(\tau_2,\tau_1)}{\teq(\tau_2)}, \quad
\f{\p D(\tau_2,\tau_1)}{\p \tau_1}  =  \f{D(\tau_2,\tau_1)}{\teq(\tau_1)},
\label{Damp2}
\end{eqnarray}
and converges to unity if the two arguments are the same, 
\begin{eqnarray}
D(\tau,\tau)=1. 
\label{Damp21}
\end{eqnarray}
These properties imply the identity~\cite{Florkowski:2014txa}
\begin{eqnarray}
1 = D(\tau,\tau_0) + \int\limits_{\tau_0}^\tau \f{d\tau^\prime}{\teq(\tau^\prime)}
D(\tau,\tau^\prime).
\label{Damp3}
\end{eqnarray}
For a constant relaxation time used in this work Eq.~(\ref{Damp1}) reduces to a simple exponential function
\begin{equation}
D(\tau_2,\tau_1)=\exp\lsb-\f{\tau_2-\tau_1}{\teq}\rsb .
\label{Damp2a}
\end{equation} 

\subsection{Initial distributions}

In what follows we assume that the initial distributions $f_\s^0(w,p_T)$ are given by the anisotropic RS forms $f_{\s,\an}^0(w,p_T)$ which follow from Eqs.~(\ref{Qa}) and (\ref{Ga}),
\begin{eqnarray}
&& f_{\Qpm, \an} \tiwpt = 
h^{+}_\eq \lp\f{\sqrt{\lp1+\xi_\Q^0\rp\lp\frac{w}{\tau_0}\rp^2+  m^2+\pT^2   } \mp \lambda ^0 }{\Lambda_\Q^0}   \rp  ,
\label{BIQeq} \\
&& f_{\G, \an} \tiwpt = h^{-}_\eq\lp\f{\sqrt{\lp1+\xi_\G^0\rp\lp\frac{w}{\tau_0}\rp^2+\pT^2   }}{\Lambda_\G^0} \,  \rp . \nn 
\label{BIGeq}
\end{eqnarray}
Here $\xi_\s^0\equiv\xi_\s(\tau_0)$, $\Lambda_\s^0\equiv\Lambda_\s(\tau_0)$,  and $\lambda^0\equiv\lambda(\tau_0)$ are initial parameters.

One may ask a question how much restrictive our assumption about the initial conditions is. By looking at the formal solution of the kinetic equation \rfn{formsolQ} and the form of the damping function \rfn{Damp2a} we conclude that the contribution from the initial condition is exponentially damped. Hence, the initial conditions seem to have little impact on the equilibration process, although they may affect the early hydrodynamization stage. In~order to check such effects we perform calculations for various initial anisotropies. Below we shall consider three cases for quark and gluon initial distributions: oblate-oblate, prolate-oblate, and prolate-prolate (respectively for quarks and gluons); see Section ~\ref{sec:adistr}.
\subsection{Moments of the exact solution and their tensor structure}
%
Knowledge about the forms of the exact distribution functions \rfn{formsolQ} allows us to calculate their moments using \EQSM{ncurrs}{thetatensors} and their tensor decompositions using \EQSM{eq:nfdecomp}{eq:thetadecompcoefgen}. The symmetry of the momentum integrals of \rfn{formsolQ}, in our case, is dictated by the $SO(2)$ symmetry in momentum space of the anisotropic RS forms of the initial distribution functions (\ref{BIQeq})--(\ref{BIGeq}). As a result  the exact energy-momentum tensor and particle number flux have the same tensor structure as the ones for the anisotropic RS distributions (\ref{ncurran})--(\ref{emtensoran}), namely
\begin{eqnarray} 
N_{\s}^\mu(x)&=&  {\cal N}^{\s} U^\mu, \label{ncurrex} \\
T_{\s}^{\mu\nu}(x)&=& {\cal E}^{\s} U^\mu U^\nu -{\cal P}^{\s}_T \Delta_T^{\mu\nu} +{\cal P}^{\s}_L Z^\mu Z^\nu,
\label{emtensorex} 
\end{eqnarray} 
where the thermodynamic quantities
\beal{exactvars}
{\cal N}^{\s} &=& n_{U}^{\s}, \quad {\cal E}^{\s} =  t_{UU}^{\s},  
\quad {\cal P}^{\s}_T = t_{XX}^{\s}= t_{YY}^{\s}, \quad {\cal P}^{\s}_L = t_{ZZ}^{\s},
\eeal
are defined in the App.~\ref{ss:esotke}.
%
\section{Baryon number conservation}
\label{sect:BNcon}

We turn now to a discussion of the conservation laws for baryon number, energy, and momentum. This is a crucial element of our approach. Using the expression for the baryon number current (\ref{bcurr}) and decompositions  (\ref{ncurreq}) and (\ref{ncurrex}) one may rewrite \EQ{baryonLMC} as 
\begin{equation}
{\cal B}^\eq = {\cal B},
\label{BIbaryonLMC}
\end{equation} 
where we define the equilibrium and exact baryon number densities as
\begin{equation}
{\cal B}^\eq = \f{1}{3} \left( {\cal N}^{\Qp, \eq} -{\cal N}^{\Qm, \eq} \right), 
\qquad
{\cal B}  = \f{1}{3} \left( {\cal N}^{\Qp } -{\cal N}^\Qm \right),
\label{BIbaryonBdensity}
\end{equation}
respectively.
The explicit formula for ${\cal B}^\eq(\tau)$ is derived in \APP{ss:ie}, see Eq.~(\ref{BeqApp}),
\begin{equation}
{\cal B}^\eq(\tau) = \f{16 \pi k_\Q T^3}{3} \sinh\lp\frac{\mu}{T}\rp\,{\cal H}_{\cal B}\lp \frac{m}{T},   \frac{\mu}{T}\rp,
\label{FIRST-EQUATION-0}
\end{equation}
where the function ${\cal H}_{\cal B}$ is defined by Eq.~(\ref{HB}). The formula for ${\cal B}(\tau)$ is more complicated and is given in \APP{ss:esotke}, see Eq.~(\ref{BApp}). It contains an integral over the time history of the functions $T' \equiv T(\tau')$ and $\mu'\equiv\mu(\tau')$ in the range $\tau_0 \leq \tau' \leq \tau$. Consequently, Eq.~(\ref{BIbaryonLMC}) becomes
an integral equation 
\begin{eqnarray}
T^3 \sinh\lp\frac{\mu}{T}\rp\,{\cal H}_{\cal B}\lp \frac{m}{T},   \frac{\mu}{T}\rp  
&=&   \f{\tau_0 \lp\Lambda_\Q^0\rp^3}{\tau \sqrt{1+\xi_\Q^0}} \sinh\lp\frac{\lambda^0}{\Lambda_\Q^0}\rp\,{\cal H}_{\cal B}\lp \frac{m}{\Lambda_\Q^0},\frac{\lambda^0}{\Lambda_\Q^0} \rp D(\tau,\tau_0) \nn \\
&& \hspace{-1.5cm} + \int\limits_{\tau_0}^{\tau} \f{d \tau'}{\teq^\prime}\  D(\tau,\tau')  \f{\tau^\prime \lp T^\prime\rp^3}{\tau  } \sinh\lp\frac{\mu^\prime}{T^\prime}\rp\,{\cal H}_{\cal B}\lp \frac{m}{T^\prime}, \frac{\mu^\prime}{T^\prime}\rp .
\label{FIRST-EQUATION}
\end{eqnarray} 
 Equation (\ref{FIRST-EQUATION}) is a single equation for two functions, $T(\tau)$ and $\mu(\tau)$. The second necessary equation required 
 for their determination is obtained from the Landau matching condition for the energy, which we discuss in the next section.

Meanwhile, it is interesting to notice that Eq.~(\ref{FIRST-EQUATION}) can be rewritten as an integral equation for the function ${\cal B}(\tau)$, namely
\begin{eqnarray}
{\cal B}(\tau) = \f{\tau_0}{\tau} {\cal B}(\tau_0) D(\tau,\tau_0)
+ \int\limits_{\tau_0}^{\tau} \f{d \tau'}{\teq^\prime} \f{\tau'}{\tau}  {\cal B}(\tau') \, D(\tau,\tau').
\label{FIRST-EQUATION-1}
\end{eqnarray}
By differentiating (\ref{FIRST-EQUATION-1}) with respect to $\tau$ we get
\begin{equation}
\f{d{\cal B}(\tau)}{d\tau} +\f{{\cal B}(\tau)}{\tau} = 0,
\label{FIRST-EQUATION-2}
\end{equation}
which is nothing else but the form of baryon number conservation law valid for the Bjorken geometry (in the original Bjorken paper \cite{Bjorken:1982qr}
the same equation was obtained for the conserved entropy current). Equation~(\ref{FIRST-EQUATION-2}) has a scaling solution 
\begin{equation}
{\cal B}(\tau) = \f{\tau_0}{\tau} \, {\cal B}(\tau_0).
\label{FIRST-EQUATION-3}
\end{equation}
Combining   (\ref{BIbaryonLMC}) and (\ref{FIRST-EQUATION-0}) with (\ref{FIRST-EQUATION-3}) we find the equation
\begin{equation}
\sinh\lp\frac{\mu}{T}\rp\,{\cal H}_{\cal B}\lp \frac{m}{T},   \frac{\mu}{T}\rp 
= \f{3 \tau_0 {\cal B}(\tau_0)}{16 \pi k_\Q \tau T^3},
\label{FIRST-EQUATION-4}
\end{equation}
which allows determining $\mu$ in terms of $T$ and $\tau$ for a given initial baryon number density. Unfortunately, in the general case we study (Fermi-Dirac statistics for quarks), Eq.~(\ref{FIRST-EQUATION-4}) is an implicit equation for $\mu$. The situation simplifies in the case of classical statistics, where the function ${\cal H}_{\cal B}$ becomes independent of $\mu$. In this case, \rf{FIRST-EQUATION-4} can be used directly to determine $\mu$.

\section{Energy-momentum conservation}

Using the expression for the energy-momentum tensor (\ref{totemtensor}) and   decompositions (\ref{emtensoreq}) and  (\ref{emtensorex})  one may rewrite \EQ{emLMC} as 
\begin{equation}
{\cal E}^\eq = {\cal E},
\label{energyLMC}
\end{equation} 
where ${\cal E}^\eq$ and ${\cal E}$ contain contributions from quarks, antiquarks, and gluons
\begin{equation}
{\cal E}^\eq =  {\cal E}^{\Qp, \eq} + {\cal E}^{\Qm, \eq}  + {\cal E}^{G, \eq} ,
\end{equation}
\begin{equation}
{\cal E}  = {\cal E}^{\Qp } +{\cal E}^\Qm  + {\cal E}^G.
\end{equation}
Using Eqs.~(\ref{eq:eqeq}), (\ref{eq:egeq}), (\ref{eq:eq}), and (\ref{eq:eg}) we obtain
\bigskip
\begin{eqnarray}
&&  T^4 \lsb \tilde{{\cal H}}^+\lp 1, \frac{m}{T}, - \frac{\mu}{T}\rp +
 \tilde{{\cal H}}^+\lp 1, \frac{m}{T}, + \frac{\mu}{T}\rp  + r  \tilde{{\cal H}}^-\lp 1, 0,0\rp \rsb \nn \\
&&  =     \lp\Lambda_\Q^0\rp^4 
\lsb \tilde{{\cal H}}^+\lp \f{\tau_0}{\tau \sqrt{1+\xi_\Q^0}}, \frac{m}{\Lambda_\Q^0}, - \frac{\lambda^0}{\Lambda_\Q^0}\rp 
+ \tilde{{\cal H}}^+\lp \f{\tau_0}{\tau \sqrt{1+\xi_\Q^0}}, \frac{m}{\Lambda_\Q^0}, + \frac{\lambda^0}{\Lambda_\Q^0}\rp \rsb
D(\tau,\tau_0) \nn \\
&& +    \int\limits_{\tau_0}^{\tau} \f{d \tau'}{\teq^\prime}\  D(\tau,\tau') \lp T^\prime\rp^4
\lsb \tilde{{\cal H}}_{\cal  }^+\lp \f{\tau^\prime}{\tau  }, \frac{m}{T^\prime}, - \frac{\mu^\prime}{T^\prime}\rp  +
\tilde{{\cal H}}_{\cal  }^+\lp \f{\tau^\prime}{\tau  }, \frac{m}{T^\prime}, + \frac{\mu^\prime}{T^\prime}\rp   \rsb
\label{SECOND-EQUATION} \\
&& +   r  \lsb  \lp\Lambda_\G^0\rp^4 \tilde{{\cal H}}^-\lp \f{\tau_0}{\tau \sqrt{1+\xi_\Q^0}},  0,0\rp D(\tau,\tau_0) 
+   \int\limits_{\tau_0}^{\tau} \f{d \tau'}{\teq^\prime}\  D(\tau,\tau') \lp T^\prime\rp^4 \tilde{{\cal H}}_{\cal  }^-\lp \f{\tau^\prime}{\tau  }, 0,0\rp\rsb, \nn
\end{eqnarray}
where the functions $\tilde{{\cal H}}^\pm$ are defined by Eqs.~(\ref{Ht}) and $r$ is the ratio of the degeneracy factors
\begin{equation}
r = \f{k_\G }{k_\Q} = \f{g_\G }{g_\Q} = \f{4}{3}.
\end{equation}

Equations (\ref{FIRST-EQUATION}) and (\ref{SECOND-EQUATION}) are two integral equations that are sufficient to determine the proper-time dependence of the functions $T(\tau)$ and $\mu(\tau)$. This is done usually by the iterative method~\cite{Banerjee:1989by}. The two initial, to large extent arbitrary input functions  $T_{\rm in}(\tau)$ and $\mu_{\rm in}(\tau)$ are used on the right-hand sides of (\ref{FIRST-EQUATION}) and (\ref{SECOND-EQUATION}) and the new values  $T_{\rm out}(\tau)$ and $\mu_{\rm out}(\tau)$ are calculated from the left-hand sides. In the next iteration the new values are used as $T_{\rm in}(\tau)$ and $\mu_{\rm in}(\tau)$ on the right-hand sides to get updated values of  $T_{\rm out}(\tau)$ and $\mu_{\rm out}(\tau)$. Such procedure is repeated until the updated values agree well with the initial values. We have found that the results converge with about 50 iterations if the final proper time is 5.0 fm. The time of the calculations grows quadratically with the final proper time. 

Our use of the two coupled integral equations is similar to the case studied previously in~\cite{Florkowski:2016qig}. We find that it is more straightforward than using (\ref{SECOND-EQUATION}) together with (\ref{FIRST-EQUATION-4}). However, the situation is different in the case of classical statistics, where (\ref{FIRST-EQUATION-4}) can be used to determine $\mu/T$ analytically. In this case, the expression for $\mu/T$ obtained from (\ref{FIRST-EQUATION-4}) may be substituted into (\ref{SECOND-EQUATION}) and we are left with a single integral equation for the function $T(\tau)$.

One may check, using \rfn{dH2} and \rfn{dtH2}, that \rf{SECOND-EQUATION} is consistent with the formula
\bel{SECOND-EQUATION-1}
\f{d{\cal E}  }{d\tau} = -\f{{\cal E}  + {\cal P}_L  }{\tau},
\eel
where ${\cal P}_L  = {\cal P}_L^{\Qp } +{\cal P}_L^\Qm  + {\cal P}_L^G$ is the total longitudinal pressure of the system. Equation \rfn{SECOND-EQUATION-1} holds in general for the Bjorken expansion. It follows directly from the conservation law in the form $\partial_\mu T^{\mu\nu} = 0$.

\section{Basic observables}
\label{sec:basobs}

If the functions $T(\tau)$ and $\mu(\tau)$ are known, one can calculate all interesting observables using the formal expression for the solution of the kinetic equations. Above, we used the moments corresponding to baryon number and energy-momentum conservation. We have also introduced the longitudinal pressure ${\cal P}_L$ in \rf{SECOND-EQUATION-1}. In a similar way one can calculate the transverse pressure ${\cal P}_T$. The difference between the longitudinal and transverse pressures shows how far our system is from local equilibrium. Analogous information can be obtained from the difference between the equilibrium, $\peq$, and average, $(2{\cal P}_T +{\cal P}_L)/3$, pressures. Analysis of such quantities will be presented in the following chapters in order to gain information about hydrodynamization and thermalization processes taking place in a non-equilibrium mixture. 
%
%
%
%
%
%
%
\chapter{Numerical results with exact solutions}
\label{sect:numresexact}

In this chapter, we present our main results obtained with exact solutions of the kinetic equations. In particular, we discuss here the hydrodynamization process and scaling properties of the ratio of longitudinal and transverse pressures. 

\begin{figure}[t]
\centering
\includegraphics[angle=0,width=0.925 \textwidth]{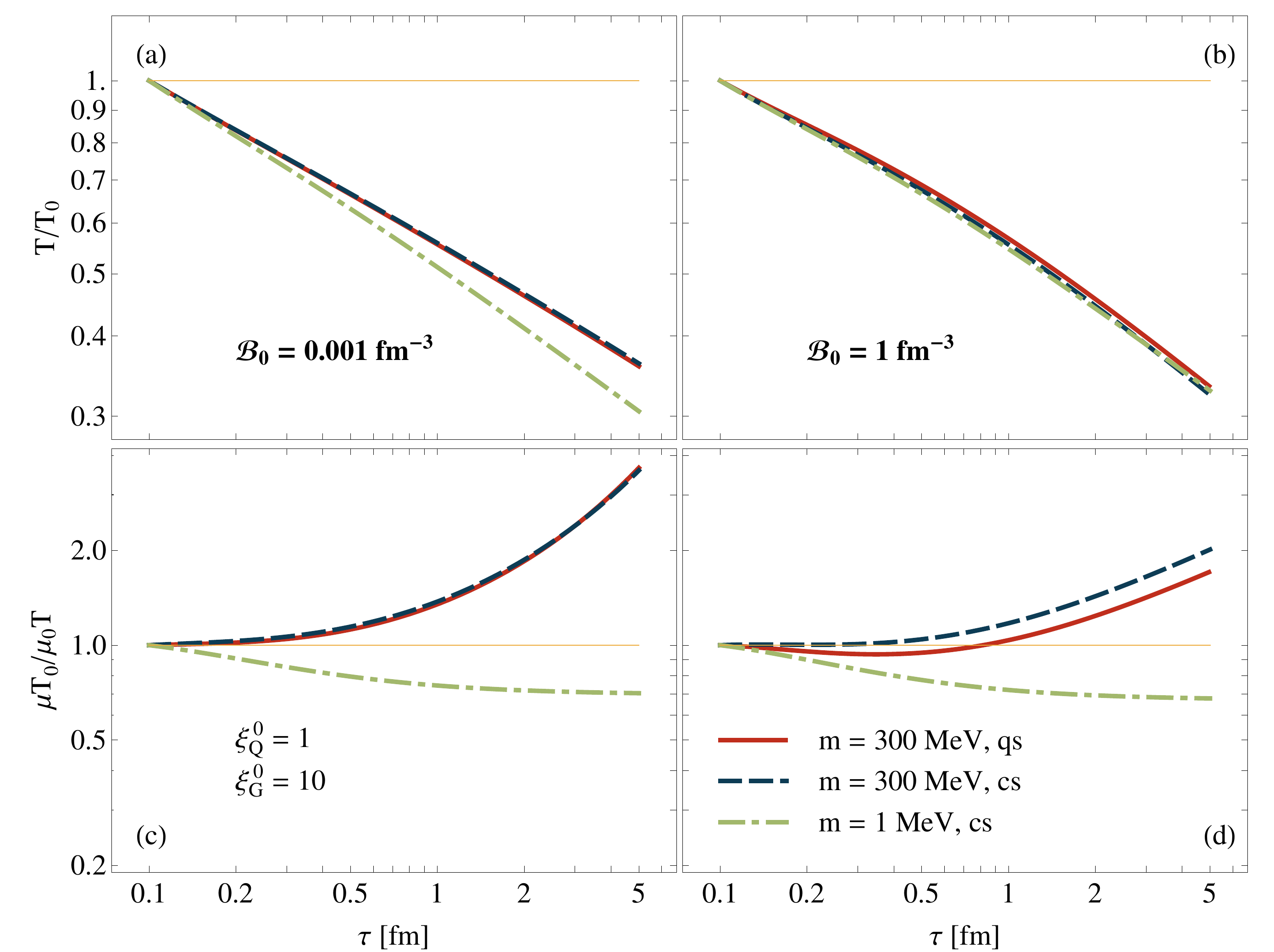} 
\caption{\small (Color online) Effective temperature $T$ (upper panels) and $\mu/T$ ratio (lower panels), shown as functions of the proper time $\tau$ and normalized to unity at the initial proper time $\tau=\tau_0$. Results correspond to the initial oblate-oblate configuration with the values of anisotropy parameters given in the figure. Three different types of lines correspond to three different choices of the statistics and the quark mass (the label ``cs'' denotes classical statistics used for both quarks and gluons, while the label ``qs'' denotes the use of Fermi-Dirac and Bose-Einstein statistics for quarks and gluons, respectively). Other parameters of the calculations are shown in the figure and discussed in the text.} 
\label{fig:TMu_oo}
\end{figure}   
%
\begin{figure}[t]
\centering
\includegraphics[angle=0,width=0.925\textwidth]{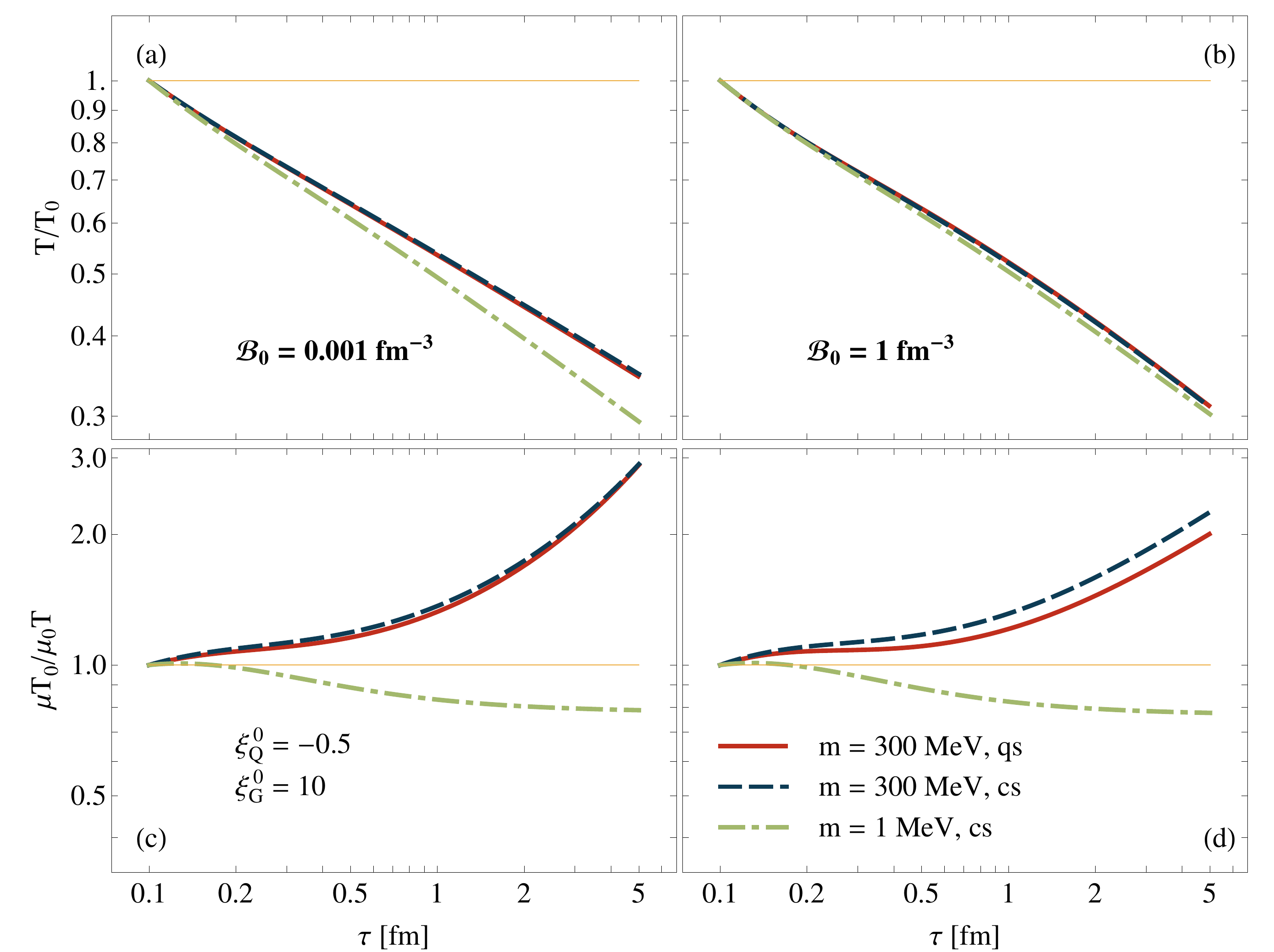}
\caption{\small (Color online) Same as Fig.~\ref{fig:TMu_oo} but for the initial prolate-oblate configuration with the parameters given in the figure.}
\label{fig:TMu_op}
\end{figure}
%
\begin{figure}[t]
\centering
\includegraphics[angle=0,width=0.925\textwidth]{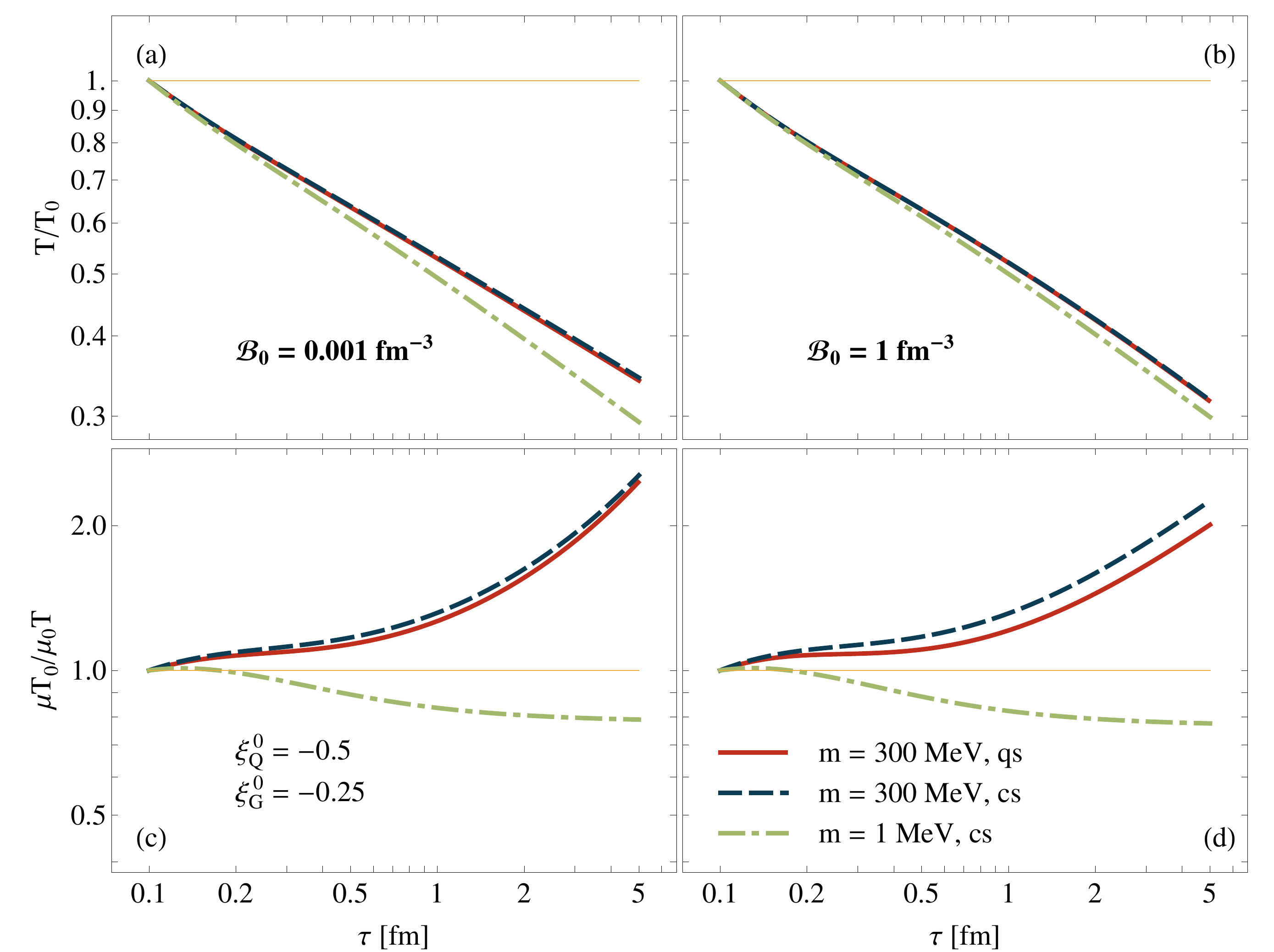} 
\caption{\small (Color online) Same as Figs.~\ref{fig:TMu_oo} and \ref{fig:TMu_op} but for the initial prolate-prolate configuration with the parameters given in the figure. }
\label{fig:TMu_pp}
\end{figure}
%

\section{Initial conditions}
\label{sect:res}

In all the cases presented in this work, we use a constant equilibration time $\tau_{\rm eq}=0.25$~fm, which is the same for quark and gluon components.~\footnote{As mentioned above, the main motivation here comes from saving the computational time. A  popular case used in conformal theories,  where $\teq$ is inversely proportional to the effective temperature $T$, leads to much longer calculations due to the additional integral in  Eq.~(\ref{Damp1}).} The starting proper time is $\tau_0=0.1$~fm and the evolution continues till $\tau_f=5.0$~fm (or $\tau_f=10$~fm in several cases). The initial transverse momentum scales of quarks and gluons are taken identical: $\Lambda_{\Q}^{0}=\Lambda_{\G}^{0}=1$~GeV. The initial non-equilibrium  chemical potential $\lambda_0$ is chosen in such a way that the initial baryon number density is either ${\cal B}_0~=$~0.001~fm$^{-3}$ or ${\cal B}_0=$~1~fm$^{-3}$, see Eq.~(\ref{BApp}). 

Other initial conditions define different values of the anisotropy parameters. We use three sets of the values for $\xi_{\Q}^{0}$ and $\xi_{\G}^{0}$: i) $\xi_{\Q}^{0}=1$ and $\xi_{\G}^{0}=10$, ii)  $\xi_{\Q}^{0}=-0.5$ and $\xi_{\G}^{0}=10$, and iii)  $\xi_{\Q}^{0}=-0.5$ and $\xi_{\G}^{0}=-0.25$. They correspond to oblate-oblate, prolate-oblate, and prolate-prolate initial momentum distributions of quarks and gluons, respectively. Such initial values for $\xi_{\Q}^{0}$ and $\xi_{\G}^{0}$ were used previously in Ref.~\cite{Florkowski:2015cba}. We note that different values of $\xi_{\Q}^{0}$, $\xi_{\G}^{0}$, and $\lambda_0$ imply different initial energy and baryon number densities, hence,  due to matching conditions, also different initial values of $T_0$ and $\mu_0$. We also note that the oblate-oblate initial configuration is supported by the microscopic calculations which suggest that the initial transverse pressure is much higher than the longitudinal one~\cite{Baier:2000sb,Mrowczynski:2005ki,Lappi:2006fp,Heller:2011ju,vanderSchee:2013pia,Gelis:2013rba,Berges:2013eia}. 

We also perform our calculations for three different choices of the particle statistics and the quark mass: in the first case both quarks and gluons are described by the classical, Boltzmann statistics~\footnote{In this case the $\pm$ sign in \rfn{feq} is neglected and $h^{\pm}_\eq(a) = \exp(-a)$.} and the quark mass is equal to 1 MeV~\footnote{Since this value of mass is much smaller than the considered temperature values, we refer sometimes to this case as to the ``massless'' one.}, in the second case we use again the classical statistics but the quark mass is 300 MeV, finally, in the third case the quarks are described by the Fermi-Dirac statistics and have the mass of 300 MeV, while the gluons are described by the Bose-Einstein statistics. The gluon mass is always set equal to zero. The case with classical statistics, ${\cal B}_0=$~0.001~fm$^{-3}$,  and negligibly small quark mass of 1~MeV agrees well with the exact massless case studied in Ref.~\cite{Florkowski:2015cba}.\linebreak 
This agreement is used as one of the checks of our present approach.  

We note that the values of the initial conditions used in this work are to large extent arbitrary, as we want to analyze here only general features of the solutions of~Eqs.~\rfn{kineq} within a large span of the parameter space. With more specific systems in mind, one can freely amend the values of the  initial parameters. 

\begin{figure}[t]
\centering
\includegraphics[angle=0,width=0.5\textwidth]{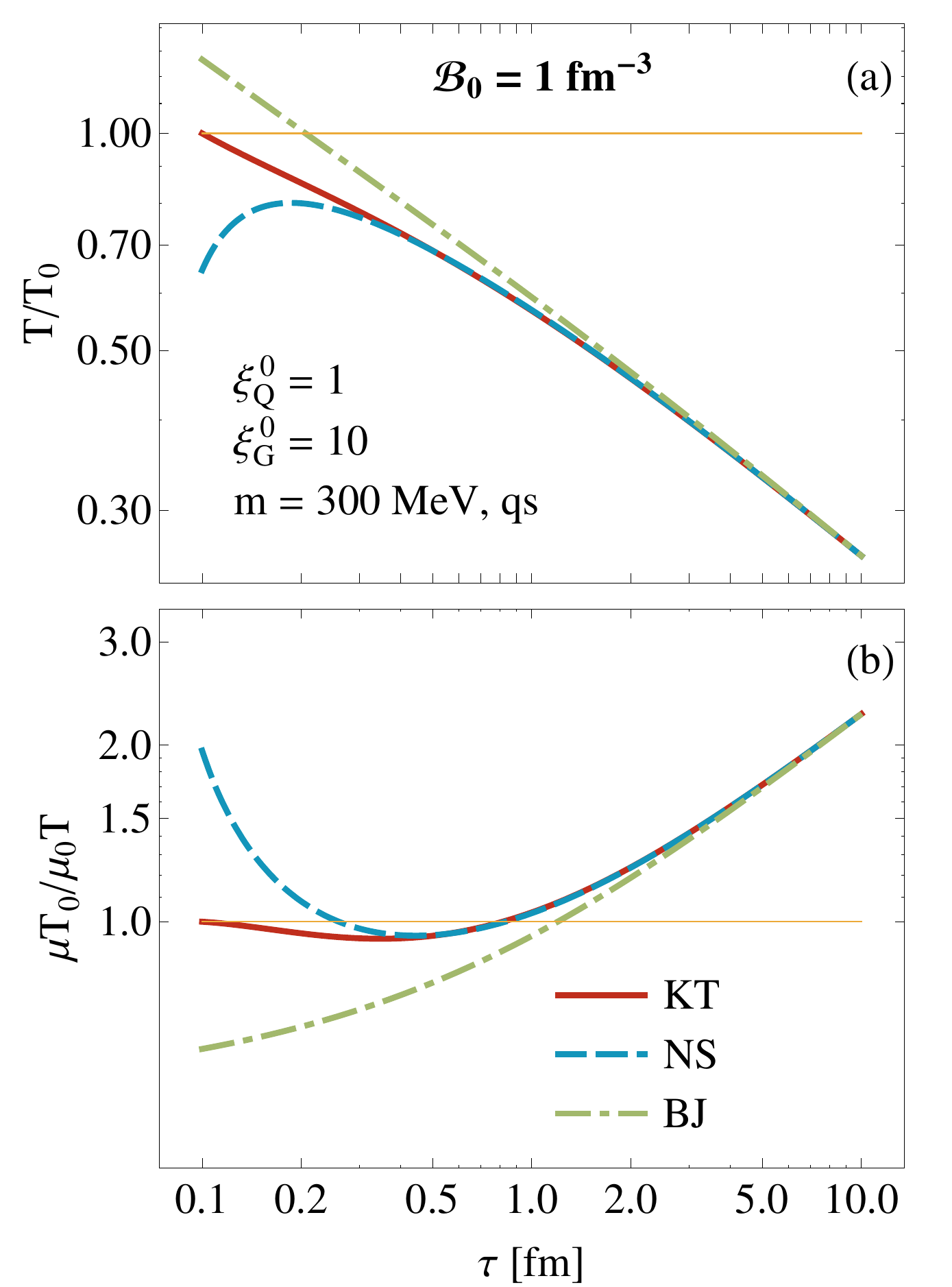}  \\
\caption{\small (Color online) Proper-time dependence of the ratios $T/T_0$ (a) and $\mu T_0/(\mu_0 T)$ (b) obtained in the range $\tau_0 < \tau < 10$~fm from: kinetic theory (red solid lines), perfect-fluid hydrodynamics (green dot-dashed lines), and Navier-Stokes hydrodynamics (blue dashed lines). All results are normalized to the initial values $T_0$ and $\mu_0$ used in the kinetic theory. The initial values of temperature and chemical potential in the hydrodynamic calculations are chosen in such a way that the final values of $T$ and $\mu$ agree with the values found in the kinetic-theory calculation. The calculations are done for the oblate-oblate initial conditions with a finite quark mass of 300 MeV, quantum statistics, and ${\cal B}_0=$~1~fm$^{-3}$.}
\label{fig:MuT_10fm_NS}
\end{figure} 
%
\begin{figure}[!t]
\centering
\includegraphics[angle=0,width=0.6\textwidth]{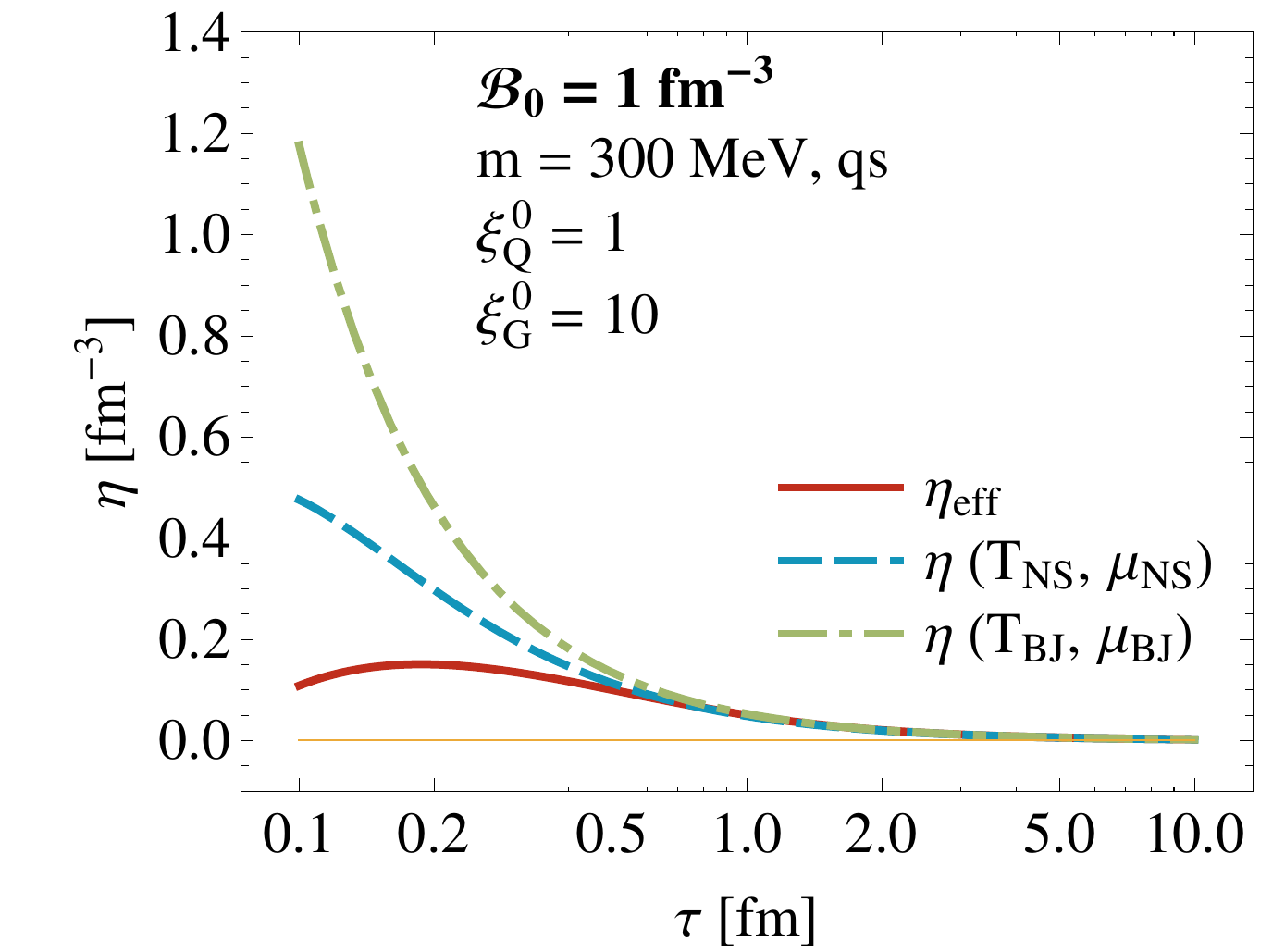} 
\caption{\small (Color online) Effective shear viscosity $\eta_{\rm eff}$ defined by \rf{pi1} (red solid line) and the shear viscosity coefficients $\eta$ calculated using \rf{visceta} for the two $T(\tau)$ and $\mu(\tau)$ profiles, found from the perfect-fluid hydrodynamics (green dot-dashed line) and from the Navier-Stokes equations (blue dashed line).  The effective shear viscosity agrees well with the standard definition of $\eta$ for $\tau >  0.5$~fm. The initial conditions are the same as in Fig.~\ref{fig:MuT_10fm_NS}.}
\label{fig:MuT_10fm_Eta}
\end{figure}
%

%
\begin{figure}[!t]
\centering
\includegraphics[angle=0,width=0.6 \textwidth]{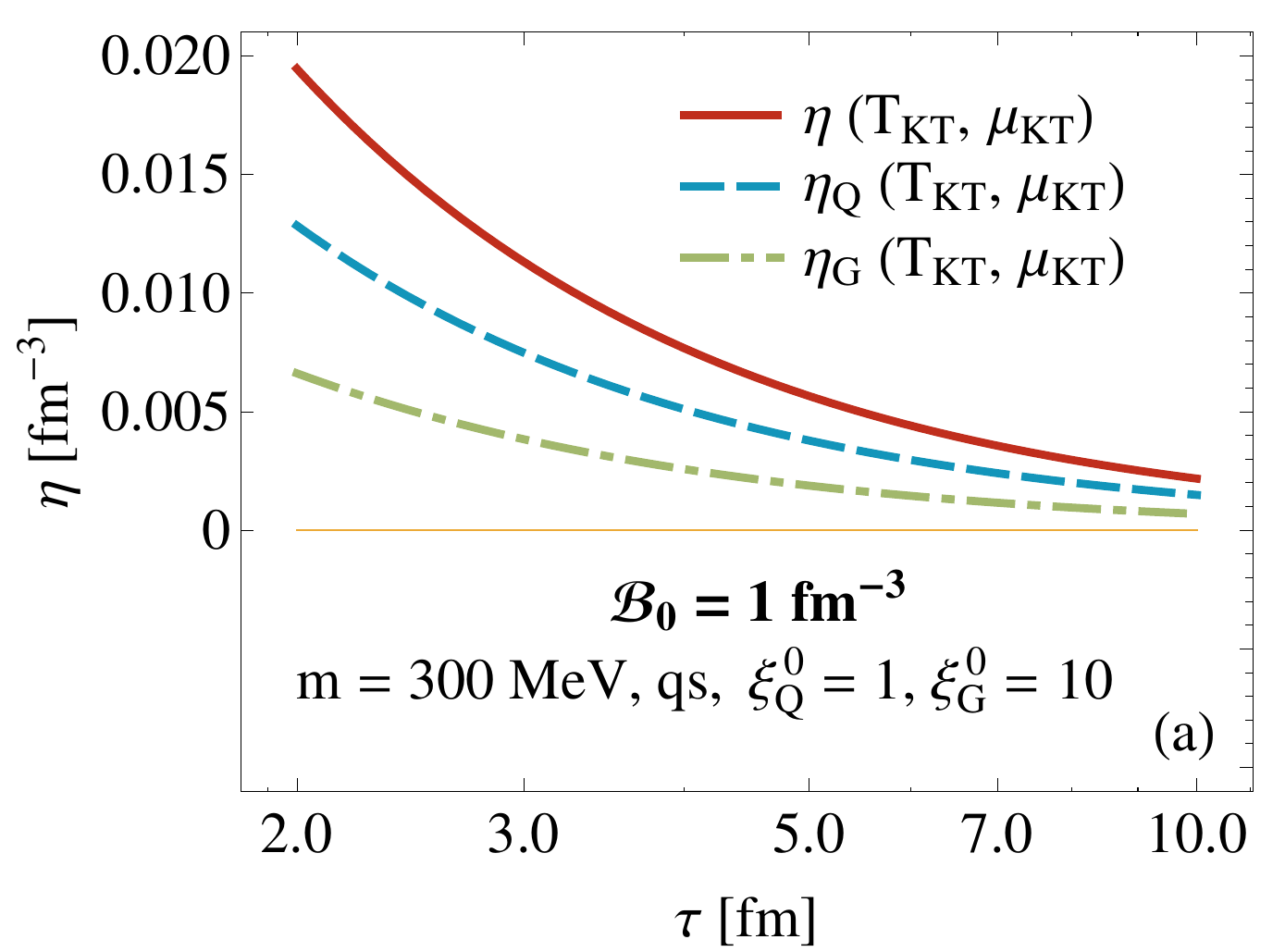} 
\caption{\small (Color online) Proper-time dependence of the shear viscosity of the mixture (red solid line), of the quark component (blue dashed line), and of the gluon component (green dot-dashed line), see Eqs.~\rfn{visceta}, \rfn{etaQ} and \rfn{etaG}, respectively, calculated within the kinetic theory. The initial conditions are the same as in Fig.~\ref{fig:MuT_10fm_NS}.  }
\label{fig:MuT_10fm_EtaQG}
\end{figure}
%

\section{Proper-time dependence of \texorpdfstring{$T$}{T}  and 
\texorpdfstring{$\mu/T$}{mu over T}}
\label{sect:resTmu}

Figures \ref{fig:TMu_oo}, \ref{fig:TMu_op} and \ref{fig:TMu_pp} show the proper-time dependence of the effective temperature $T$ and $\mu/T$ ratio, which are normalized to unity at the initial time $\tau=\tau_0$. The two upper panels, (a) and (b), show temperature profiles, while the two lower panels, (c) and (d), show $\mu/T$. The two left panels, (a) and (c), correspond to the case ${\cal B}_0=$~0.001~fm$^{-3}$, and the two right panels, (b) and (d), describe the case ${\cal B}_0=$~1~fm$^{-3}$. The three figures correspond to three different initial conditions specified by the initial anisotropy parameters. Figures~ \ref{fig:TMu_oo}, \ref{fig:TMu_op}, and \ref{fig:TMu_pp} illustrate the effects of the finite mass and quantum statistics on the time evolution of $T$ and $\mu/T$. 

We observe that the inclusion of the finite mass (for either classical or quantum statistics) has an important effect on the $\mu/T$ ratio. For $m=300$~MeV it asymptotically increases with time, while in the $m=1$~MeV case it approaches a constant, which is expected for the massless system in the Bjorken model assuming local equilibrium.  
The finite mass has a significant effect on the time dependence of the effective temperature. The latter decreases more slowly in the massive cases (especially in the ${\cal B}_0=$~0.001~fm$^{-3}$ case). The effects of quantum statistics are most visible in the $\mu/T$ proper-time dependence. 

To analyze the proper-time dependence of $T$ and $\mu/T$ in more detail, in Fig.~\ref{fig:MuT_10fm_NS} we compare the kinetic-theory (KT) results for the quantum, massive, and oblate-oblate case with hydrodynamic calculations. The latter are performed for the Bjorken perfect-fluid (BJ) and Navier-Stokes (NS) versions, see Appendix~\ref{s:NS} for definitions of these frameworks. The initial values of temperature and chemical potential in the hydrodynamic calculations are chosen in such a way that the final values of $T$ and $\mu$ agree with the values found in the kinetic-theory calculation. Although such matching is required only for the last moment of the time evolution, we see that the hydrodynamic calculations approximate very well the kinetic-theory results within a few last fermis of the time evolution. As expected, we see that the Navier-Stokes approach reproduces better the exact  kinetic-theory result, compared to the perfect-fluid calculation.

\begin{figure}[!t]
\centering
\includegraphics[angle=0,width=0.55\textwidth]{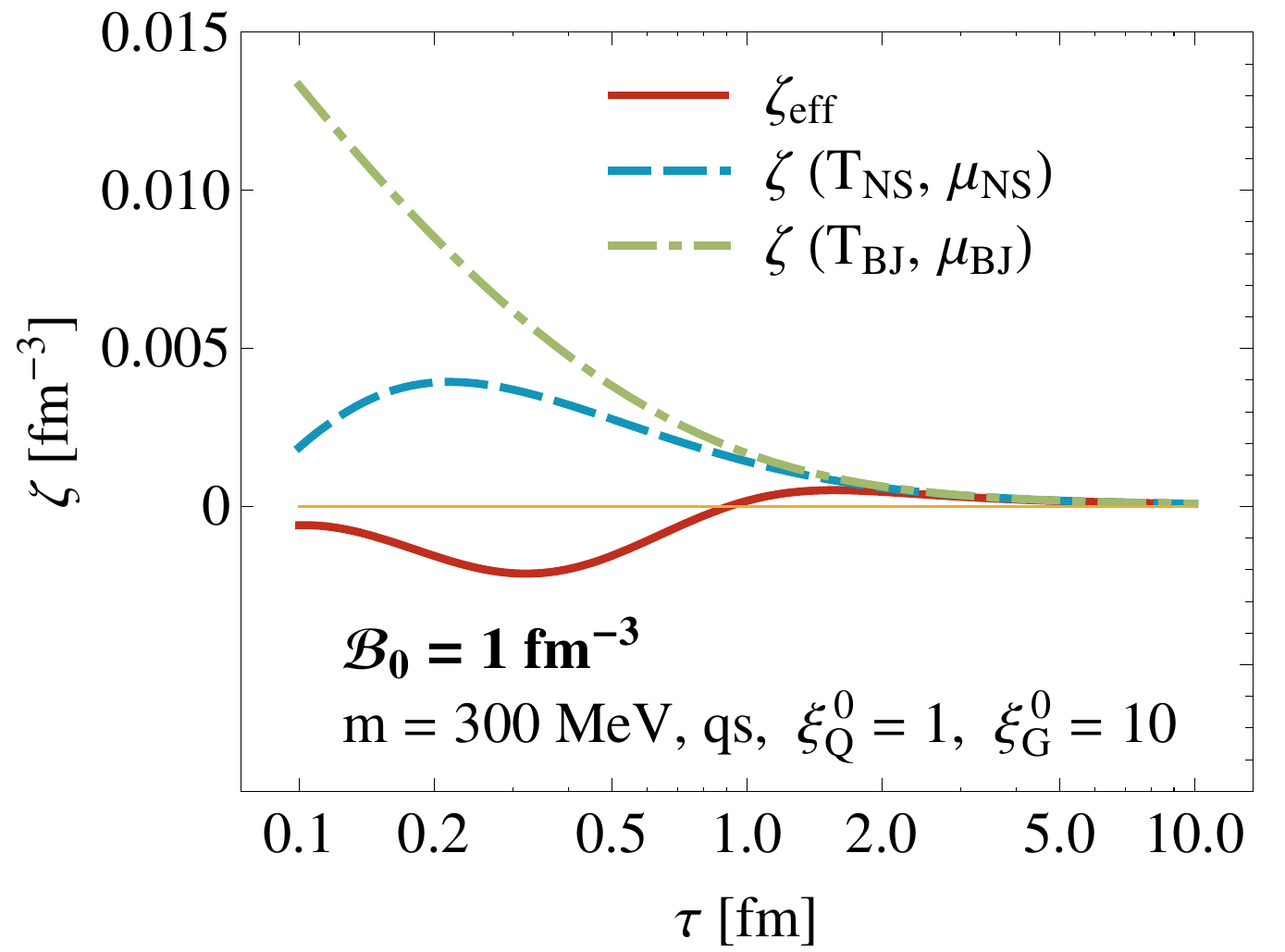}  \\
\caption{\small (Color online) Effective bulk viscosity $\zeta_{\rm eff}$ defined by \rf{Pi1} (red line) and the bulk viscosity coefficient $\zeta$ calculated with the help of \rf{zeta} for the two $T(\tau)$ and $\mu(\tau)$ profiles, found from the perfect-fluid hydrodynamics (green dot-dashed line) and from the Navier-Stokes equations (blue dashed line).  We find that the effective bulk viscosity agrees with the standard definition of $\zeta$ for $\tau > 2$~fm. The initial conditions are the same as in Fig.~\ref{fig:MuT_10fm_NS}. }
\label{fig:MuT_10fm_Zeta}
\end{figure}
%
\begin{figure}[!t]
\centering
\includegraphics[angle=0,width=0.55\textwidth]{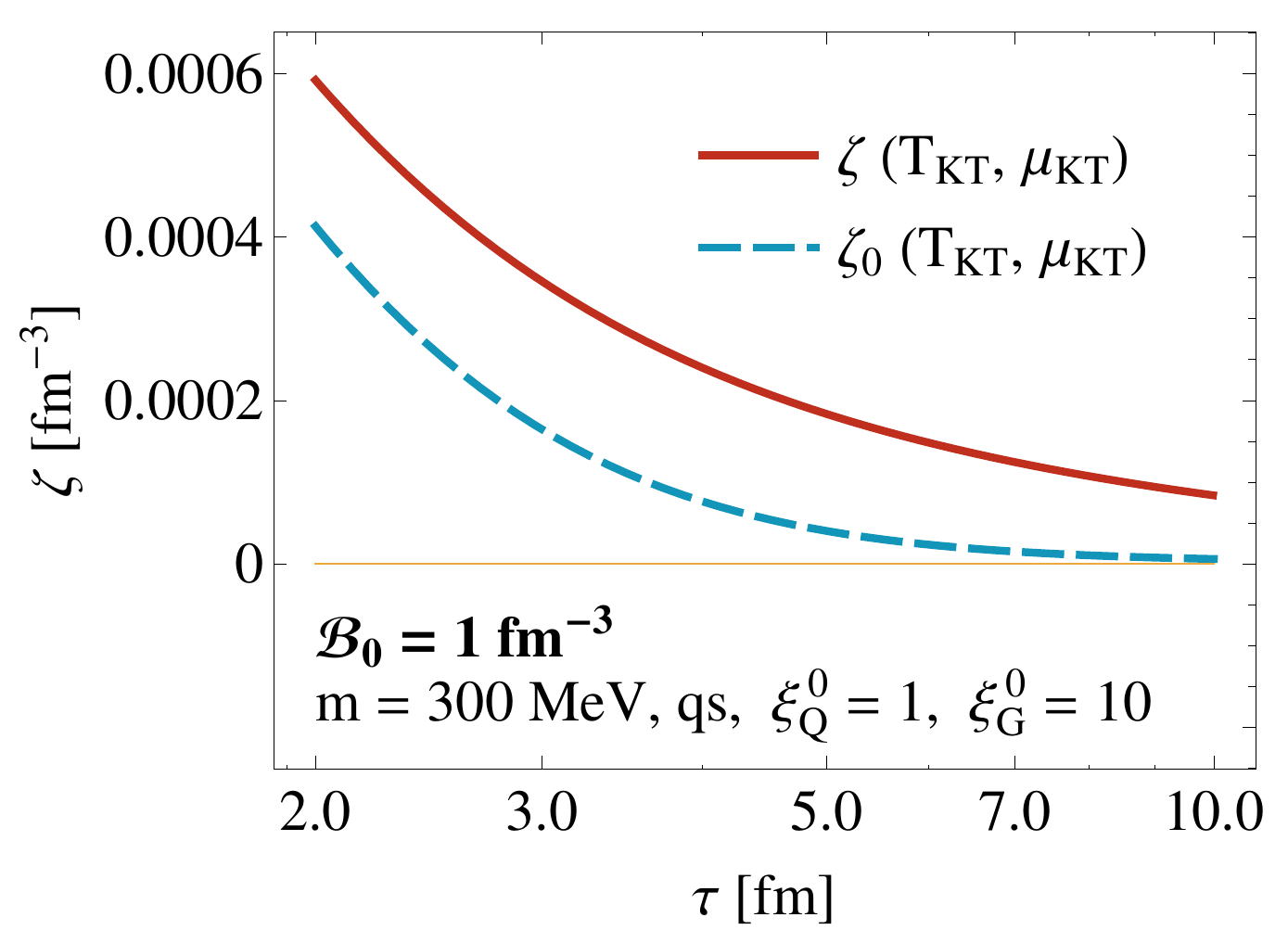}  \\
\caption{\small (Color online) Bulk viscosity coefficient $\zeta$ calculated with the help of \rf{zeta}  with the thermodynamic coefficients $\kappa_1$ and $\kappa_2$ determined for the whole quark-gluon system (red solid line) and the coefficient $\zeta_0$ obtained from  \rf{zeta} with $\kappa_1$ and $\kappa_2$ determined only for the quark component (blue dashed line). The initial conditions are the same as in Fig.~\ref{fig:MuT_10fm_NS}. }
\label{fig:MuT_10fm_Zeta0}
\end{figure}
%

\section{Hydrodynamization}
\label{sect:hydrodynamization}

\subsection{Shear sector}

The results shown in Fig.~\ref{fig:MuT_10fm_NS} suggest that the non-equilibrium dynamics of the system enters
rather fast the hydrodynamic regime described by the NS equations (at the stage where deviations from local
equilibrium are still substantial). Such a phenomenon was identified first in the context
of AdS/CFT calculations \cite{Heller:2011ju} and is known now as the \textit{hydrodynamization} process. To illustrate this
behavior  in our case, we show in Fig.~\ref{fig:MuT_10fm_Eta} the proper-time dependence of the \textit{effective shear viscosity}  coefficient
$\eta_{\rm eff}$ defined by the expression~\cite{Florkowski:2013lya}~\footnote{We use the notation where calligraphic symbols such as ${\cal E}$, ${\cal P}_T $ or ${\cal P}_L$ refer to exact values obtained from the kinetic theory. In the situations where the system is close to equilibrium and described by the Navier-Stokes hydrodynamics we add the subscript $NS$. The standard kinetic coefficients describe the systems close to equilibrium, hence, the shear viscosity 
is defined by the formula $\eta = \f{\tau}{2}  \left({\cal P}_T - {\cal P}_L \right)_{\rm NS}$ and the bulk viscosity by $\zeta = -  \f{\tau}{3} \left({\cal P}_L + 2 {\cal P}_T - 3 {\cal P}^{\rm eq}  \right)_{\rm NS}$, see App.~\ref{s:NS}. If we use the exact  kinetic-theory values
on the right-hand sides of these definitions, we deal with effective values, which should agree with the standard definitions for systems being close to local equilibrium. }

\bel{pi1}
\eta_{\rm eff} = \f{\tau}{2}  \left({\cal P}_T - {\cal P}_L \right) ,
\eel
see Eqs.~\ref{PLPTns}.
The effective shear viscosity (solid red line in Fig.~\ref{fig:MuT_10fm_Eta}) is compared with the standard shear viscosity coefficient, $\eta$, valid for the system close to equilibrium. 
\linebreak
For the quark-gluon mixture the latter is defined as the sum of the quark and gluon coefficients,~\footnote{For general collision kernels, the total shear viscosity of the system (although written formally as a sum of the individual contributions) may not be a simple sum of {\it independent} terms, for example, see~\cite{Itakura:2007mx}.}
\bel{visceta}
\eta = \eta_\Q + \eta_\G,
\eel
where following~\cite{Sasaki:2008fg}, see also \cite{Bozek:2009dw,Chakraborty:2010fr,Bluhm:2010qf}, we use
\bel{etaQ}
\eta_\Q = \f{g_\Q \teq}{15 T} \int\limits_0^\infty \f{dp\, p^6}{2 \pi^2 (m^2+ p^2)} 
\left[ f_{\Qp, \eq} \left(1-f_{\Qp, \eq} \right) + f_{\Qm, \eq} \left(1-f_{\Qm, \eq} \right) \right],
\eel
\bel{etaG}
\eta_\G = \f{g_\G \teq}{15 T} \int\limits_0^\infty \f{dp\, p^4}{2 \pi^2} f_{\G, \eq} \left(1+f_{\G, \eq} \right).
\eel
\pagebreak 
The coefficient $\eta$ is calculated as a function of  $T$ and $\mu$ obtained either from the perfect-fluid  
(green dot-dashed line in Fig.~\ref{fig:MuT_10fm_Eta}) or  NS hydrodynamic calculation 
(blue dashed line in Fig.~\ref{fig:MuT_10fm_Eta}). In the two cases, 
we find that $\eta_{\rm eff}$ agrees very well with $\eta$ for $\tau > 0.5$~fm which is about two times the 
relaxation time. Thus, in the shear sector, we observe a very fast approach to the hydrodynamic NS regime. It is important to notice that 
the agreement with the NS description is reached when $\eta$ is significantly different from zero, which supports the idea 
that the hydrodynamic description becomes appropriate before the system thermalizes, i.e., before the state of local thermal 
equilibrium with ${\cal P}_T \approx {\cal P}_L$ is reached.

In Fig.~\ref{fig:MuT_10fm_EtaQG} we show the proper-time dependence of the shear viscosity of the mixture (red solid line) using the kinetic theory results
and compare it with the shear viscosity of the quark component (blue dashed line) and the gluon component (green dot-dashed line), see Eqs.~\rfn{visceta}, \rfn{etaQ} and \rfn{etaG}, respectively. The initial conditions are the same as in Fig.~\ref{fig:MuT_10fm_NS}. The results shown in Fig.~\ref{fig:MuT_10fm_EtaQG}
show that the shear viscosity of the mixture is dominated by the shear viscosity of quarks.

\begin{figure}[t]
\centering
\includegraphics[angle=0,width=0.925\textwidth]{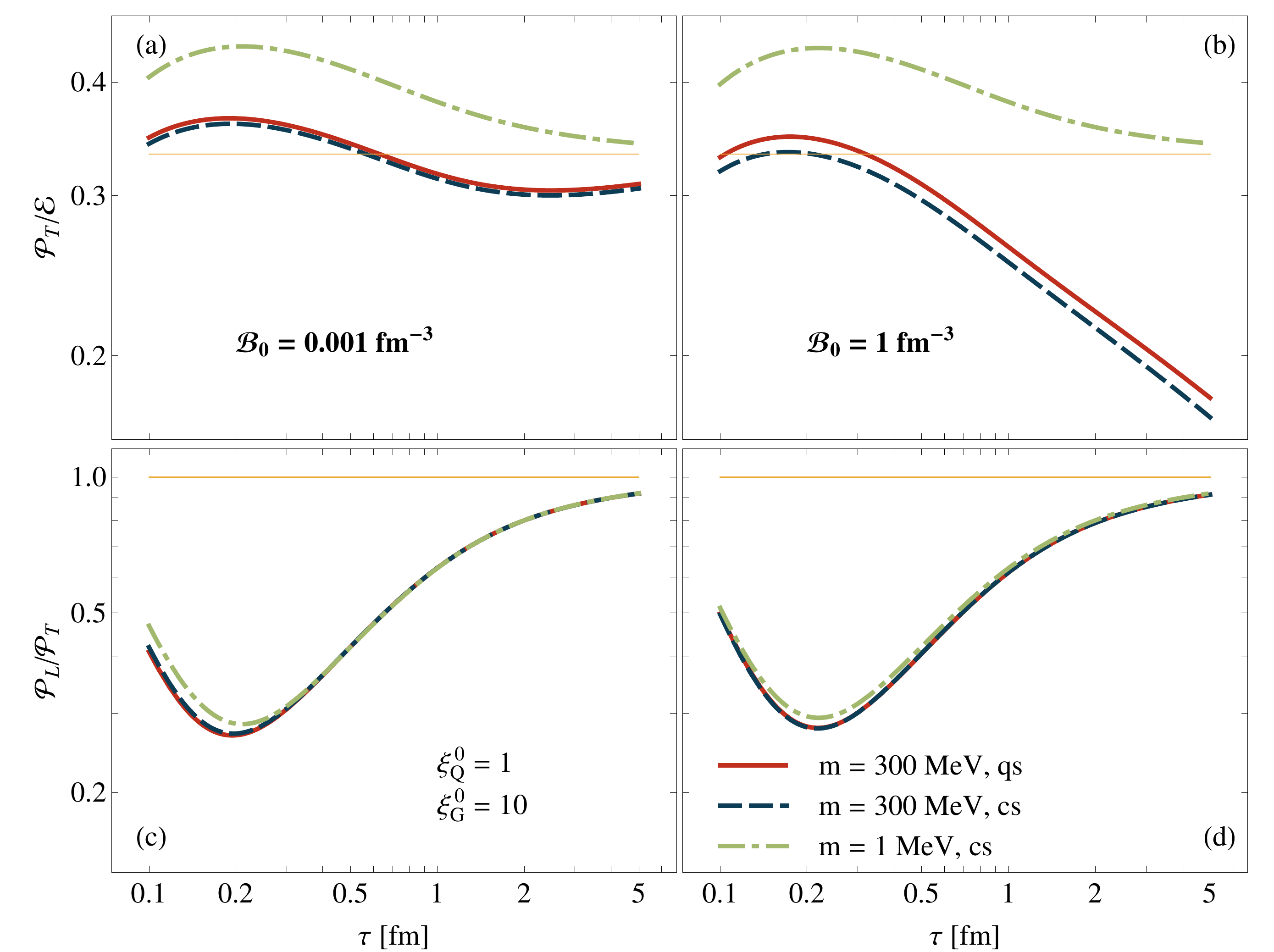} 
\caption{\small (Color online) ${\cal P}_T/{\cal E}$ (upper panels) and ${\cal P}_L/{\cal P}_T$ (lower panels) for the initially oblate-oblate system. Green dot-dashed lines correspond to ``massless'' quarks and classical distribution functions, navy blue dashed lines correspond to the massive quarks and classical distribution functions, while red solid lines are for massive quarks and quantum distributions. Left (right) panels describe the results for ${\cal B}_0=$~0.001~fm$^{-3}$ (${\cal B}_0=$~1~fm$^{-3}$).
 }
\label{fig:ooPP}
\end{figure}
%
\begin{figure}[t]
\centering
\includegraphics[angle=0,width=0.925\textwidth]{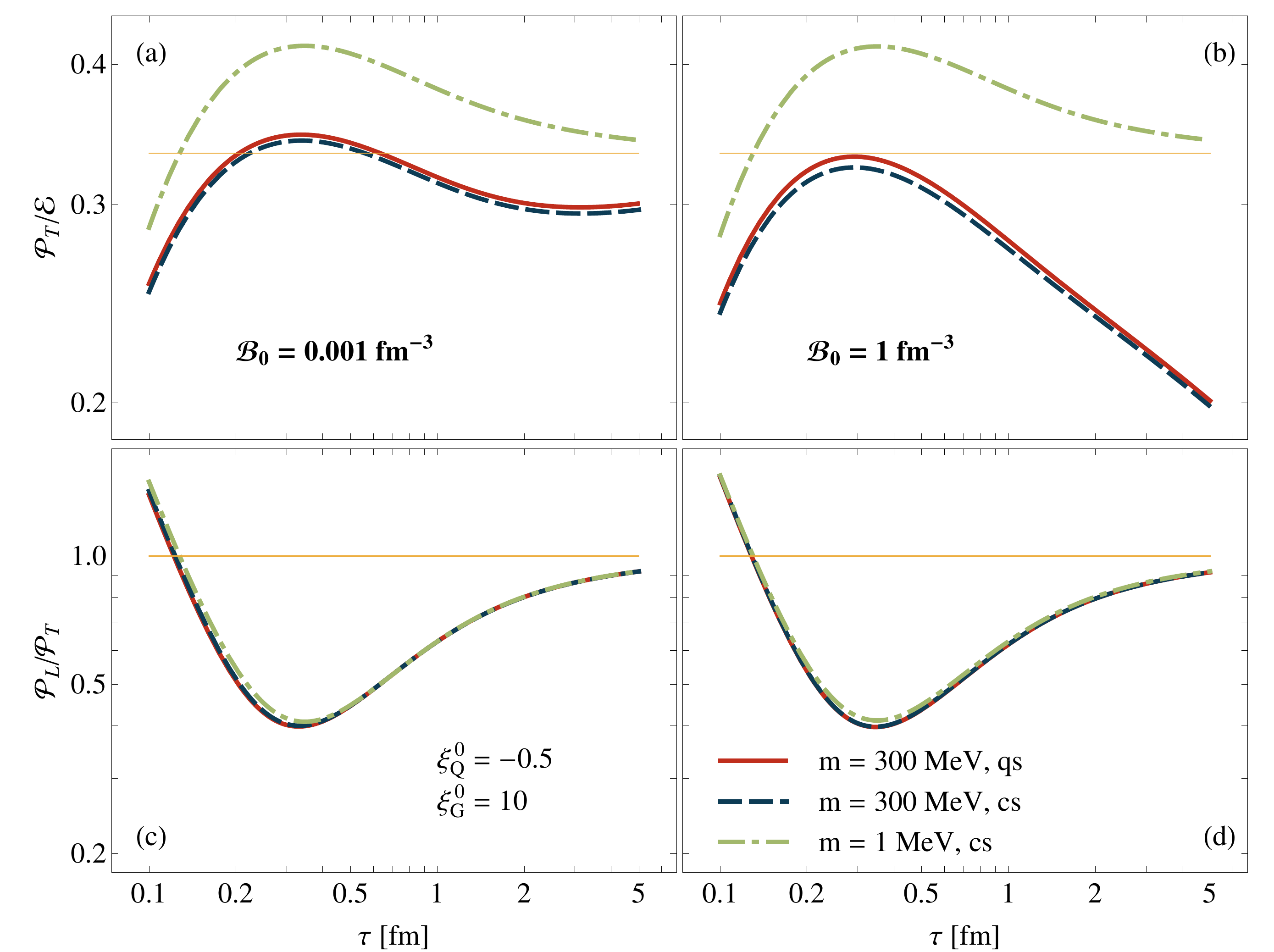} 
\caption{\small (Color online) Same as Fig. \ref{fig:ooPP} but for the initially prolate-oblate system.}
\label{fig:opPP}
\end{figure} 
%
\begin{figure}[t]
\centering
\includegraphics[angle=0,width=0.925\textwidth]{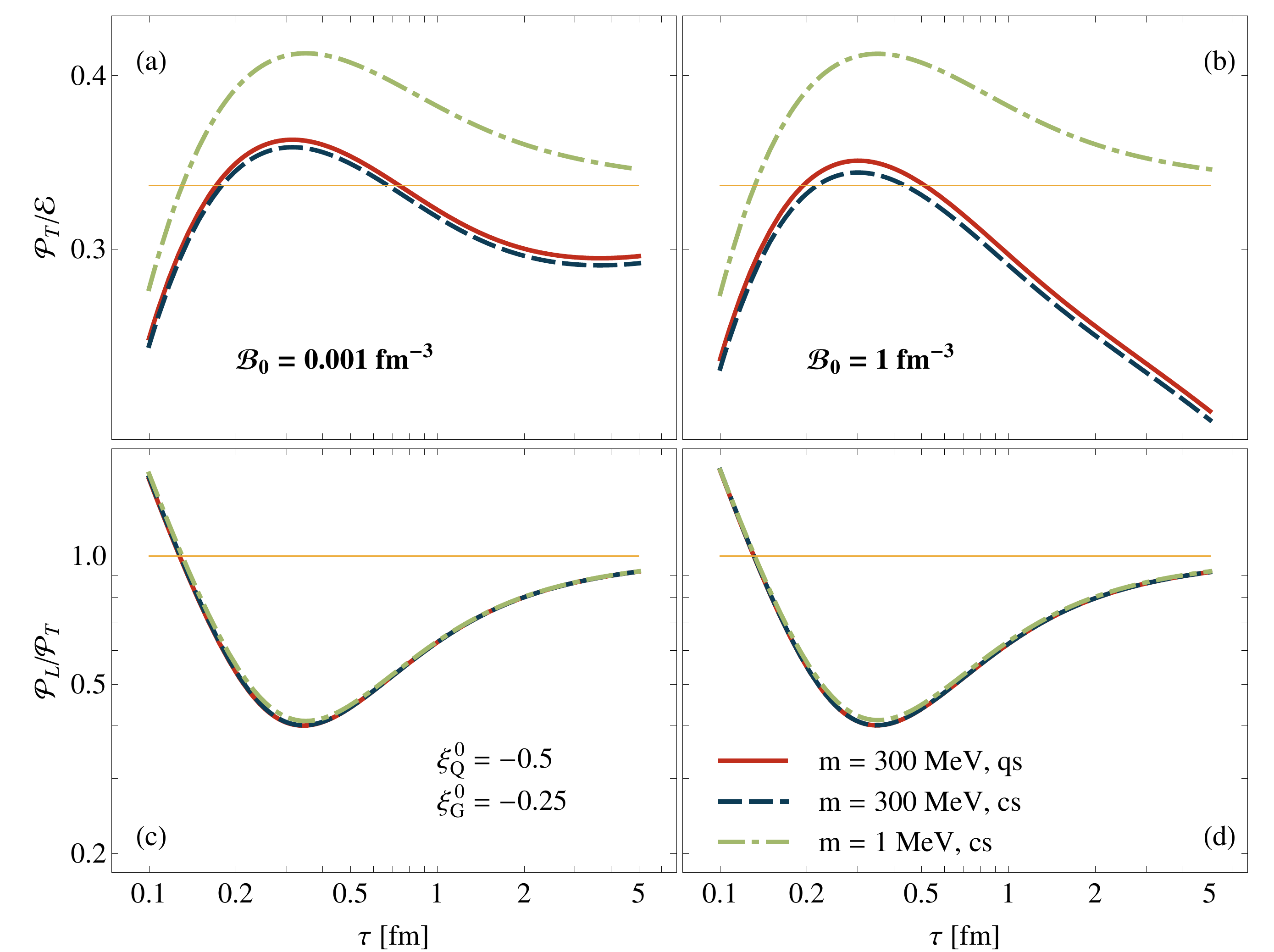} 
\caption{\small (Color online) Same as Fig. \ref{fig:ooPP} but for the initially prolate-prolate system. }
\label{fig:ppPP}
\end{figure}  

\subsection{Bulk sector}  
Similarly to the shear-viscosity effects, we can analyze the bulk sector, where we define the \textit{effective bulk viscosity} by the expression
\bel{Pi1}
\zeta_{\rm eff} = - \tau \Pi,
\eel
where $\Pi$ is the exact bulk pressure
\bel{Pi2}
\Pi = \f{1}{3} \left({\cal P}_L + 2 {\cal P}_T - 3 {\cal P}^{\rm eq}  \right).
\eel
In Fig.~\ref{fig:MuT_10fm_Zeta} the time dependence of the effective bulk viscosity is compared  with the time dependence of the bulk viscosity coefficient
given by the expression
\begin{eqnarray}
\label{zeta}
\zeta &=& \f{g_\Q m^2 \teq}{3 T} \int\limits_0^\infty \f{dp\, p^2}{2 \pi^2} 
\left[ \left( f_{\Qp, \eq} \left(1-f_{\Qp, \eq} \right) +  f_{\Qm, \eq} \left(1-f_{\Qm, \eq} \right) \right)
\left( \kappa_1 - \f{p^2}{3(m^2+p^2)} \right) \right. \nn \\
&& \hspace{3.5cm}  \left. +
\left( f_{\Qp, \eq} \left(1-f_{\Qp, \eq} \right) -  f_{\Qm, \eq} \left(1-f_{\Qm, \eq} \right) \right) 
\f{\kappa_2}{\sqrt{m^2 +p^2}} \right],
\end{eqnarray}
where $\kappa_1$ and $\kappa_2$ are defined by the thermodynamic derivatives
\bel{kappas}
\kappa_1(T,\mu) =  \left( \f{\partial {\cal P}^{\eq} }{\partial {\cal E}^{\eq}} \right)_{{\cal B}^{\eq}}, \qquad
\kappa_2(T,\mu) =  \f{1}{3} \left( \f{\partial {\cal P}^{\eq} }{\partial {\cal B}^{\eq} } \right)_{{\cal E}^{\eq}}.
\eel
For a simple fluid with zero baryon density, the coefficient $\kappa_1$ becomes equal to the sound velocity squared. 
The steps leading to \rf{zeta} are described in more detail in Appendix~\ref{s:shearbulk}. The form of  \rfn{zeta} agrees with that
given in \cite{Sasaki:2008fg} for fermions. There is, however, one important difference between our approach and that of
\cite{Sasaki:2008fg}. In Ref.~\cite{Sasaki:2008fg} a simple system of fermions is considered and \rfn{zeta} includes 
the derivatives \rfn{kappas} where only fermionic thermodynamic functions appear. In our case, we deal with
a mixture and we have checked that \rfn{kappas} should include the total thermodynamic functions being the
sums of quark and gluon contributions. Thus, although the bulk viscosity of a quark-gluon mixture is given by
the formula known for massive quarks (and $\zeta=0$ if $m=0$), the use of the full thermodynamic functions 
in Eqs.~\rfn{kappas} means that although gluons are considered  massless  they  contribute to the bulk viscosity of the full system.

Similarly, as in the shear sector, we can see in Fig.~\ref{fig:MuT_10fm_Zeta}  that $\zeta_{\rm eff}(\tau)$ 
approaches $\zeta(\tau)$, however, the agreement is reached 
for significantly larger times, $\tau > 2$~fm. This means that the hydrodynamization of the bulk sector is slower and takes place after the 
hydrodynamization of the shear sector.  Observations that the hydrodynamization in the shear sector may happen before
the hydrodynamization in the bulk sector have been done recently in  Ref.~\cite{Attems:2017zam} within the \textit{gauge/gravity correspondence} with the breaking of conformality, where the
hydrodynamization in the bulk sector has been dubbed the \textit{EoSization} process \cite{Attems:2016tby}. In this scenario first ${\cal P}_L$ and ${\cal P}_T$
tend to a common value ${\bar {\cal P}} \neq {\cal P}^{\rm eq}$ and, subsequently, ${\bar {\cal P}}$ approaches ${\cal P}^{\rm eq}$, which signals establishing equation of state (EOS) of the system.

To visualize the importance of the gluon degrees of freedom in expressions \rfn{kappas} for the bulk viscosity of the mixture in Fig.~\ref{fig:MuT_10fm_Zeta0} we show the bulk viscosity coefficient $\zeta$ and compare it with the coefficient $\zeta_0$
that has been calculated in the same way as $\zeta$ except that the thermodynamic coefficients $\kappa_1$ and $\kappa_2$ of the former
were calculated only for the quark component. We find that neglecting the gluon contribution in $\kappa_1$ and $\kappa_2$
changes substantially the values of $\zeta$ making it significantly smaller. This finding indicates that gluons, although, massless,
contribute to the bulk viscosity of a quark-gluon mixture. The necessary requirement for this effect is, however, that quarks
are massive.

\subsection{ \texorpdfstring{${\cal P}_T/{\cal E}$}{PT over E} and 
\texorpdfstring{${\cal P}_L/{\cal P}_T$ }{PL over PT  ratios}}

 Figures \ref{fig:ooPP}, \ref{fig:opPP}, and \ref{fig:ppPP} correspond to Figs.  \ref{fig:TMu_oo}, \ref{fig:TMu_op}, 
and \ref{fig:TMu_pp}, respectively, and show the time dependence of the ratios ${\cal P}_T/{\cal E}$ (upper panels)
and ${\cal P}_L/{\cal P}_T$ (lower panels). In the case of quarks with a very small mass (green dot-dashed lines)
the ratios ${\cal P}_T/{\cal E}$ tend to 1/3 as expected for massless systems approaching local equilibrium. The ratios 
${\cal P}_L/{\cal P}_T$ in all studied cases tend to unity which again reflects equilibration of the system. 
Interestingly, the ratios ${\cal P}_L/{\cal P}_T$ very weakly depend on the quark mass and the choice of the
statistics.

\begin{figure}[!ht]
\centering
\includegraphics[angle=0,width=0.925\textwidth]{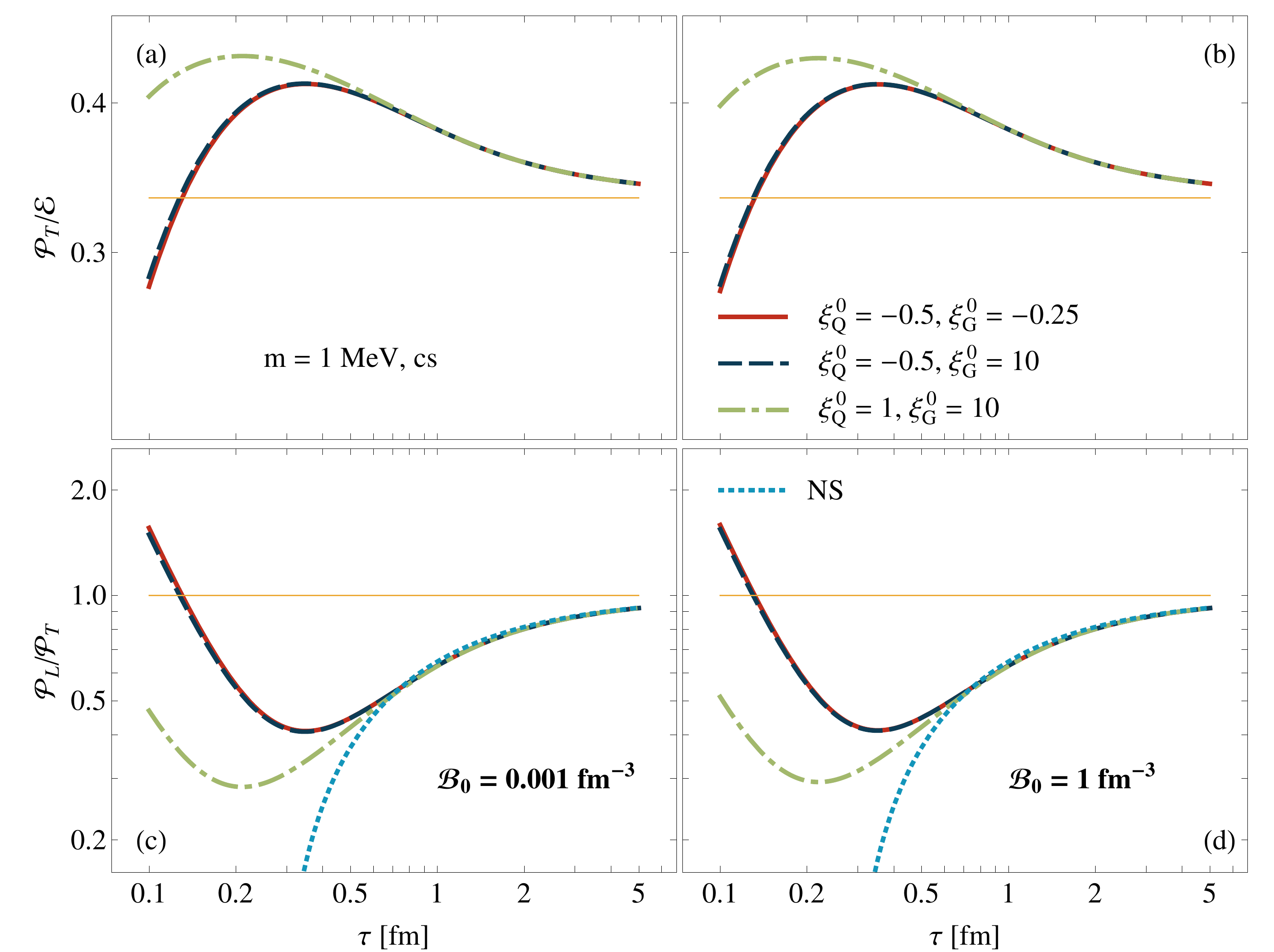} 
\caption{\small (Color online) ${\cal P}_T/{\cal E}$ (upper panels) and ${\cal P}_L/{\cal P}_T$ (lower  panels) for ``massless'' quarks and classical distribution functions. Green dot-dashed, navy blue dashed, and red solid lines describe the results for the oblate-oblate, prolate-oblate, and prolate-prolate initial conditions.
The blue dotted line describes $({\cal P}_L/{\cal P}_T)_{\rm NS}$ obtained from the Navier-Stokes hydrodynamics.  Left (right) panels describe the results for ${\cal B}_0=$~0.001~fm$^{-3}$ (${\cal B}_0=$~1~fm$^{-3}$).
}
\label{fig:m1cl_oo_op_pp}
\end{figure}
%
\begin{figure}[!ht]
\centering
\includegraphics[angle=0,width=0.925\textwidth]{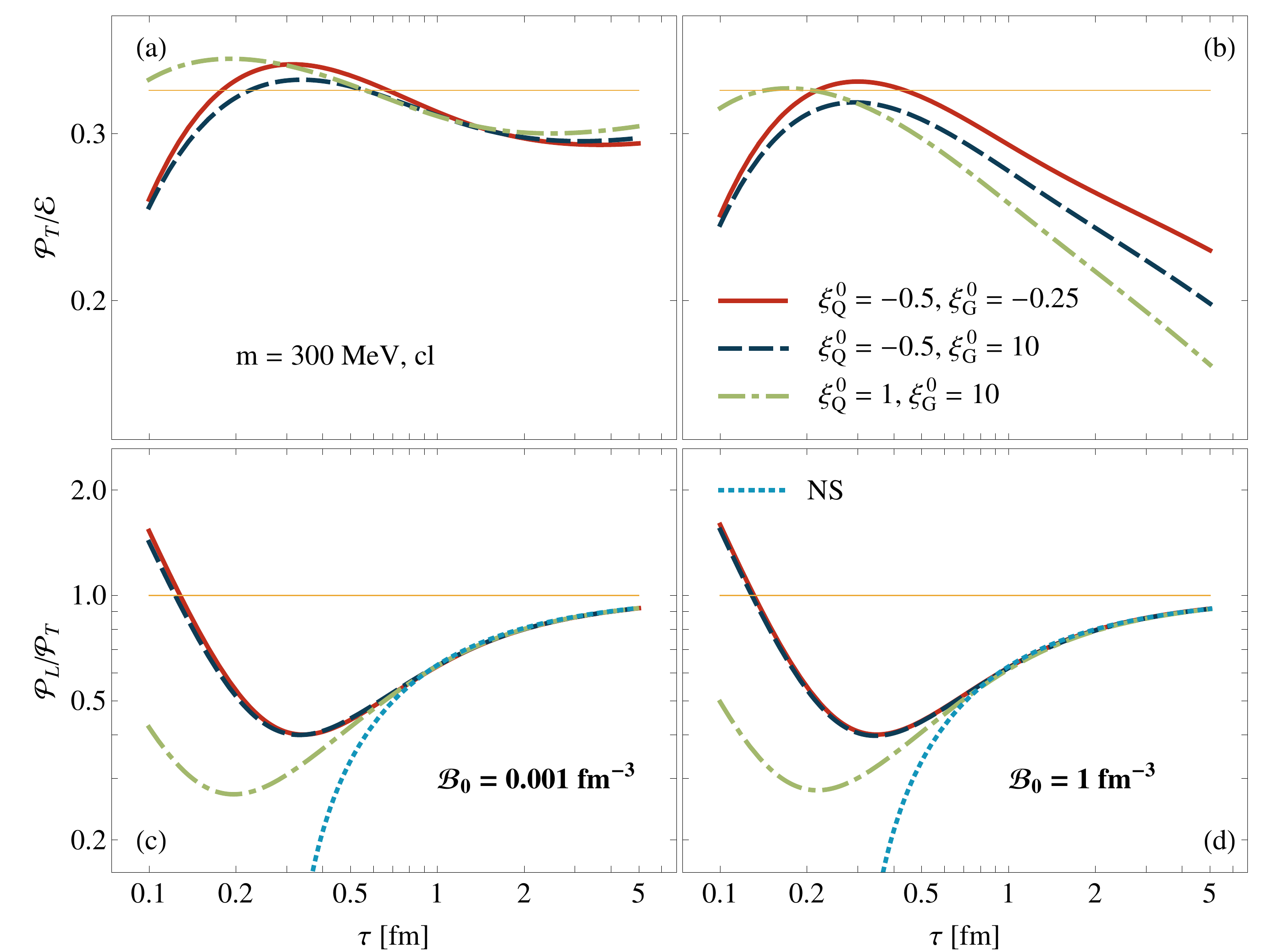} 
\caption{\small (Color online) Same as Fig.~\ref{fig:m1cl_oo_op_pp} but for massive quarks and classical statistics.
}
\label{fig:m300cl_oo_op_pp}
\end{figure} 

%
\begin{figure}[!ht]
\centering
\includegraphics[angle=0,width=0.925\textwidth]{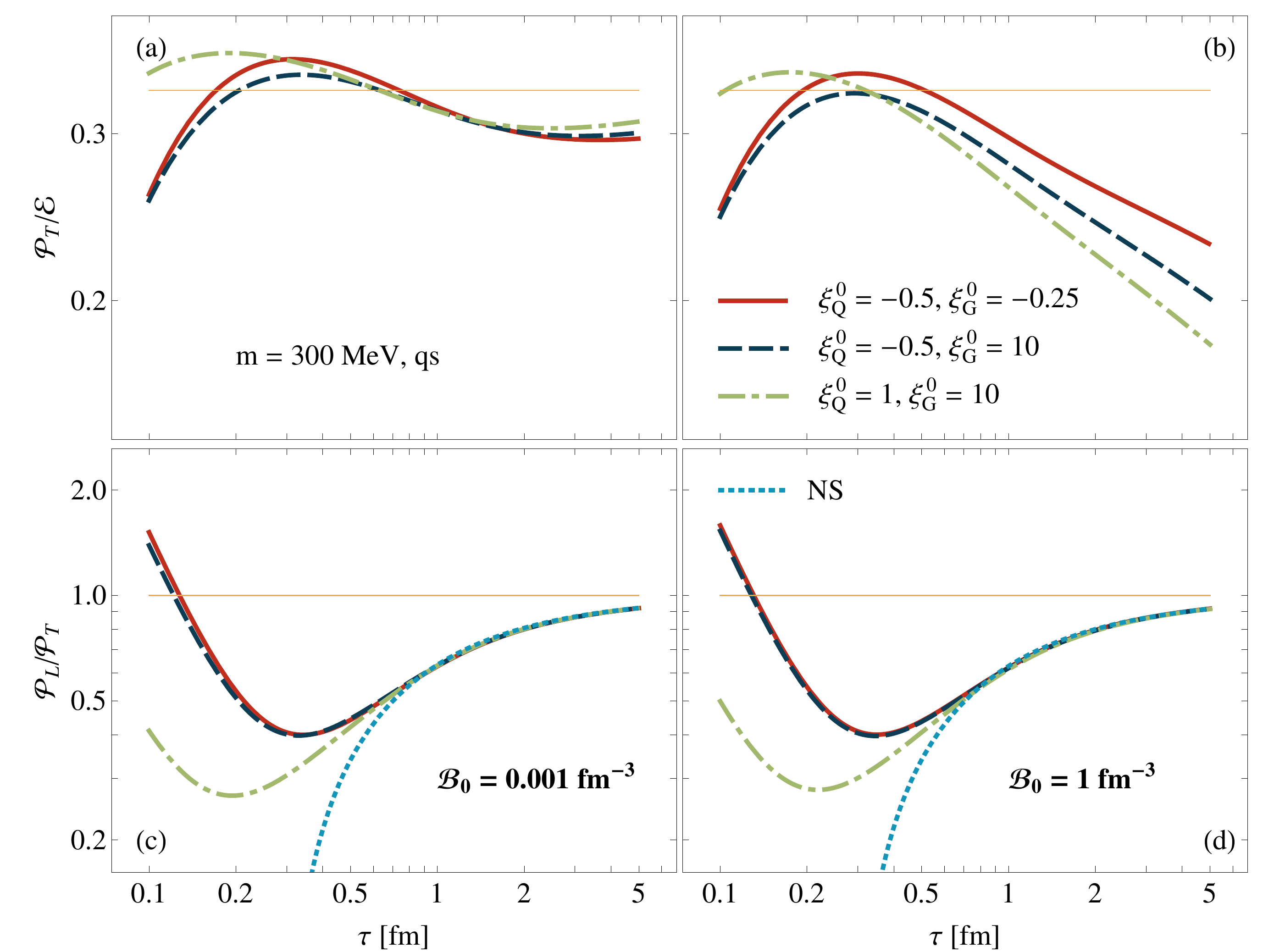} 
\caption{\small (Color online) Same as Fig.~\ref{fig:m1cl_oo_op_pp} but for massive quarks and quantum statistics.
}
\label{fig:m300qs_oo_op_pp}
\end{figure} 

\newpage
\section{Scaling properties}
\label{sect:scaling}

Each panel of Figs. \ref{fig:ooPP}, \ref{fig:opPP}, and \ref{fig:ppPP} shows our results obtained for different values of the quark 
mass and particle statistics but for the same initial anisotropies. In Figs. \ref{fig:m1cl_oo_op_pp}, \ref{fig:m300cl_oo_op_pp}, 
and \ref{fig:m300qs_oo_op_pp} we rearrange this information showing in each panel our results obtained for different initial 
anisotropies, i.e., for oblate-oblate, prolate-oblate, and prolate-prolate initial quark and gluon distributions. Figures~\ref{fig:m1cl_oo_op_pp}, 
\ref{fig:m300cl_oo_op_pp}, and \ref{fig:m300qs_oo_op_pp}  collect the results for different mass and statistics.
The most striking feature of our results presented in these figures is that the ${\cal P}_L/{\cal P}_T$ ratios
(shown in lower panels) converge to the same values, although they describe the system evolutions starting
from completely different initial conditions.

The origin of this behavior can be found if we analyze the NS formula for the ${\cal P}_L/{\cal P}_T$ ratio.
Let us first consider the massless case where we may neglect the bulk viscosity and write
\bel{PLPTNS}
\left( \f{{\cal P}_L}{{\cal P}_T}  \right)_{\rm NS} =
\f{ {\cal P}^{\Q, \rm eq} -4 \eta_Q/(3 \tau)+ {\cal P}^{\G, \rm eq} -4 \eta_G/(3 \tau) }
{{\cal P}^{\Q, \rm eq} +2 \eta_Q/(3 \tau)+ {\cal P}^{\G, \rm eq} +2 \eta_G/(3 \tau)}.
\eel
Assuming in addition that the baryon number density is zero, we may use the following relations connecting 
the shear viscosity with equilibrium pressure~\footnote{See our discussion below \rf{shear4}.}: 
\bel{etaP}
\eta_Q = \f{4}{5} \teq {\cal P}^{\Q, \rm eq}, \quad \eta_G = \f{4}{5} \teq {\cal P}^{\G, \rm eq}.
\eel
It is interesting to note that the coefficient 4/5 is the same for quarks and gluons, hence
\bel{PLPTNSa}
\left( \f{{\cal P}_L}{{\cal P}_T}  \right)_{\rm NS}  =
\f{ 1 - 16 \teq/(15 \tau)}{1+ 8 \teq/(15 \tau)},
\eel
which explains the late-time dependence of ${\cal P}_L/{\cal P}_T$ on the proper time only, observed in panel (c) 
of Fig. \ref{fig:m1cl_oo_op_pp}. We note that if the relaxation time is inversely proportional to the temperature, 
\rf{PLPTNSa} indicates that  $\left({\cal P}_L/{\cal P}_T\right)_{\rm NS}$ depends on the product of $\tau$ and $T$, which is expected for 
conformal systems and related to the existence of a hydrodynamic attractor for such systems~\cite{Heller:2015dha,Romatschke:2016hle,Spalinski:2016fnj,Romatschke:2017vte,Spalinski:2017mel,Strickland:2017kux}.  It turns out that the inclusion of the finite mass and baryon chemical potential 
(with the values studied in this work) affects very little Eqs.~\rfn{etaP} connecting the shear viscosity with pressure. The main difference is that the
coefficient 4/5 is slightly changed. It should be replaced by an effective value obtained for the studied range of $T$ and~$\mu$.

\begin{figure}[t]
\begin{center}
\centering
\subfigure{\includegraphics[angle=0,width=0.47\textwidth]{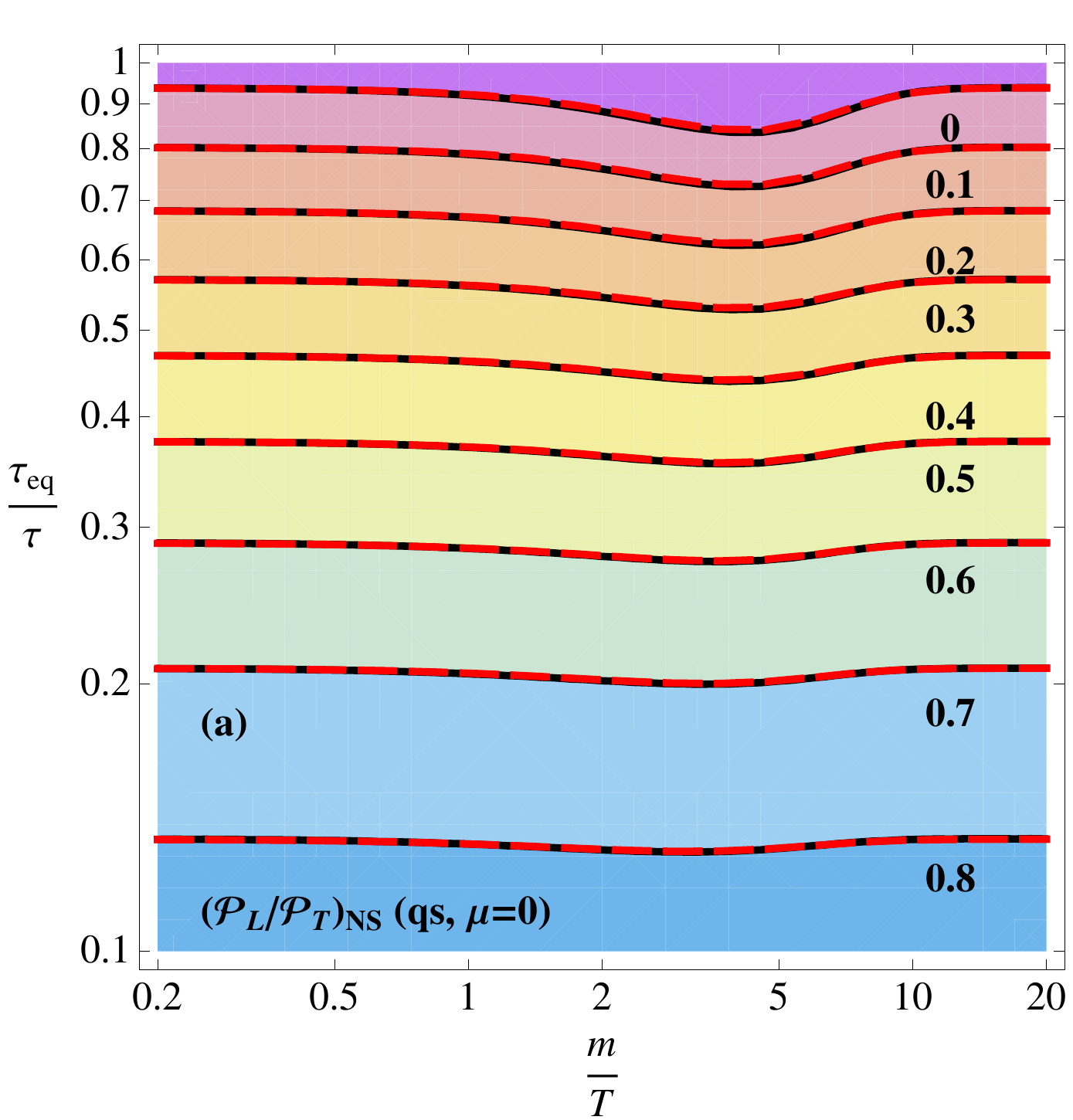}} 
\subfigure{\includegraphics[angle=0,width=0.47\textwidth]{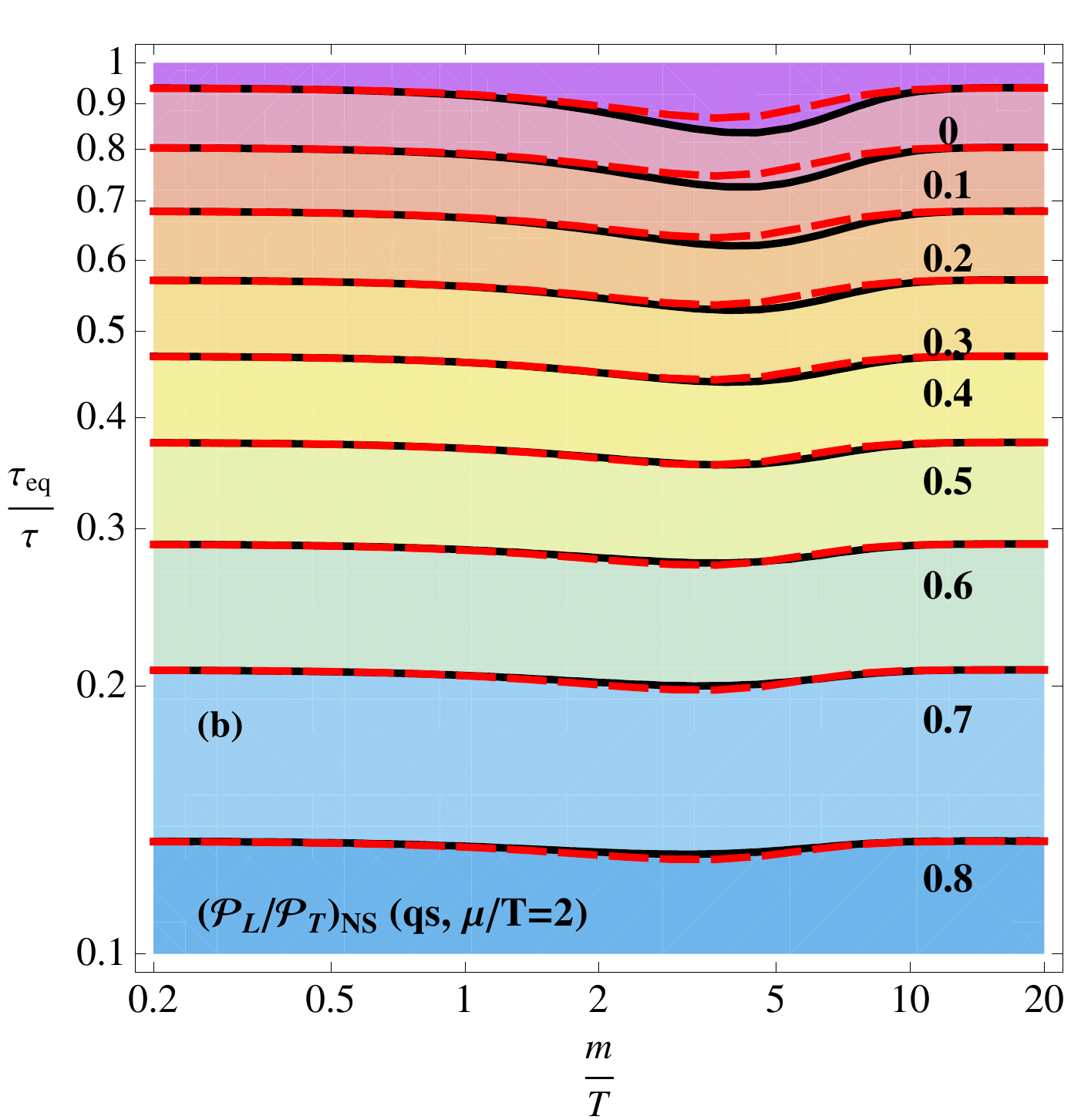}}
\end{center}
\caption{\small (Color online) Contour plots of $({\cal P}_L/{\cal P}_T)_{\rm NS}$  obtained for the Navier-Stokes hydrodynamics for $\mu = 0$ (a) and $\mu/T=2$ (b). In the two cases quarks and gluons are described by quantum statistics. The solid black lines together with the contour shading represent the classical baryon-free system.}
\label{fig:scaling}
\end{figure}

To analyze the $\left({\cal P}_L/{\cal P}_T\right)_{\rm NS}$ ratio in a general case in Fig.~\ref{fig:scaling} we plot it as a function of  two variables, 
$\teq/\tau$ and $m/T$, for a fixed value of $\mu$.
The left panel of Fig.~\ref{fig:scaling} shows the contour plot of $\left({\cal P}_L/{\cal P}_T\right)_{\rm NS}$ 
in the case where quantum statistics are used and $\mu=0$.
The fact that the contour (red dashed) lines have horizontal shapes indicates that $\left({\cal P}_L/{\cal P}_T\right)_{\rm NS}$ 
depends effectively only on $\teq/\tau$ ratio
(except for the region where $\tau \approx \teq$ and $T \approx m/5$). The red dashed lines overlap with solid black lines corresponding to the result for 
the case of classical statistics. It shows that quantum statistics  have a negligible effect on $\left({\cal P}_L/{\cal P}_T\right)_{\rm NS}$ 
in the  studied, rather broad range of $\teq/\tau$ and $m/T$. These observations explain similarities of the close-to-equilibrium behavior of $\left({\cal P}_L/{\cal P}_T\right)_{\rm NS}$ in the left panels of Figs. \ref{fig:ooPP}, \ref{fig:opPP}, and \ref{fig:ppPP}. The right panel of 
Fig.~\ref{fig:scaling} shows the contour plot of $\left({\cal P}_L/{\cal P}_T\right)_{\rm NS}$ for $\mu/T=2$. 
In this case, we find again a weak dependence on $m/T$ as compared to the case of classical statistics and $\mu=0$ represented by the solid black lines. Again this helps to understand the similarities of the right and left panels of Figs. \ref{fig:ooPP}, \ref{fig:opPP}, and \ref{fig:ppPP}.

\begin{figure}[t]
\begin{center}
\includegraphics[angle=0,width=0.6\textwidth]{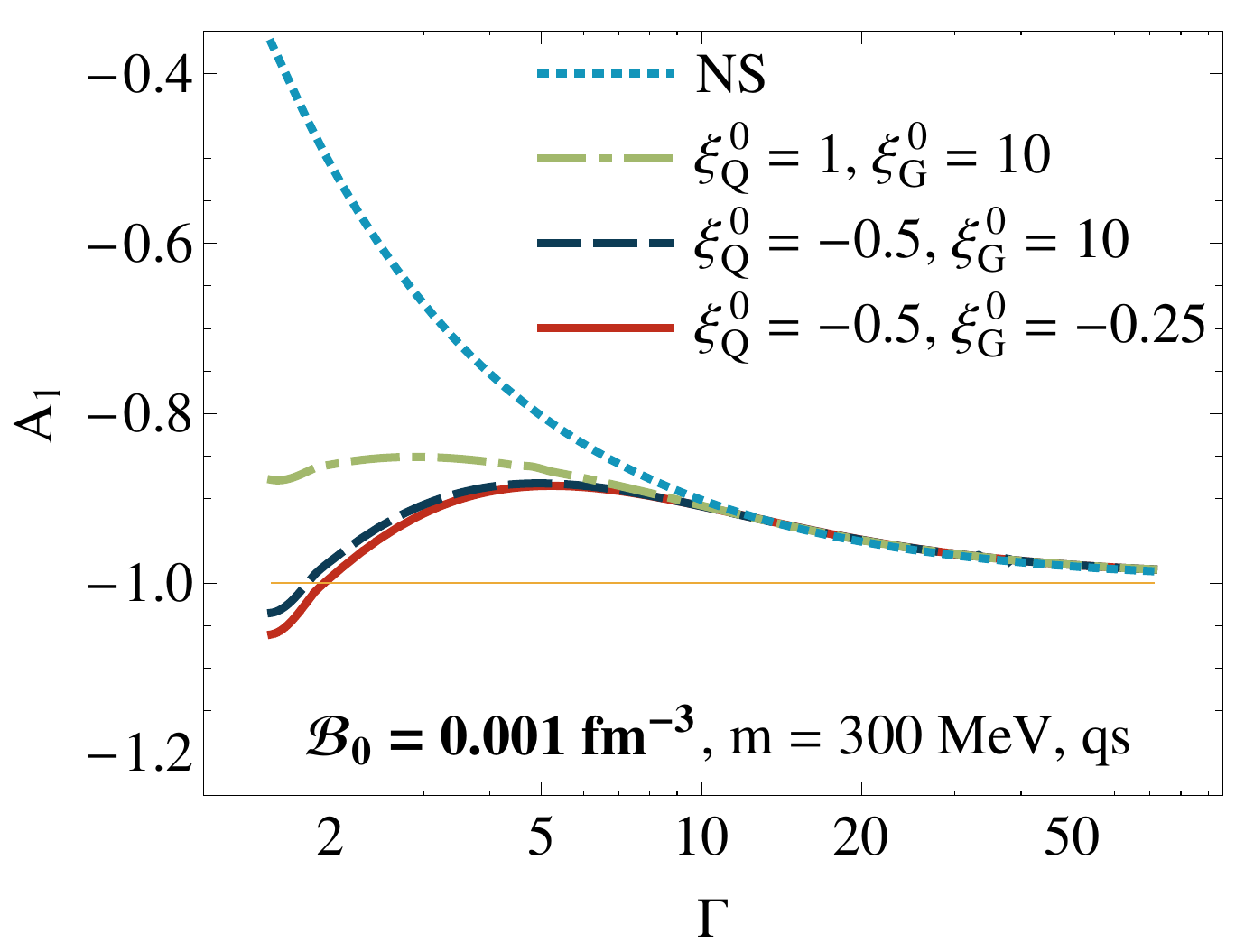}
\end{center}
\caption{\small (Color online) The quantity $A_1$ plotted as a function of $\Gamma$ for three different initial anisotropies, finite quark mass, and quantum statistics. }
\label{fig:A1}
\end{figure}

\begin{figure}[t]
\begin{center}
\includegraphics[angle=0,width=0.6\textwidth]{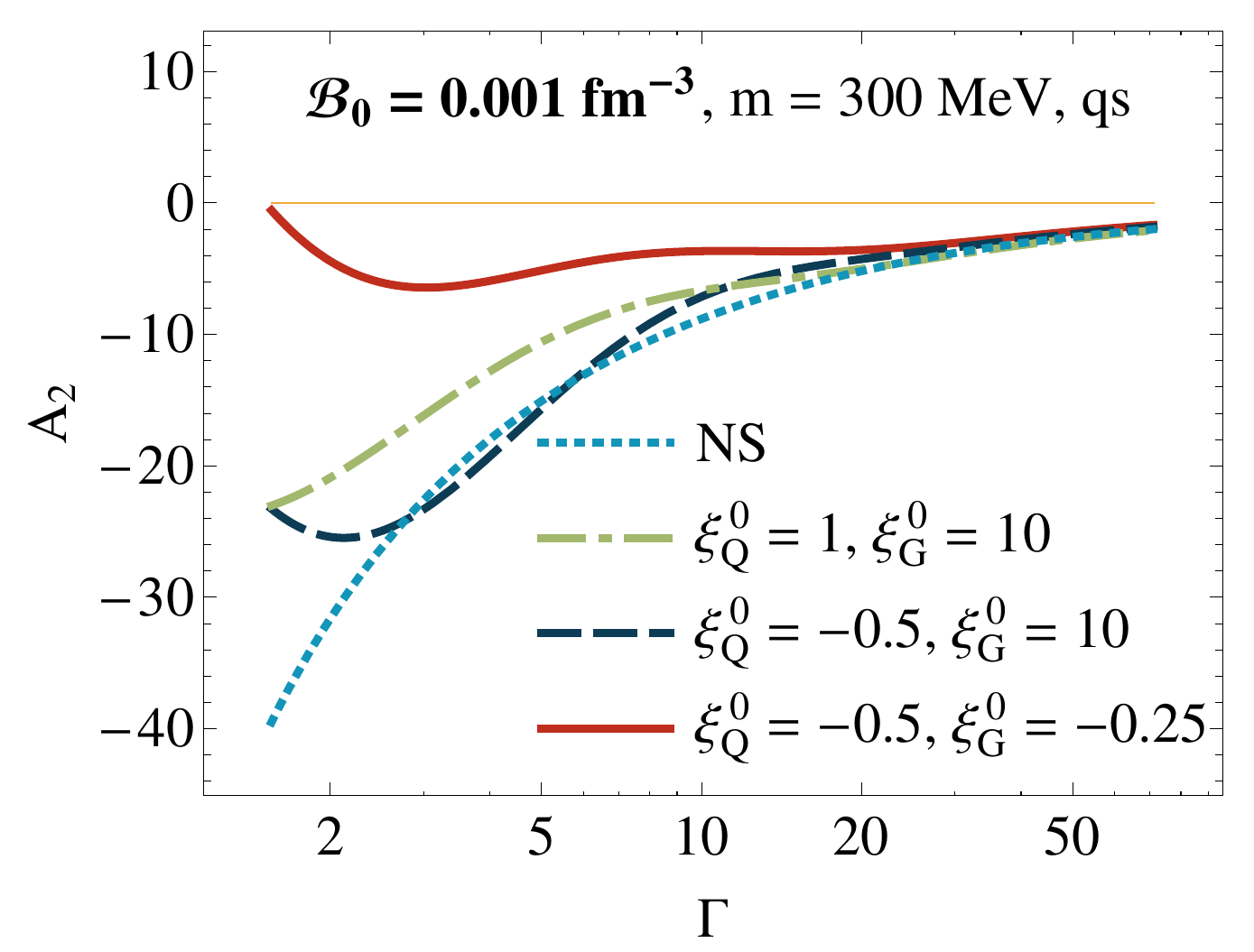}
\end{center}
\caption{\small (Color online) Same as Fig.~\ref{fig:A1} but for $A_2$ vs. $\Gamma$. }
\label{fig:A2}
\end{figure}

\subsection{Remarks on non-conformal attractors}
\label{sect:nonconfat}

In a very recent paper \cite{Romatschke:2017acs} it has been suggested by Romatschke to look for attractor behavior by studying the quantities
\bel{A1}
A_1 = \f{\tau d{\cal E}}{({\cal E}+ {\cal P}_{\rm eq})d\tau}
\eel
and
\bel{A2}
A_2 = \f{2 {\cal P}_T + {\cal P}_L - 3 {\cal P}_{\rm eq}}{\zeta T}
\eel
as functions of the variable
\bel{Gamma}
\Gamma = \tau \left[ \f{4}{3} \f{\eta}{{\cal E}+ {\cal P}_{\rm eq}} + \f{\zeta}{{\cal E}+ {\cal P}_{\rm eq}} \right]^{-1}.
\eel
Note that in Eqs. \rfn{A1}--\rfn{Gamma} we used boost invariance to simplify our notation.

In Fig.~\ref{fig:A1} we show the function $A_1(\Gamma)$ obtained for three different initial anisotropies studied in this 
work. To get the connection with \cite{Romatschke:2017acs} we consider the case with negligible baryon number density. Otherwise, we include the 
finite mass of quarks and quantum statistics. Figure ~\ref{fig:A1} shows that first the lines corresponding to three different initial
conditions converge and later approach the Navier-Stokes line. This observation supports the existence of a non-conformal attractor
for $A_1$ in our system.

Figure~\ref{fig:A2} shows similar results as Fig.~\ref{fig:A1} but for $A_2(\Gamma)$. In this case, the lines corresponding to
different initial conditions converge with each other only in the NS regime. Hence, our present results are insufficient to 
demonstrate the existence of an attractor for $A_2$.  Further study of this behavior is planned for our future investigations.

\chapter{Anisotropic hydrodynamics for a~mixture}
\label{chap:ahydromix}

\section{Anisotropic-hydrodynamics concept}
\label{sect:ahydro}

In this Thesis, following the main ideas of anisotropic hydrodynamics (aHydro) \cite{Florkowski:2010cf,Martinez:2010sc}, we make an assumption that the exact solutions, $f_{\s}(x,p)$, of the kinetic equations \rfn{kineq} are very well approximated by the RS anisotropic distributions~\cite{Romatschke:2003ms}.

The use of the RS ansatz for the quark and gluon distributions means that we deal with seven unknown functions: $\xi_\Q, \Lq, \xi_\G, \Lg, \lambda, \mu$, and $T$. Their time dependence will be determined by using a properly selected set of moments of Eqs.~\rfn{kineq}. In this work we follow Ref.~\cite{Florkowski:2015cba} and use two equations constructed from the zeroth moments, one from the first moment, and two from the second moments. In addition, we use the Landau matching conditions that guarantee the baryon number and energy-momentum conservation~\footnote{We note that the Landau matching conditions for baryon number and four-momentum follow also from appropriate combinations of the zeroth and first moments of kinetic equations, respectively.}.

\section{Zeroth moments}
\label{sect:zmotke}

The zeroth moments of the kinetic equations \rfn{kineq} give three scalar equations
\beal{0mq}
\p_\mu  \left( {\cal N}_{\Qpm, \rm a} U^\mu \right)   &=& \frac{{\cal N}_{\Qpm, \rm eq} -{\cal N}_{\Qpm, \rm a} }{\teq},
\eeal
\beal{0mg}
\p_\mu  \left( {\cal N}_{\G, \rm a} U^\mu \right)   &=& \frac{{\cal N}_{\G, \rm eq} -{\cal N}_{\G, \rm a} }{\teq}.
\eeal
To construct the hydrodynamic framework, we cannot use all the equations listed in (\ref{0mq}) and (\ref{0mg}),  since this would lead to the overdetermined system \footnote{For a discussion of this point see \cite{Florkowski:2015cba}.}. Therefore, following Ref.~\cite{Florkowski:2015cba},  we use only two equations constructed as linear combinations of (\ref{0mq}) and (\ref{0mg}). The first equation is obtained from the difference of the quark  ($\Qp$) and antiquark ($\Qm$) components in Eqs.~(\ref{0mq}),
\beal{0m1}
\f{d}{d\tau}  \left( {\cal N}_{\Qp, \rm a} - {\cal N}_{\Qm, \rm a}  \right)   + \f{{\cal N}_{\Qp, \rm a} - {\cal N}_{\Qm, \rm a}
}{\tau}
&=& \frac{{\cal N}_{\Qp, \rm eq} -{\cal N}_{\Qm, \rm eq} - \left( {\cal N}_{\Qp, \rm a} - {\cal N}_{\Qm, \rm a} \right) }{\teq},
\eeal
while the second equation is a linear combination of Eqs.~(\ref{0mq})~and~(\ref{0mg}),
\begin{eqnarray}
&& \alpha \left( \f{d {\cal N}_{Q, \rm a}}{d\tau} +  \f{{\cal N}_{\Q, \rm a}}{\tau} \right) 
+ (1-\alpha) \left( \f{d {\cal N}_{G, \rm a}}{d\tau} +  \f{ {\cal N}_{\G, \rm a}}{\tau} \right) \nn \\
&& = \alpha  \,  \frac{{\cal N}_{\Q, \rm eq}  -  {\cal N}_{\Q, \rm a}  }{\teq}
+ (1-\alpha) \,  \frac{{\cal N}_{\G, \rm eq}  -  {\cal N}_{\G, \rm a}  }{\teq}, \label{0m2}
\end{eqnarray}
where the parameter $\alpha$ is a constant taken from the range $0 \leq \alpha \leq 1$. To denote the sums of the contributions from quarks and antiquarks we use the symbol $\Q$, for example
\bel{Q}
N_\Q^\mu = N_\Qm^\mu + N_\Qp^\mu, \quad {\cal N}_{\Q} = {\cal N}_{\Qm} + {\cal N}_{\Qp}.
\eel

By doing comparisons between
the predictions of the kinetic theory and the results obtained with aHydro one can check which value of $\alpha$ is optimal.
In Ref.~\cite{Florkowski:2015cba} we found that the best cases corresponded to either $\alpha=0$ or $\alpha=1$. One may
understand this behavior, since such values of $\alpha$ do not introduce any direct coupling between the quark and gluon sectors
except for that included by the energy-momentum conservation, which is accounted for by the first moment --- such a situation
takes place in the case where kinetic equations are treated exactly. Our present investigations of more complex systems also 
favor the values $\alpha=0$ and $\alpha=1$. We return to this discussion in Chapter~\ref{sec:numresahyd}.

%
\subsection{Baryon number conservation}
\label{sect:bnc}

Equation \rfn{0m1} leads directly to the constraint on the baryon number density
\bel{Ba1}
\f{d{\cal B}_{\rm a}(\tau)}{d\tau} +\f{{\cal B}_{\rm a}(\tau)}{\tau} = \f{{\cal B}_{\rm eq} - {\cal B}_{\rm a}}{\teq}.
\eel
The conservation of the baryon number requires that both the left- and the right-hand sides of \rfn{Ba1} vanish.
This leads to two equations
\bel{Ba2}
{\cal B}_{\rm a}(\tau) = \f{{\cal B}_{\rm 0} \tau_0}{\tau}
\eel
and
\bel{Ba3}
{\cal B}_{\rm a}(\tau)  = {\cal B}_{\rm eq}(\tau),
\eel
which lead to
\begin{eqnarray}
\frac{ \Lambda_\Q^3}{ \sqrt{1+\xi_\Q}}  
\sinh\lp\frac{\lambda}{\Lambda_\Q}\rp\,{\cal H}_{\cal B}\lp \frac{m}{\Lambda_\Q},   \frac{\lambda}{\Lambda_\Q}\rp
= T^3 \sinh\lp\frac{\mu}{T}\rp\,{\cal H}_{\cal B}\lp \frac{m}{T},   \frac{\mu}{T}\rp
\label{BaBeq}
\end{eqnarray}
and
\begin{eqnarray}
\frac{16 \pi k_\Q T^3}{3} \sinh\lp\frac{\mu}{T}\rp\,{\cal H}_{\cal B}\lp \frac{m}{T},   \frac{\mu}{T}\rp = \f{{\cal B}_{\rm 0} \tau_0}{\tau}.
\label{BaB0}
\end{eqnarray}
The function ${\cal H}_{\cal B}$ is defined explicitly in \APP{ss:ae}. 

The last two equations can be  used to determine $\lambda$ and $\mu$ in terms of ${\cal B}_{\rm 0} \tau_0/\tau$, $T$, $\Lambda_\Q$,
and $\xi_\Q$. Thus, in the following equations we may treat $\lambda$ and $\mu$ as known functions of other hydrodynamic
variables.~\footnote{In the case of classical statistics, the function ${\cal H}_{\cal B}$ becomes independent of the second argument 
and Eqs.~\rfn{BaBeq} and \rfn{BaB0} can be easily solved for $\mu$ and $\lambda$. However,  in the general case of Fermi-Dirac statistics one has to solve
 Eqs.~\rfn{BaBeq} and \rfn{BaB0} numerically, together with other hydrodynamic equations. }

%
\subsection{Sum of the zeroth moments}
\label{sect:sotzm}

Equation \rfn{0m2} can be written in the form
\begin{eqnarray}
&& \f{d}{d\tau} \left[ \alpha \Lambda_\Q^3 \left(  \tilde{{\cal H}}_{\cal N}^+\lp \f{1}{\sqrt{1+\xi_\Q}}, \frac{m}{\Lambda_\Q}, - \frac{\lambda}{\Lambda_\Q}\rp
+   \tilde{{\cal H}}_{\cal N}^+\lp \f{1}{\sqrt{1+\xi_\Q}}, \frac{m}{\Lambda_\Q}, + \frac{\lambda}{\Lambda_\Q}\rp  \right) \right. \nn \\
&& \hspace{3cm} \left. + \, (1-\alpha) \, r \,  \Lambda_\G^3 \tilde{{\cal H}}_{\cal N}^-\lp \f{1}{\sqrt{1+\xi_\G}}, 0,0\rp   \right] \nn \\
&& + \left(\f{1}{\tau} + \f{1}{\teq} \right) \left[ \alpha \Lambda_\Q^3 \left(  \tilde{{\cal H}}_{\cal N}^+\lp \f{1}{\sqrt{1+\xi_\Q}}, \frac{m}{\Lambda_\Q}, - \frac{\lambda}{\Lambda_\Q}\rp
+   \tilde{{\cal H}}_{\cal N}^+\lp \f{1}{\sqrt{1+\xi_\Q}}, \frac{m}{\Lambda_\Q}, + \frac{\lambda}{\Lambda_\Q}\rp  \right) \right. \nn \\
&& \hspace{3cm} \left. + \, (1-\alpha) \, r \,  \Lambda_\G^3 \tilde{{\cal H}}_{\cal N}^-\lp \f{1}{\sqrt{1+\xi_\G}}, 0,0\rp   \right] \nn \\
&=& \f{T^3}{\teq} \left[ \alpha \left( \tilde{{\cal H}}_{\cal N}^+\lp 1, \frac{m}{T}, - \frac{\mu}{T}\rp
+   \tilde{{\cal H}}_{\cal N}^+\lp 1, \frac{m}{T}, + \frac{\mu}{T}\rp \right) + \, (1-\alpha) \, r \,  \tilde{{\cal H}}_{\cal N}^-\lp 1, 0,0\rp   \right] .
\label{ZM3}
\end{eqnarray}
Here we have introduced again the ratio of the internal degeneracies
\begin{equation}
r = \f{k_\G }{k_\Q} = \f{g_\G }{g_\Q} = \f{4}{3}.
\end{equation}
The functions $ \tilde{{\cal H}}_{\cal N}^\pm$ are defined explicitly in \APP{ss:ae}.

\section{First moments}
\label{sect:fmotke}
By considering sum over ``s'' of the first moments of the kinetic equations \rfn{kineq} we have 

\beal{KEfirsthmom3}
\p_\mu  {{T}}^{\mu\nu}_{\rm a}  &=& U_\mu   \frac{{T}^{\mu\nu}_\eq -{T}^{\mu\nu}_{\rm a}}{\teq}.
\eeal
The energy-momentum conservation requires that the right-hand side of Eq.~(\ref{KEfirsthmom3}) vanishes, which leads to the Landau matching for the energy density 
\begin{equation}
{\cal E}_{\rm a}   = {\cal E}^\eq,
\label{energyLMC2}
\end{equation} 
where ${\cal E}_{\rm a}  = {\cal E}^{\Q,\an}  + {\cal E}^{\G, \an}$  and ${\cal E}^\eq = {\cal E}^{\Q, \eq} + {\cal E}^{G, \eq}$ contain contributions from quarks, antiquarks, and gluons. In the explicit notation, we obtain
\begin{eqnarray}
 &&  \Lambda_\Q^4  \left( \tilde{{\cal H}}^+\lp \f{1}{\sqrt{1+\xi_\Q}}, \frac{m}{\Lambda_\Q}, - \frac{\lambda}{\Lambda_\Q}\rp
 +  \tilde{{\cal H}}^+\lp \f{1}{\sqrt{1+\xi_\Q}}, \frac{m}{\Lambda_\Q}, + \frac{\lambda}{\Lambda_\Q}\rp \right) \nn \\
 && \hspace{2cm} + \, r \, \Lambda_\G^4 \tilde{{\cal H}}^-\lp \f{1}{\sqrt{1+\xi_\G}}, 0,0\rp \nn \\
 &&   =  T^4 \left( \tilde{{\cal H}}^+\lp 1, \frac{m}{T}, - \frac{\mu}{T}\rp +  \tilde{{\cal H}}^+\lp 1, \frac{m}{T}, + \frac{\mu}{T}\rp
 + r \,  \tilde{{\cal H}}^-\lp 1, 0,0\rp   \right), 
 \label{FM1}
\end{eqnarray}
where the functions $ \tilde{{\cal H}}^\pm$ are defined explicitly in \APP{ss:ae}.

On the other hand, the left-hand side of Eq.~(\ref{KEfirsthmom3}) leads to the equation expressing the energy-momentum conservation of the form
\bel{SECOND-EQUATION-1A}
\f{d{\cal E}_{\rm a}   }{d\tau} = -\f{{\cal E}_{\rm a}   + {\cal P}^{\rm a} _{L}  }{\tau},
\eel
where ${\cal P}^{\rm a}_L  = {\cal P}_L^{\Q,\an} + {\cal P}_L^{G,\an}$ is the total longitudinal momentum of the system. This leads to the equation
\begin{eqnarray}
&& \f{d}{d\tau} \left[ \Lambda_\Q^4  \left( \tilde{{\cal H}}^+\lp \f{1}{\sqrt{1+\xi_\Q}}, \frac{m}{\Lambda_\Q}, - \frac{\lambda}{\Lambda_\Q}\rp
 +  \tilde{{\cal H}}^+\lp \f{1}{\sqrt{1+\xi_\Q}}, \frac{m}{\Lambda_\Q}, + \frac{\lambda}{\Lambda_\Q}\rp \right) \right. \nn \\
&& \left. \hspace{2cm}  + \, r \, \Lambda_\G^4 \tilde{{\cal H}}^-\lp \f{1}{\sqrt{1+\xi_\G}}, 0,0\rp \right] \nn \\
&& = -\f{1}{\tau}  \left[ \Lambda_\Q^4  \left( \tilde{{\cal H}}^+\lp \f{1}{\sqrt{1+\xi_\Q}}, \frac{m}{\Lambda_\Q}, - \frac{\lambda}{\Lambda_\Q}\rp
 +  \tilde{{\cal H}}^+\lp \f{1}{\sqrt{1+\xi_\Q}}, \frac{m}{\Lambda_\Q}, + \frac{\lambda}{\Lambda_\Q}\rp \right) \right. \nn \\
 &&  \hspace{1.2cm} +  \Lambda_\Q^4  \left( \tilde{{\cal H}}^+_L \lp \f{1}{\sqrt{1+\xi_\Q}}, \frac{m}{\Lambda_\Q}, - \frac{\lambda}{\Lambda_\Q}\rp
 +  \tilde{{\cal H}}^+_L \lp \f{1}{\sqrt{1+\xi_\Q}}, \frac{m}{\Lambda_\Q}, + \frac{\lambda}{\Lambda_\Q}\rp \right)  \nn \\
&& \left. \hspace{2cm}  + \, r \, \Lambda_\G^4 \left( \tilde{{\cal H}}^-\lp \f{1}{\sqrt{1+\xi_\G}}, 0,0\rp
+ \tilde{{\cal H}}^-_L \lp \f{1}{\sqrt{1+\xi_\G}}, 0,0\rp \right) \right], \label{FM2} 
\end{eqnarray}
where the functions $ \tilde{{\cal H}}_L^\pm$ are again defined in \APP{ss:ae}. We note that Eqs.~\rfn{FM1} and \rfn{FM2} couple the quark and gluon hydrodynamic parameters in the way similar to that known from the exact treatment of the kinetic equations.

\section{Second moments}
\label{sect:2nd}

In order to close the system of dynamical equations, we finally consider the second moments of  the kinetic equations \rfn{kineq}, 

\beal{KEsecondhmom3}
\p_\lambda  {{\Theta}}_{\s, \an}^{\lambda\mu\nu}  &=& U_\lambda   \frac{{\Theta}^{\lambda\mu\nu}_{\s,\eq} -{\Theta}^{\lambda\mu\nu}_{\s, \an}}{\teq}.
\eeal
Using tensor decompositions \rfn{thetatensoreq} and \rfn{thetatensoran} and performing the projections of Eqs.~(\ref{KEsecondhmom3}) on the basis four-vectors, one obtains, however, an overdetermined system of equations. A possible remedy to this problem was proposed in Ref.~\cite{Tinti:2013vba} where a selection rule for equations was proposed, see also \cite{Nopoush:2014pfa}.
In this work, we select the same combinations of the second moments of the Boltzmann equations as in Ref.~\cite{Florkowski:2015cba}, which follows the methodology of Refs.~\cite{Tinti:2013vba, Nopoush:2014pfa}. This implies that we use the equations of the form
\begin{eqnarray}
&& \frac{d}{d\tau}\ln\vartheta_T^{\Q,\an}-\frac{d}{d\tau}\ln\vartheta_L^{\Q,\an}-\frac{2}{\tau}
=\frac{\vartheta^{\Q,\eq}}{\tau_{\rm eq}}\left[\frac{1}{\vartheta_T^{\Q,\an}}-\frac{1}{\vartheta_L^{\Q,\an}}  \right] , \label{sumXq} \\
&& \frac{d}{d\tau}\ln\vartheta_T^{\G,\an}-\frac{d}{d\tau}\ln\vartheta_L^{\G,\an}-\frac{2}{\tau}
=\frac{\vartheta^{\G,\eq}}{\tau_{\rm eq}}\left[\frac{1}{\vartheta_T^{\G,\an}}-\frac{1}{\vartheta_L^{\G,\an}}  \right] .  \label{sumXg}
\end{eqnarray}
It has been demonstrated in Ref.~\cite{Tinti:2013vba} that such forms are consistent with the Israel-Stewart theory for systems being close to local equilibrium.
 The explicit expressions for the functions $\vartheta$ are given in the Appendix \ref{s:sm}.
Using~Eqs.~\rfn{sumXq} and \rfn{sumXg} we find

\begin{eqnarray}
&& \f{1}{1+\xq} \f{d\xq}{d\tau} -\frac{2}{\tau}
= - \frac{\xq  (1+\xq)^{1/2}}{\tau_{\rm eq}} \frac{T^5 }{\Lq^5} \f{\tilde{{\cal H}}_{\vartheta}^+\lp  \frac{m}{T}, - \frac{\mu}{T}\rp +\tilde{{\cal H}}_{\vartheta}^+\lp  \frac{m}{T}, + \frac{\mu}{T}\rp} {\tilde{{\cal H}}_{\vartheta}^+\lp  \frac{m}{\Lq}, - \frac{\lambda}{\Lq}\rp +\tilde{{\cal H}}_{\vartheta}^+\lp  \frac{m}{\Lq}, + \frac{\lambda}{\Lq}\rp}
, \label{sumXqgen} \\
&&  \f{1}{1+\xg} \f{d\xg}{d\tau} -\frac{2}{\tau}
= - \frac{\xg  (1+\xg)^{1/2}}{\tau_{\rm eq}} \f{T^5} {\Lg^5}, \label{sumXggen}
\end{eqnarray} 
with the function $\tilde{{\cal H}}_{\vartheta}^+$ defined by \rf{Htheta}.

%
\chapter{Numerical results for anisotropic hydrodynamics}
\label{sec:numresahyd} 
%
\begin{figure}[t]
\centering
\includegraphics[angle=0,width=0.6\textwidth]{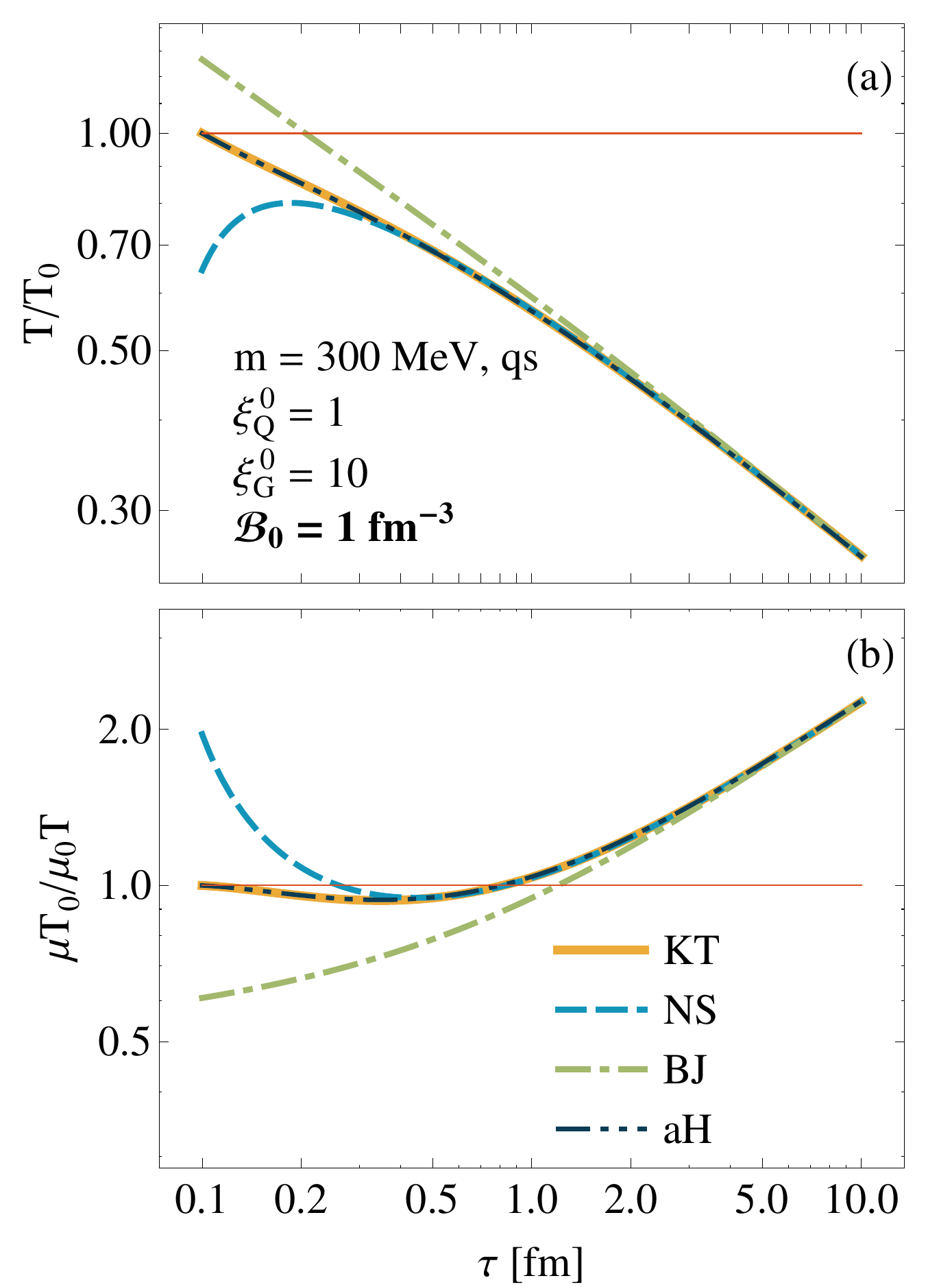} 
\caption{\small (Color online)  Proper-time dependence of the effective temperature (a) and the effective baryon chemical potential (b).  The exact kinetic-theory result
(KT, brown solid line) is compared with the aHydro (aH, navy blue double-dot-dashed line), Navier-Stokes (NS, blue dashed line), and perfect-fluid (BJ, green dot-dashed line) results, respectively. The kinetic and aHydro calculations start with the same initial conditions. The NS and BJ calculations are adjusted to reproduce the late time behavior of the KT calculation. }
\label{fig:MuT_oo_ex_Bj_NS_aH}
\end{figure}
%
\begin{figure}[!ht]
\centering
\includegraphics[angle=0,width=0.925\textwidth]{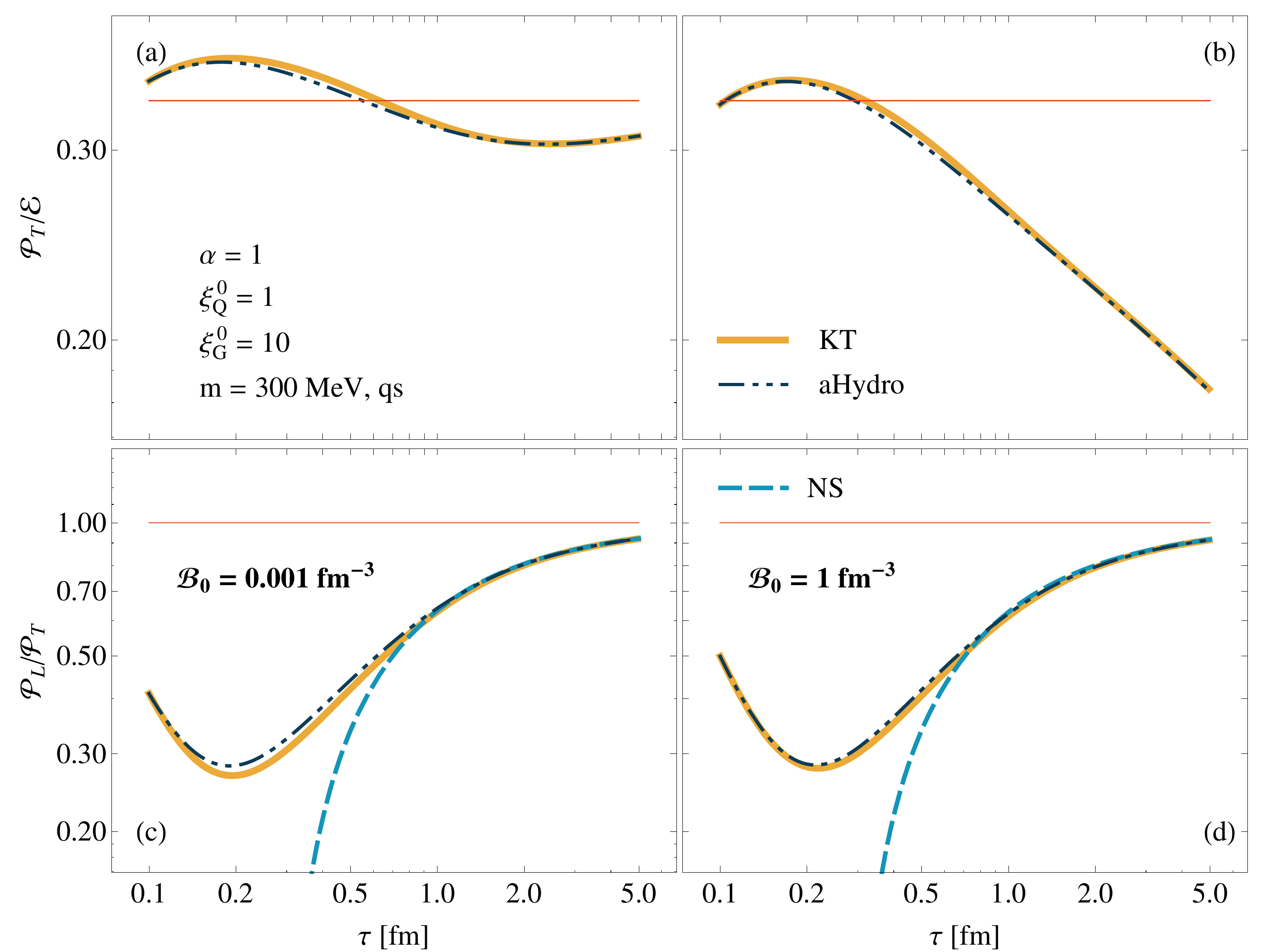} 
\caption{\small (Color online) Proper-time dependence of the ${\cal P}_T/{\cal E}$ (a) and ${\cal P}_L/{\cal P}_T$ (b) ratios for the initial oblate-oblate configuration. Notation the same as in Fig.~\ref{fig:MuT_oo_ex_Bj_NS_aH}.}
\label{fig:P_m300qs_NS_KT_aH}
\end{figure}
%
\begin{figure}[!t]
\centering
\includegraphics[angle=0,width=0.925\textwidth]{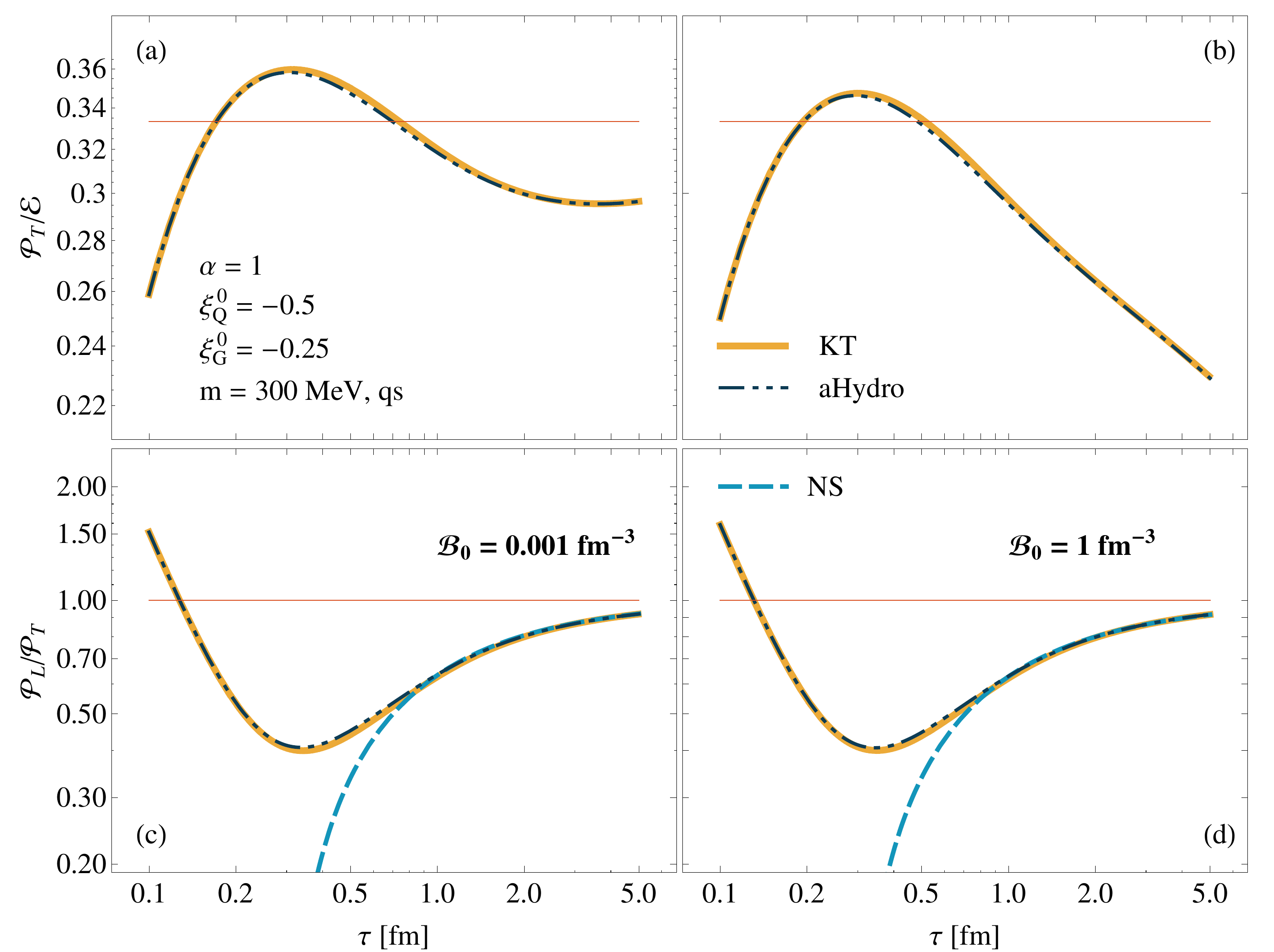} 
\caption{\small (Color online) Same as Fig.~\ref{fig:P_m300qs_NS_KT_aH} but for the prolate-prolate initial configuration. }
\label{fig:P_m300qs_pp_NS_KT_aH}
\end{figure}
%

Equations \rfn{BaBeq}, \rfn{BaB0}, \rfn{ZM3}, \rfn{FM1}, \rfn{FM2}, \rfn{sumXqgen} and \rfn{sumXggen} are seven equations for
seven unknown functions of the proper time: $\xi_\Q(\tau), \Lq(\tau), \xi_\G(\tau), \Lg(\tau), \lambda(\tau), \mu(\tau)$, and $T(\tau)$.
Three equations [\rfn{BaBeq}, \rfn{BaB0} and \rfn{FM1}] are algebraic but one can differentiate them with respect to proper time and
use them together with the remaining four equations [\rfn{ZM3},  \rfn{FM2}, \rfn{sumXqgen} and \rfn{sumXggen}] as a system of 
seven ordinary differential equations of the first order. Such equations require initial values which we choose in a similar way as
in~\cite{Florkowski:2017jnz}. The initial values of the anisotropy parameters correspond to the two options:  i) $\xi_{\Q}^{0}=1$ and $\xi_{\G}^{0}=10$, and 
ii)~$\xi_{\Q}^{0}=-0.5$ and $\xi_{\G}^{0}=-0.25$. Such values correspond to oblate-oblate and prolate-prolate initial momentum distributions of quarks and gluons, respectively. The same initial values for $\xi_{\Q}^{0}$ and $\xi_{\G}^{0}$ were used before also in Ref.~\cite{Florkowski:2015cba}. The initial transverse momentum scales of quarks and gluons are assumed to be the same and equal to $\Lambda_{\Q}^{0}=\Lambda_{\G}^{0}=1$~GeV.  The gluons are treated as massless, while quarks have a finite mass of 300~MeV. The initial non-equilibrium  chemical potential $\lambda_0$ is chosen in such a~way that the initial baryon number density is ${\cal B}_0=$~1~fm$^{-3}$.  The initial proper time is $\tau_0=$~0.1~fm and the relaxation time is $\teq=$~0.25~fm. The results shown in this section were obtained with $\alpha=0$ in \rf{0m2}. We comment on this choice below.

In Fig.~\ref{fig:MuT_oo_ex_Bj_NS_aH} we show the proper-time dependence of the effective temperature (a) and the effective baryon chemical potential (b).  The exact kinetic-theory result (KT, brown solid line) is compared with the aHydro (aH, navy blue double-dot-dashed line), Navier-Stokes (NS, blue dashed line), and perfect-fluid (BJ, green dot-dashed line) results, respectively. The kinetic and aHydro calculations start with the same initial conditions corresponding to oblate-oblate configuration defined above. The NS and BJ calculations are adjusted in such a way as to reproduce the late time behavior of the KT calculation.
We observe that the KT and aHydro results agree very well during the whole evolution process. On~the~other hand, the NS and BJ results can reproduce
the KT result only when the system approaches local equilibrium. As expected, the NS framework coincides with the KT result much earlier than~BJ.

If the functions $\xi_\Q(\tau), \Lq(\tau), \xi_\G(\tau), \Lg(\tau), \lambda(\tau), \mu(\tau)$, and $T(\tau)$ are known, we may determine  the anisotropic RS distribution functions for quarks and gluons and, subsequently, use these distributions to calculate various physical observables. In Fig.~\ref{fig:P_m300qs_NS_KT_aH} we show the proper-time dependence of the ${\cal P}_T/{\cal E}$ (a) and ${\cal P}_L/{\cal P}_T$ (b) ratios for the initial oblate-oblate configuration, for which the functions $T(\tau)$ and $\mu(\tau)$ are shown in Fig.~\ref{fig:MuT_oo_ex_Bj_NS_aH}. We find that aHydro reproduces very well the kinetic-theory results.

Similar, very good agreement between the KT and aHydro results is shown in Fig.~\ref{fig:P_m300qs_pp_NS_KT_aH} for the prolate-prolate
initial conditions with $\xi_{\Q}^{0}=-0.5$ and $\xi_{\G}^{0}=-0.25$. Other studied cases, not shown here, also confirm very good performance of anisotropic hydrodynamics as an approximation for the kinetic theory (as far as the observables studied in this work are considered). We note that we have achieved excellent agreement with the parameter $\alpha$ set equal to zero in  \rf{0m2}. Similar consistent results can be  found also in the case $\alpha =1$. On the other hand, choosing a finite value of $\alpha$ from the range $0 < \alpha < 1$ spoils the agreement. This behavior was already observed in  \cite{Florkowski:2015cba}  where the case with massless particles and classical statistics was studied. Apparently, the values $0 < \alpha < 1$ introduce a redundant coupling between quarks and gluons, which is absent in the RTA kinetic theory.

\chapter{Summary}
\label{sec:sumcon}

In this work, we have solved a system of coupled kinetic equations for quarks, antiquarks and gluons in the relaxation time
approximation. We have generalized previous results by including the finite quark mass, the quantum statistics for both
quarks and gluons, and the finite baryon number. We have compared the results of the numerical calculations with the
first-order hydrodynamic calculations to demonstrate the hydrodynamization process. We have found that
equalization of the longitudinal and transverse pressures takes place earlier than equalization of the average and equilibrium pressures.
We have determined the shear and bulk viscosities of a mixture and find that the shear viscosity is a sum of the quark and gluon shear viscosities, 
while the bulk viscosity of a mixture is given by the formula known for a massive quark gas. However, the bulk viscosity depends 
on thermodynamic coefficients characterizing the whole mixture rather than quarks alone, which means that massless gluon
do contribute to the bulk viscosity (if quarks are massive).

\appendix
%

%
\chapter{Generalized thermodynamic functions}
\label{s:tf}
%
In this section, we present explicit expressions for various physical quantities such as the particle and energy densities or the transverse  and longitudinal pressures. These expressions are obtained with the use of different distribution functions which not necessarily correspond to local equilibrium. Thus, we call them generalized thermodynamic functions --- in local equilibrium they become standard thermodynamic functions satisfying well known thermodynamic identities.  We start with the anisotropic RS distributions, as other cases can be easily worked out if the results for the RS distributions are known.
%
\section{Anisotropic distributions}
\label{ss:ae}
%
The forms of the generalized thermodynamic functions  for anisotropic distributions (\ref{Qa}) and (\ref{Ga}) are given by the following integrals:
\begin{eqnarray}
{\cal N}^{\s,\an}  \equiv n_{U}^{\s, \an} &=&  k_\s   \int \!dP \, \lp p\cdot U \rp  f_{\s, \an}\lsb\VP\,  p \cdot U, \, p \cdot Z \rsb   ,
\label{eq:nans}\\
{\cal E}^{\s,\an} \equiv t_{UU}^{\s, \an} &=& k_\s\int dP \, \lp p\cdot U \rp^2 f_{\s, \an}\lsb\VP\,  p \cdot U,\, p \cdot Z\rsb ,
\label{eq:eans}\\
{\cal P}^{\s,\an}_T \equiv t_{AA}^{\s, \an} 
&=& k_\s\int  dP \, \lp p\cdot X \rp^2 f_{\s, \an}\lsb\VP\,  p \cdot U,\, p \cdot Z\rsb  \qquad (A\neq U, Z) \nn\\
&=& k_\s\int  dP \, \lp p\cdot Y \rp^2 f_{\s, \an}\lsb\VP\,  p \cdot U,\, p \cdot Z\rsb \nn\\
&=& - \frac{k_\s}{2}\int dP \, (p\cdot \Delta_T \cdot p)f_{\s, \an}\lsb\VP\,  p \cdot U,\, p \cdot Z\rsb ,
\label{eq:ptans}\\
{\cal P}^{\s,\an}_L \equiv t_{ZZ}^{\s, \an} &=& k_\s\int  dP \, \lp p\cdot Z \rp^2 f_{\s, \an}\lsb\VP\,  p \cdot U,\, p \cdot Z\rsb  .
\label{eq:plans}
\end{eqnarray}
The explicit calculations lead to the following expressions  for quarks and antiquarks
\begin{eqnarray}
{\cal N}^{\Qpm, \an}    &=&   4 \pi k_\Q  \Lambda_\Q^3 \tilde{{\cal H}}_{\cal N}^+\lp \f{1}{\sqrt{1+\xi_\Q}}, \frac{m}{\Lambda_\Q}, \mp \frac{\lambda}{\Lambda_\Q}\rp,
\label{eq:nan}\\
 {\cal E}^{\Qpm, \an}   &=&    2 \pi k_\Q    \Lambda_\Q^4 \tilde{{\cal H}}^+\lp \f{1}{\sqrt{1+\xi_\Q}}, \frac{m}{\Lambda_\Q}, \mp \frac{\lambda}{\Lambda_\Q}\rp,
\label{eq:ean}\\
{\cal P}^{\Qpm,\an}_T   &=&     \pi k_\Q   \Lambda_\Q^4 \tilde{{\cal H}}_T^+\lp \f{1}{\sqrt{1+\xi_\Q}}, \frac{m}{\Lambda_\Q}, \mp \frac{\lambda}{\Lambda_\Q}\rp,
\label{eq:ptan}\\
{\cal P}^{\Qpm,\an}_L   &=&   2 \pi k_\Q   \Lambda_\Q^4 \tilde{{\cal H}}_L^+\lp \f{1}{\sqrt{1+\xi_\Q}}, \frac{m}{\Lambda_\Q}, \mp \frac{\lambda}{\Lambda_\Q}\rp,
\label{eq:plan}
\end{eqnarray}
 where functions $\tilde{{\cal H}}\lp a,y,z\rp$ are defined by the integrals:
\beal{Ht}
\tilde{{\cal H}}_{\cal N}^\pm\lp a,y,z\rp &\equiv& \int\limits_0^\infty r^2  dr \, h^{^\pm}_\eq\lp \sqrt{r^2+y^2}+z\rp a,  \\
\tilde{{\cal H}} ^\pm\lp a,y,z\rp &\equiv& \int\limits_0^\infty r^3  dr \,  h^{^\pm}_\eq\lp \sqrt{r^2+y^2}+z\rp   {\cal H}_{2}\lp a,\frac{y}{r} \rp, \nn\\
\tilde{{\cal H}}_T^\pm\lp a,y,z\rp &\equiv& \int\limits_0^\infty r^3  dr \,  h^{^\pm}_\eq\lp \sqrt{r^2+y^2}+z\rp   {\cal H}_{2T}\lp a,\frac{y}{r} \rp, \nn\\
\tilde{{\cal H}}_L ^\pm\lp a,y,z\rp &\equiv& \int\limits_0^\infty r^3  dr  \, h^{^\pm}_\eq\lp \sqrt{r^2+y^2}+z\rp   {\cal H}_{2L}\lp a,\frac{y}{r} \rp, \nn
\eeal
and the functions ${\cal H}_2(a,b)$ were introduced in~\cite{Florkowski:2014sfa}:
\beal{H2}
{\cal H}_{2}\lp a,b \rp &\equiv& a  \int \limits_0^\pi   d\varphi\,\sin\varphi\sqrt{ a^2 \cos^2\varphi +\sin^2\varphi +b^2},\\ 
{\cal H}_{2T}\lp a,b \rp &\equiv& a \int\limits_0^{\pi} d\varphi\,
\f{\sin^3\varphi  }{\sqrt{a^2\cos^2\varphi +\sin^2\varphi +b^2}},\nn\\ 
{\cal H}_{2L}\lp a,b \rp &\equiv& a^3\int\limits_0^{\pi} d\varphi\,
\f{\sin\varphi \,\cos^2\varphi}{\sqrt{a^2\cos^2\varphi +\sin^2\varphi +b^2}}.\nn 
\eeal
The integrals in \EQB{H2} are analytic and give~\cite{Florkowski:2014sfa}:
\bea\label{eq:H2}
\hspace{-1cm} {\cal H}_2(a,b)  \!\!&=&\!\! \frac{a}{\sqrt{a^2-1}} \left( (1+b^2)
\tanh^{-1} \sqrt{\frac{a^2-1}{a^2+b^2}} + \sqrt{(a^2-1)(a^2+b^2)} \, \right),\\
\hspace{-1cm} {\cal H}_{2T}(a,b)  
\!\!&=&\!\! \frac{a}{(a^2-1)^{3/2}}
\lp\left(b^2+2a^2-1\right) 
\tanh^{-1}\sqrt{\frac{a^2-1}{a^2+b^2}}
-\sqrt{(a^2-1)(a^2+b^2)} \rp,  
\label{eq:H2T}\\
\hspace{-1cm} {\cal H}_{2L}(a,b)  
\!\!&=&\!\! \frac{a^3}{(a^2-1)^{3/2}}
\lp-(1+b^2)
\tanh^{-1}\sqrt{\frac{a^2-1}{a^2+b^2}}
+\sqrt{(a^2-1)(a^2+b^2)} \,\,\rp. 
\label{eq:H2L}
\eea
With $b=0$ the functions ${\cal H}_{2}(a,b)$, ${\cal H}_{2T}(a,b)$ and ${\cal H}_{2L}(a,b)$ reduce to the functions ${\cal H}(a)$, ${\cal H}_{T}(a)$ and ${\cal H}_{L}(a)$  used in~\cite{Florkowski:2013lza,Florkowski:2013lya}. 

For gluons one has:  
\begin{eqnarray}
{\cal N}^{\G, \an}  &=&  4 \pi k_\G  \Lambda_\G^3 \tilde{{\cal H}}_{\cal N}^-\lp \f{1}{\sqrt{1+\xi_\G}}, 0,0\rp = 8 \pi \zeta(3)  k_\G  \Lambda_\G^3 \frac{1}{\sqrt{1+\xi_\G}} ,
\label{eq:nang}\\
 {\cal E}^{\G, \an} &=&  2 \pi k_\G \Lambda_\G^4 \tilde{{\cal H}}^-\lp \f{1}{\sqrt{1+\xi_\G}}, 0,0\rp =\frac{2  \pi^5}{15} k_\G \Lambda_\G^4  {\cal H}  \lp \f{1}{\sqrt{1+\xi_\G}} \rp,
\label{eq:eang}\\
{\cal P}^{\G,\an}_T  &=&   \pi k_\G   \Lambda_\G^4 \tilde{{\cal H}}_T^-\lp \f{1}{\sqrt{1+\xi_\G}}, 0,0\rp =\frac{ \pi^5}{15} k_\G \Lambda_\G^4  {\cal H}_T  \lp \f{1}{\sqrt{1+\xi_\G}} \rp,
\label{eq:ptang}\\
{\cal P}^{\G,\an}_L  &=& 2 \pi k_\G   \Lambda_\G^4 \tilde{{\cal H}}_L^-\lp \f{1}{\sqrt{1+\xi_\G}}, 0,0\rp =\frac{ 2 \pi^5}{15} k_\G \Lambda_\G^4  {\cal H}_L  \lp \f{1}{\sqrt{1+\xi_\G}} \rp,
\label{eq:plang}
\end{eqnarray}
where $\zeta$ is the Riemann zeta function~\footnote{The coefficient $\zeta(3)$ is known as Ap\'ery's constant.}. The expressions on the right-hand sides of
Eqs.~\rfn{eq:nang}--\rfn{eq:plang} hold for the Bose-Einstein statistics. Note that in the case of massless gluons the integrals \rfn{H2} are done for $b=0$ and can be factorized in~\EQS{Ht}. The~functions ${\cal H} $, ${\cal H}_T $, and ${\cal H}_L $ in Eqs.~\rfn{eq:eang}--\rfn{eq:plang} are defined in \cite{Florkowski:2013lza,Florkowski:2013lya}.

It is useful to notice that the functions ${\cal H}_{2}$ and ${\cal H}_{2L}$ are related by the expression
\bel{dH2}
\f{\partial {\cal H}_{2}\lp a,b \rp}{\partial a} = \f{{\cal H}_{2}\lp a,b \rp + {\cal H}_{2L}\lp a,b \rp}{a},
\eel
hence, we also have
\bel{dtH2}
\f{\partial \tilde{{\cal H}} ^\pm\lp a,y,z\rp }{\partial a} = \f{\tilde{{\cal H}} ^\pm\lp a,y,z\rp + \tilde{{\cal H}} ^\pm\lp a,y,z\rp}{a}.
\eel
We can use \rfn{dtH2} to derive \rfn{SECOND-EQUATION-1} from \rfn{SECOND-EQUATION}.

We close this section with the formula for the baryon number density valid for anisotropic RS systems
\begin{eqnarray}
{\cal B}^\an &=& \frac{{\cal N}^{\Qp, \an} -{\cal N}^{\Qm, \an}}{3} 
=   \frac{16 \pi k_\Q \Lambda_\Q^3}{3 \sqrt{1+\xi_\Q}}  \sinh\lp\frac{\lambda}{\Lambda_\Q}\rp\,{\cal H}_{\cal B}\lp \frac{m}{\Lambda_\Q},   \frac{\lambda}{\Lambda_\Q}\rp,
\end{eqnarray} 
where
\begin{eqnarray}
{\cal H}_{\cal B}\lp y,z\rp &\equiv&
  \frac{1}{4}\int\limits_0^\infty r^2  dr \lsb  \frac{1}{ \cosh \sqrt{r^2+y^2}+\cosh z   } \rsb .
\label{HB}
\end{eqnarray}
\pagebreak
\section{Isotropic distributions}
\label{ss:ie}
%
The forms of the thermodynamic functions for the isotropic equilibrium state are commonly known, nevertheless, we quote them here for completeness. They are given by the formulas
\begin{eqnarray}
{\cal N}^{\s, \eq} \equiv n_{U}^{\s, \eq} &=&  k_\s   \int \! dP \, \lp p\cdot U \rp f_{\s, \eq}(p\cdot U)   ,
\label{eq:neq}\\
 {\cal E}^{\s, \eq}\equiv t_{UU}^{\s, \eq} &=& k_\s\int dP \, \lp p\cdot U \rp^2 f_{\s, \eq}(p\cdot U), 
\label{eq:eeq}\\
{\cal P}^{\s, \eq}\equiv t_{AA}^{\s, \eq} &=& k_\s\int \!dP\, \lp p\cdot A \rp^2 f_{\s, \eq}(p\cdot U)  \nn\\
&=& -\frac{k_\s}{3}\int \!dP\, (p\cdot \Delta \cdot p) f_{\s, \eq}(p\cdot U),  \quad (A\neq U) .
\label{eq:peq}
\end{eqnarray}
The explicit expressions for quarks and antiquarks may be obtained  from  \EQSM{eq:nan}{eq:plan} as a special case of $\xi_\s\to 0$, $\Lambda_\s\to T$, and $\lambda_\s\to \mu$,
\begin{eqnarray}
{\cal N}^{\Qpm, \eq} \!\!  &=&\!\!  4 \pi k_\Q  T^3 \tilde{{\cal H}}_{\cal N}^+\lp 1, \frac{m}{T}, \mp \frac{\mu}{T}\rp,
\label{eq:nqeq}\\
 {\cal E}^{\Qpm, \eq} \!\! &=&\!\!   2 \pi k_\Q    T^4 \tilde{{\cal H}}^+\lp 1, \frac{m}{T}, \mp \frac{\mu}{T}\rp,
\label{eq:eqeq}\\
{\cal P}^{\Qpm,\eq}_T  \!\! &=&\!\!    \pi k_\Q   T^4 \tilde{{\cal H}}_T^+\lp 1, \frac{m}{T}, \mp \frac{\mu}{T}\rp,
\label{eq:ptqeq}\\
{\cal P}^{\Qpm,\eq}_L  \!\! &=&\!\!  2 \pi k_\Q   T^4 \tilde{{\cal H}}_L^+\lp 1, \frac{m}{T}, \mp \frac{\mu}{T}\rp.
\label{eq:plqeq}
\end{eqnarray}
Note that ${\cal H}_{2L}(1,b)=2/\lp 3\sqrt{1+b^2} \rp$ and ${\cal H}_{2T}(1,b)=4/\lp 3\sqrt{1+b^2} \rp$ which means that ${\cal P}^{\Qpm,\eq}_T ={\cal P}^{\Qpm,\eq}_L \equiv {\cal P}^{\Qpm,\eq}$, as expected for the isotropic state.  
%
Analogous results may be obtained for gluons  
\begin{eqnarray}
{\cal N}^{\G, \eq}  \!\!&=&\!\!  4 \pi k_\G  T^3 \tilde{{\cal H}}_{\cal N}^-\lp 1, 0,0\rp = 8  \pi \zeta(3) k_\G  T^3 ,
\label{eq:ngeq}\\
 {\cal E}^{\G, \eq} \!\!&=&\!\!  2 \pi k_\G T^4 \tilde{{\cal H}}^-\lp 1, 0,0\rp=\frac{4  \pi^5}{15} k_\G T^4 ,
\label{eq:egeq}\\
{\cal P}^{\G,\eq}_T  \!\!&=&\!\!   \pi k_\G   T^4 \tilde{{\cal H}}_T^-\lp 1, 0,0\rp=\frac{4  \pi^5}{45} k_\G T^4,
\label{eq:ptgeq}\\
{\cal P}^{\G,\eq}_L  \!\!&=&\!\! 2 \pi k_\G   T^4 \tilde{{\cal H}}_L^-\lp 1, 0,0\rp=\frac{4  \pi^5}{45} k_\G T^4,
\label{eq:plgeq}
\end{eqnarray}
where to get the last expressions on the right-hand sides we again assumed the Bose-Einstein statistics. Here, similarly as for quarks ${\cal P}^{\G,\eq}_T ={\cal P}^{\G,\eq}_L \equiv {\cal P}^{\G,\eq}$. 
Similarly to the anisotropic case the baryon number density is
\begin{eqnarray}
{\cal B}^\eq &=& \frac{{\cal N}^{\Qp, \eq} -{\cal N}^{\Qm, \eq}}{3} 
= \frac{16 \pi k_\Q T^3}{3} \sinh\lp\frac{\mu}{T}\rp\,{\cal H}_{\cal B}\lp \frac{m}{T},   \frac{\mu}{T}\rp.
\label{BeqApp}
\end{eqnarray} 
with ${\cal H}_{\cal B}\lp y,z\rp$ given in \EQ{HB}.
%
\section{Exact distributions}
\label{ss:esotke}
%
For the solutions of \EQS{BIke} of the form \EQB{formsolQ} the thermodynamic variables have the forms
\begin{eqnarray}
{\cal N}^{\s}  &=&  k_\s   \int \!dP \, \lp p\cdot U \rp  f_{\s}\lsb\VP\,  p \cdot U, p \cdot Z \rsb   \nn,
\label{eq:nane}\\
{\cal E}^{\s}  &=& k_\s\int dP \, \lp p\cdot U \rp^2 f_{\s}\lsb\VP\,  p \cdot U, p \cdot Z\rsb ,
\label{eq:eane}\\
{\cal P}^{\s}_T  &=& k_\s\int  dP \, \lp p\cdot A \rp^2 f_{\s}\lsb\VP\,  p \cdot U, p \cdot Z\rsb  \qquad  (A\neq U,Z), \nn
\label{eq:ptane}\\
{\cal P}^{\s}_L  &=& k_\s\int  dP \, \lp p\cdot Z \rp^2 f_{\s}\lsb\VP\,  p \cdot U, p \cdot Z\rsb. \nn
\label{eq:plane}
\end{eqnarray}
Using the above definitions and repeating the calculation from \SEC{ss:ae}, gives 
\begin{eqnarray}
{\cal N}^{\Qpm}  &=&  4 \pi k_\Q \lsb  \lp\Lambda_\Q^0\rp^3 \tilde{{\cal H}}_{\cal N}^+\lp \f{\tau_0}{\tau \sqrt{1+\xi_\Q^0}}, \frac{m}{\Lambda_\Q^0}, \mp \frac{\lambda^0}{\Lambda_\Q^0}\rp D(\tau,\tau_0) \right. \label{eq:nq} \\
&&  \left. \hspace{5cm} + \int\limits_{\tau_0}^{\tau} \f{d \tau'}{\teq^\prime}\  D(\tau,\tau') \lp T^\prime\rp^3 \tilde{{\cal H}}_{\cal N}^+\lp \f{\tau^\prime}{\tau  }, \frac{m}{T^\prime}, \mp \frac{\mu^\prime}{T^\prime}\rp\rsb, \nn
\end{eqnarray}
\begin{eqnarray}
 {\cal E}^{\Qpm} &=&  2 \pi k_\Q \lsb  \lp\Lambda_\Q^0\rp^4 \tilde{{\cal H}}^+\lp \f{\tau_0}{\tau \sqrt{1+\xi_\Q^0}}, \frac{m}{\Lambda_\Q^0}, \mp \frac{\lambda^0}{\Lambda_\Q^0}\rp D(\tau,\tau_0) \right. \label{eq:eq} \\
&&  \left. \hspace{5cm} + \int\limits_{\tau_0}^{\tau} \f{d \tau'}{\teq^\prime}\  D(\tau,\tau') \lp T^\prime\rp^4 \tilde{{\cal H}}_{\cal  }^+\lp \f{\tau^\prime}{\tau  }, \frac{m}{T^\prime}, \mp \frac{\mu^\prime}{T^\prime}\rp\rsb, \nn
\end{eqnarray}
\begin{eqnarray}
{\cal P}^{\Qpm}_T  &=&   \pi k_\Q  \lsb \lp\Lambda_\Q^0\rp^4 \tilde{{\cal H}}_T^+\lp \f{\tau_0}{\tau \sqrt{1+\xi_\Q^0}}, \frac{m}{\Lambda_\Q^0}, \mp \frac{\lambda^0}{\Lambda_\Q^0}\rp D(\tau,\tau_0) \right. \label{eq:ptq} \\
&&  \left. \hspace{5cm} + \int\limits_{\tau_0}^{\tau} \f{d \tau'}{\teq^\prime}\  D(\tau,\tau')\lp T^\prime\rp^4 \tilde{{\cal H}}_{  T}^+\lp \f{\tau^\prime}{\tau  }, \frac{m}{T^\prime}, \mp \frac{\mu^\prime}{T^\prime}\rp \rsb, \nn
\end{eqnarray}
\begin{eqnarray}
{\cal P}^{\Qpm}_L  &=& 2 \pi k_\Q  \lsb \lp\Lambda_\Q^0\rp^4 \tilde{{\cal H}}_L^+\lp \f{\tau_0}{\tau \sqrt{1+\xi_\Q^0}}, \frac{m}{\Lambda_\Q^0}, \mp \frac{\lambda^0}{\Lambda_\Q^0}\rp D(\tau,\tau_0) \right. \label{eq:plq} \\
&&  \left. \hspace{5cm} + \int\limits_{\tau_0}^{\tau} \f{d \tau'}{\teq^\prime}\  D(\tau,\tau') \lp T^\prime\rp^4 \tilde{{\cal H}}_{L}^+\lp \f{\tau^\prime}{\tau  }, \frac{m}{T^\prime}, \mp \frac{\mu^\prime}{T^\prime}\rp\rsb, \nn
\end{eqnarray}
for quarks and 
\begin{eqnarray}
{\cal N}^{\G}  &=&  4 \pi k_\G \lsb  \lp\Lambda_\G^0\rp^3 \tilde{{\cal H}}_{\cal N}^-\lp \f{\tau_0}{\tau \sqrt{1+\xi_\G^0}}, 0,0\rp D(\tau,\tau_0)+ \int\limits_{\tau_0}^{\tau} \f{d \tau'}{\teq^\prime}\  D(\tau,\tau') \lp T^\prime\rp^3 \tilde{{\cal H}}_{\cal N}^-\lp \f{\tau^\prime}{\tau  }, 0,0\rp\rsb, 
\nn \\
\label{eq:ng}
\end{eqnarray}
\begin{eqnarray}
 {\cal E}^{\G} &=&  2 \pi k_\G \lsb  \lp\Lambda_\G^0\rp^4 \tilde{{\cal H}}^-\lp \f{\tau_0}{\tau \sqrt{1+\xi_\G^0}},  0,0\rp D(\tau,\tau_0)+ \int\limits_{\tau_0}^{\tau} \f{d \tau'}{\teq^\prime}\  D(\tau,\tau') \lp T^\prime\rp^4 \tilde{{\cal H}}_{\cal  }^-\lp \f{\tau^\prime}{\tau  }, 0,0\rp\rsb,
 \nn \\
\label{eq:eg}
\end{eqnarray}
\begin{eqnarray}
{\cal P}^{\G}_T  &=&   \pi k_\G  \lsb \lp\Lambda_\G^0\rp^4 \tilde{{\cal H}}_T^-\lp \f{\tau_0}{\tau \sqrt{1+\xi_\G^0}},  0,0\rp D(\tau,\tau_0)+ \int\limits_{\tau_0}^{\tau} \f{d \tau'}{\teq^\prime}\  D(\tau,\tau')\lp T^\prime\rp^4 \tilde{{\cal H}}_{  T}^-\lp \f{\tau^\prime}{\tau  }, 0,0\rp \rsb,
\nn \\
\label{eq:ptg}
\end{eqnarray}
\begin{eqnarray}
{\cal P}^{\G}_L  &=& 2 \pi k_\G  \lsb \lp\Lambda_\G^0\rp^4 \tilde{{\cal H}}_L^-\lp \f{\tau_0}{\tau \sqrt{1+\xi_\G^0}},  0,0\rp D(\tau,\tau_0)+ \int\limits_{\tau_0}^{\tau} \f{d \tau'}{\teq^\prime}\  D(\tau,\tau') \lp T^\prime\rp^4 \tilde{{\cal H}}_{L}^-\lp \f{\tau^\prime}{\tau  }, 0,0\rp\rsb,
\nn \\
\label{eq:plg}
\end{eqnarray}
for gluons.
We define the baryon number density for the exact solution of the kinetic equation as follows
\begin{eqnarray}
{\cal B}  &=&  \frac{16 \pi k_\Q}{3} \lsb   \f{\tau_0 \lp\Lambda_\Q^0\rp^3}{\tau \sqrt{1+\xi_\Q^0}} \sinh\lp\frac{\lambda^0}{\Lambda_\Q^0}\rp\,{\cal H}_{\cal B}\lp \frac{m}{\Lambda_\Q^0},\frac{\lambda^0}{\Lambda_\Q^0} \rp D(\tau,\tau_0) \right. \label{BApp} \\
&& \left. \hspace{5cm}
+ \int\limits_{\tau_0}^{\tau} \f{d \tau'}{\teq^\prime}\  D(\tau,\tau')  \f{\tau^\prime \lp T^\prime\rp^3}{\tau  } \sinh\lp\frac{\mu^\prime}{T^\prime}\rp\,{\cal H}_{\cal B}\lp \frac{m}{T^\prime}, \frac{\mu^\prime}{T^\prime}\rp\rsb  . \nn 
\end{eqnarray}

\chapter{Second moments of the distribution function}
\label{s:sm}
%
The second moment of the distribution function is expressed by the momentum integral
\bea  
\Theta_\s^{\lambda\mu\nu}(x) &=&   k_\s \int \!dP\, p^{\lambda} p^{\mu}p^{\nu} f_\s(x,p).
\label{app:thetatensors} 
\eea
The latter may be decomposed in the tensorial basis contructed using the tensor products of the basis four-vectors $A^\mu_{(\alpha)}=\{U^\mu, X^\mu, Y^\mu, Z^\mu\}$,
\beal{app:thetadecomp}
\Theta_\s^{\lambda\mu\nu}(x) &=& \sum_{A,B,C} c_{ABC}^{\s}  A^\lambda B^\mu C^\nu,
\eeal
where the coefficients $c_{ABC}^{\s}$ are defined through the expression 
\beal{app:thetadecomp2}
 c_{ABC}^{\s} &=& A_{\lambda}^{\,} B_{\mu}^{\,} C_{\nu}^{\,} \Theta_\s^{\lambda\mu\nu}(x) \, A^2  B^2 C^2.
\eeal
Using \EQ{app:thetatensors} one thus has
\bea  
c_{ABC}^{\s} &=& k_\s\int \!dP\,  \lp p \cdot A \rp \lp p \cdot B \rp \lp p \cdot C \rp A^2 B^2 C^2 f_\s(x,p).
\label{app:thetadecompcoefgen}
\eeal
For the distribution functions specified in \EQSTWO{Qeq}{Geq} and \EQSTWO{Qa}{Ga}, due to the symmetry arguments, the  only non-vanishing coefficients $c_{ABC}^{\s}$ out of those in \EQ{app:thetadecompcoefgen} are the ones with  an even number of each spatial index of $p^{\mu}$ which means that  \EQ{app:thetadecomp} may be expressed as follows
\beal{app:thetadecomp3}
\Theta_\s^{\lambda\mu\nu}(x) &=& c_{UUU}^{\s}  U^\lambda U^\mu U^\nu   +\sum_{A} c_{UAA}^{\s}  (U^\lambda A^\mu A^\nu+ A^\lambda U^\mu A^\nu+ A^\lambda A^\mu U^\nu).
\eeal
%
\section{Anisotropic distributions}
\label{ssec:appad}
%
For the anisotropic distribution functions, one may exploit the SO(2) symmetry   of the Eqs.~(\ref{Qa})-(\ref{Ga}) in transverse momentum plane and write
\bea
\Theta_{\s, \an}^{\lambda\mu\nu} &=& \vartheta_U^{\s, \an}\,  U^{\lambda} U^{\mu} U^{\nu}    \\ \,&-&\,  \vartheta_T^{\s, \an}\, \left( U^{\lambda} \Delta^{\mu\nu}_T + U^{\mu} \Delta^{\lambda\nu}_T+    U^{\nu}\Delta ^{\lambda \mu}_T\right)\\
\,&+&\,  \vartheta_L^{\s, \an}\, \left( U^{\lambda}Z^{\mu} Z^{\nu} + U^{\mu}Z^{\lambda} Z^{\nu}+U^{\nu} Z^{\lambda} Z^{\mu} \right),
 \label{app:thetatensoran}
\eea  
where $\vartheta_U^{\s, \an} \equiv c_{UUU}^{\s, \an}$, $
\vartheta_T^{\s, \an} \equiv c_{UXX}^{\s, \an}=c_{UYY}^{\s, \an}$, and $ 
\vartheta_L^{\s, \an} \equiv c_{UZZ}^{\s, \an}$.

For quarks explicit calculation gives 
\begin{eqnarray}
\vartheta_U^{\Qpm, \an}    &=&   \frac{4\pi k_\Q \Lambda_\Q^5}{3}   \frac{(3+2\xi_\Q)}{(1+\xi_\Q)^{3/2}} \tilde{{\cal H}}_{\vartheta}^+\lp  \frac{m}{\Lambda_\Q}, \mp \frac{\lambda}{\Lambda_\Q}\rp +m^2 {\cal N}^{\Qpm, \an} ,
\label{app:thetaU}\\
\vartheta_T^{\Qpm, \an}   &=&   \frac{4\pi k_\Q \Lambda_\Q^5}{3}   \frac{1}{\sqrt{1+\xi_\Q}} \tilde{{\cal H}}_{\vartheta}^+\lp  \frac{m}{\Lambda_\Q}, \mp \frac{\lambda}{\Lambda_\Q}\rp,
\label{app:thetaT}\\
\vartheta_L^{\Qpm,\an}    &=&   \frac{4\pi k_\Q \Lambda_\Q^5}{3}   \frac{1}{(1+\xi_\Q)^{3/2}} \tilde{{\cal H}}_{\vartheta}^+\lp  \frac{m}{\Lambda_\Q}, \mp \frac{\lambda}{\Lambda_\Q}\rp,
\label{app:thetaL}
\end{eqnarray}
where 
\beal{Htheta}
\tilde{{\cal H}}_{\vartheta}^\pm\lp y,z\rp &\equiv& \int\limits_0^\infty r^4  dr \, h^{^\pm}_\eq\lp \sqrt{r^2+y^2}+z\rp.
\eeal
Analogous expressions hold for gluons where the integral in \EQ{Htheta} yields $\tilde{{\cal H}}_{\vartheta}^\pm\lp 0,0\rp  = 24 \zeta(5)$ with $\zeta$ being the Riemann zeta function.
\section{Isotropic distributions}
\label{ssec:ed}
%
For the equilibrium distribution functions  the SO(3) symmetry of the Eqs.~(\ref{Qeq})-(\ref{Geq}) in the momentum space allows one to write
\bea
\Theta_{\s, \eq}^{\lambda\mu\nu} &=& \vartheta_U^{\s, \eq}\,  U^{\lambda} U^{\mu} U^{\nu}   - \vartheta^{\s, \eq}\, \left( U^{\lambda} \Delta^{\mu\nu} + U^{\mu} \Delta^{\lambda\nu}+    U^{\nu}\Delta ^{\lambda \mu}\right)
 \label{app:thetatensoreq}
\eea  
where $\vartheta_U^{\s, \eq} \equiv c_{UUU}^{\s, \an}$, and $
\vartheta^{\s, \eq} \equiv c_{UXX}^{\s, \eq}=c_{UYY}^{\s, \eq} = c_{UZZ}^{\s, \eq}$ are obtained from expressions (\ref{app:thetaU})-(\ref{app:thetaL}) taking the limit $\xi_s\to 0$ where $\Lambda_s\to T$.

%
\chapter{Navier-Stokes hydrodynamics}
\label{s:NS}
%

The results of our kinetic-theory calculations are compared with the viscous hydrodynamic results obtained by solving 
the Navier-Stokes (NS) hydrodynamic equations. The latter have the form
\begin{eqnarray}
\f{d}{d\tau} \left( {\cal E}^{\Q, \eq}  +  {\cal E}^{\G, \eq}  \right)
&=& - \f{{\cal E}^{\Q, \eq} +  {\cal E}^{\G, \eq} + {\cal P}^{\Q,\eq}  +  {\cal P}^{\G,\eq}  + \Pi_{\rm NS} - \pi_{\rm NS}  }{\tau}, 
\label{NSapp1} \\
\f{d{\cal B}^{\eq}}{d\tau} +\f{{\cal B}^{\eq}}{\tau} &=& 0.
\label{NSapp2}
\end{eqnarray}
Here ${\cal E}^{\Q, \eq}  = {\cal E}^{\Qp, \eq} +  {\cal E}^{\Qm, \eq}$ is the equilibrium energy density of quarks and antiquarks, ${\cal P}^{\Q,\eq} = {\cal P}^{\Qp,\eq} + {\cal P}^{\Qm,\eq}$ is the equilibrium pressure of quarks and antiquarks, $\Pi_{\rm NS}$ is the bulk pressure, and $\pi_{\rm NS}$ is the shear pressure (both used in the close-to-equilibrium limit). All the functions appearing in \rfn{NSapp1} and \rfn{NSapp2} depend on $T$ and $\mu$, hence  Eqs.~\rfn{NSapp1} and \rfn{NSapp2} are two coupled equations that can be used to determine $T(\tau)$ and $\mu(\tau)$.  
One can easily notice that Eq.~\rfn{NSapp1} may be written in the form of Eq.~\rfn{SECOND-EQUATION-1} once we identify
\beal{PLPTns}
 {\cal P}_L   &=&  {\cal P}^{\rm eq} - \pi_{\rm NS} + \Pi_{\rm NS},\\
 {\cal P}_T  &=&  {\cal P}^{\rm eq} + \frac{1}{2}\pi_{\rm NS} + \Pi_{\rm NS}, \nn
\eeal
where ${\cal P}^{  \eq}  = {\cal P}^{\Q, \eq} +  {\cal P}^{\G, \eq}$.
Within NS approach, the shear and bulk pressures  are expressed by the kinetic coefficients $\eta$ and $\zeta$,
\bel{pi2}
\pi_{\rm NS} = \f{4 \eta}{3\tau} = \f{4 (\eta_\Q+\eta_G)}{3\tau},
\eel
\bel{Zeta1}
\Pi_{\rm NS} = -\f{\zeta}{\tau}.
\eel
The expressions for $\eta_Q$, $\eta_G$, and $\zeta$ are given by Eqs.~\rfn{etaQ}, \rfn{etaG}, and \rfn{zeta}.

\medskip
For the moment let us denote $T(\tau)$ and $\mu(\tau)$ obtained from the kinetic theory as $T_{\rm KT}(\tau)$  and $\mu_{\rm KT}(\tau)$, while those obtained from the NS hydrodynamics as $T_{\rm NS}(\tau)$ and $\mu_{\rm NS}(\tau)$. We expect that $T_{\rm KT}(\tau)$ and  $\mu_{\rm KT}(\tau)$ agree well with $T_{\rm NS}(\tau)$ and $\mu_{\rm NS}(\tau)$ in the late stages of the evolution when the system approaches local equilibrium. To check this behavior  we choose such initial conditions for hydrodynamic equations  \rfn{NSapp1} and \rfn{NSapp2} that for the final time $\tau=\tau_f$ we match the temperature and chemical potential in the two approaches: $T_{\rm NS}(\tau_f)=T_{\rm KT}(\tau_f)$,  $\mu_{\rm NS}(\tau_f)=\mu_{\rm KT}(\tau_f)$. Then, we check if the functions  $T_{\rm NS}(\tau)$ and $\mu_{\rm NS}(\tau)$
smoothly approach $T_{\rm KT}(\tau)$  and $\mu_{\rm KT}(\tau)$ if $\tau \to \tau_f$. By neglecting the bulk and shear pressures in   \rfn{NSapp1}  we can also make a comparison with perfect fluid hydrodynamics and check if the system approaches local equilibrium.

\chapter{Shear and bulk viscosities for mixtures}
\label{s:shearbulk}

In this section, we present details of our method used to calculate the shear and bulk viscosity coefficients for a quark-gluon mixture. 
We follow the treatment of Refs.~\cite{Florkowski:2015rua,Florkowski:2015dmm}, where the bulk viscosity was obtained 
for the Gribov-Zwanziger plasma.  Analyzing a~boost-invariant system, we deal with a simple structure of hydrodynamic equations, 
which facilitates the calculations. 

\section{Landau matching conditions in the case of boost invariant geometry}
\label{s:LM-BI}

In the first order of the gradient expansion, the non-equilibrium corrections to the equilibrium distribution function
have the form
\bel{df}
\delta f_\Qpm = -\teq \f{\partial f_{\Qpm, \eq}}{\partial \tau}, \quad \delta f_\G = -\teq \f{\partial f_{\G, \eq}}{\partial \tau}.
\eel
Using the form of $f_{\Qpm, \eq}$ and $f_{\G, \eq}$ for boost-invariant geometry (\ref{BIQeq0})--(\ref{BIGeq0}) we find
\begin{eqnarray}
\delta f_\Qpm &=& -\teq f_{\Qpm, \eq} \left(1 - f_{\Qpm, \eq} \right)
 \left[ \f{w^2}{v \tau^2 T} \pm \f{d\mu}{T d\tau } + \left( \f{v}{\tau} \mp \mu \right) \f{d \ln T}{T d\tau} \right] ,
 \label{df1q} \\
 \delta f_\G &=& -\teq f_{\G, \eq} \left(1 + f_{\G, \eq} \right)
 \left[ \f{w^2}{v \tau^2 T}  + \f{v}{\tau}  \f{d \ln T}{T d\tau}  \right]. \label{df1g}
\end{eqnarray}
The Landau matching conditions for the energy and momentum read
\begin{eqnarray}
&&\int  \f{dw d^2 \pT}{v} \,\f{v^2}{\tau^2} \left[ k_\Q \left( \delta f_\Qp + \delta f_\Qm \right) + k_\G  \delta f_\G \right] = 0 \label{LMdq}, \\
&& \int \f{dw d^2 \pT}{v} \,\f{v^2}{3 \tau^2} \left[ k_\Q \left( \delta f_\Qp - \delta f_\Qm \right)  \right] = 0 \label{LMdg}.
\end{eqnarray}
Using \rfn{df1q} and \rfn{df1g} we rewrite \rfn{LMdq} and \rfn{LMdg} as
\begin{eqnarray}
&&  \int \f{dw d^2 \pT}{v} \, \f{v^2}{\tau^2} \, f_{3S} \, \f{w^2}{v \tau^2 T}
+  \int \f{dw d^2 \pT}{v} \, \f{v^2}{\tau^2} \, f_{2D} \, \f{d\mu}{T d\tau} \nn \\
&& +   \int \f{dw d^2 \pT}{v} \, \f{v^2}{\tau^2} \, f_{3S} \, \f{v}{T \tau} \, \f{d \ln T}{d\tau}
-  \int \f{dw d^2 \pT}{v} \, \f{v^2}{\tau^2} \, f_{2D} \, \f{\mu}{T} \f{d \ln T}{d\tau} = 0
\label{LMdq1}
\end{eqnarray}
and
\begin{eqnarray}
&&  \int \f{dw d^2 \pT}{v} \, \f{v^2}{\tau^2} \, f_{2D} \, \f{w^2}{v \tau^2 T}
+  \int \f{dw d^2 \pT}{v} \, \f{v^2}{\tau^2} \, f_{2S} \, \f{d\mu}{T d\tau} \nn \\
&& +   \int \f{dw d^2 \pT}{v} \, \f{v^2}{\tau^2} \, f_{2D} \, \f{v}{T \tau} \, \f{d \ln T}{d\tau}
-  \int \f{dw d^2 \pT}{v} \, \f{v^2}{\tau^2} \, f_{2S} \, \f{\mu}{T} \f{d \ln T}{d\tau} = 0,
\label{LMdg1}
\end{eqnarray}
where
\begin{eqnarray}
f_{3S} &=& k_\Q \left[ f_{\Qp, \eq} \left(1 - f_{\Qp, \eq} \right) + f_{\Qm, \eq} \left(1 - f_{\Qm, \eq} \right) \right] + k_\G f_{\G, \eq} \left(1 + f_{\G, \eq} \right), \nn \\
f_{2S} &=& k_\Q \left[ f_{\Qp, \eq} \left(1 - f_{\Qp, \eq} \right) + f_{\Qm, \eq} \left(1 - f_{\Qm, \eq} \right) \right] , \nn \\
f_{2D} &=& k_\Q \left[ f_{\Qp, \eq} \left(1 - f_{\Qp, \eq} \right) - f_{\Qm, \eq} \left(1 - f_{\Qm, \eq} \right) \right].
\end{eqnarray}
By introducing the ``averaged'' values defined as
\bel{av}
\langle ... \rangle_\alpha \equiv \int \f{dw d^2 \pT}{v} ... f_\alpha, 
\eel
where $\alpha = 3S, 2S, 2D$, we rewrite  \rfn{LMdq1} and \rfn{LMdg1} in the compact form
\begin{eqnarray}
\langle w^2 \rangle_{3S} + \langle v^2 \rangle_{3S} \f{d\ln T}{d\ln \tau} +\langle v \rangle_{2D} \tau^2 T \f{d}{d\tau} \left(\f{\mu}{T} \right) &=& 0, \nn \\
\langle \f{w^2}{v} \rangle_{2D} + \langle v \rangle_{2D}   \f{d\ln T}{d\ln \tau} +  \langle 1 \rangle_{2S} \tau^2 T \f{d}{d\tau} \left(\f{\mu}{T} \right) &=& 0.
\end{eqnarray}
To proceed further it is convenient to introduce the notation
\begin{eqnarray}
A = \langle w^2 \rangle_{3S}, \,\,B = \langle v^2 \rangle_{3S}, \,\,C = \langle v \rangle_{2D}, 
\,\,D = \langle \f{w^2}{v} \rangle_{2D}, \,\,E = \langle 1 \rangle_{2S}.
\label{ABC}
\end{eqnarray}
Then, we find the proper-time derivatives of $T$ and $\mu/T$ expressed by the coefficients \rfn{ABC}
\bel{xy}
\f{d\ln T}{d\ln \tau} = \f{A E - C D}{C^2 - B E}, \quad \tau^2 T \f{d}{d\tau} \left(\f{\mu}{T} \right) = \f{D B - A C}{C^2 - B E}.
\eel
The coefficients \rfn{ABC} can be used also to express various thermodynamic derivatives. After straightforward calculations, where $T$ and $\mu$ are treated as independent thermodynamic variables,  we find
\begin{eqnarray}
\f{\p {\cal P}^{\rm eq}}{\p T} &=& \f{\p {\cal P}^{\Qp, \rm eq}}{\p T}+\f{\p {\cal P}^{\Qm, \rm eq}}{\p T}+\f{\p {\cal P}^{\G, \rm eq}}{\p T}
= \f{A}{\tau^3 T^2} - \f{D \mu}{\tau^2 T^2}, \nn \\
\f{\p {\cal P}^{\rm eq}}{\p \mu} &=& \f{\p {\cal P}^{\Qp, \rm eq}}{\p \mu}+\f{\p {\cal P}^{\Qm, \rm eq}}{\p \mu}
=  \f{D}{\tau^2 T}, \nn \\
\f{\p {\cal E}^{\rm eq}}{\p T} &=& \f{\p {\cal E}^{\Qp, \rm eq}}{\p T}+\f{\p {\cal E}^{\Qm, \rm eq}}{\p T}+\f{\p {\cal E}^{\G, \rm eq}}{\p T}
= \f{B}{\tau^3 T^2} - \f{C \mu}{\tau^2 T^2}, \nn \\
\f{\p {\cal E}^{\rm eq}}{\p \mu} &=& \f{\p {\cal E}^{\Qp, \rm eq}}{\p \mu}+\f{\p {\cal E}^{\Qm, \rm eq}}{\p \mu}
=  \f{C}{\tau^2 T}, \nn \\
\f{\p {\cal B}^{\rm eq}}{\p T} &=& = \f{C}{3 \tau^2 T^2} - \f{E \mu}{3 \tau T^2}, \nn \\
\f{\p {\cal B}^{\rm eq}}{\p \mu} &=& =   \f{E \mu}{3 \tau T}.
\label{thermder}
\end{eqnarray}
Using \rfn{thermder} we find that
\begin{eqnarray}
\kappa_1(T,\mu) \!\!&=&\!\! \left( \f{\p{\cal P}^{\rm eq} }{\p {\cal E}^{\rm eq}} \right)_{{\cal B}^{\rm eq}} 
=   \f{\p ({\cal P}^{\rm eq},{\cal B}^{\rm eq}) }{\p ({\cal E}^{\rm eq}, {\cal B}^{\rm eq} ) } 
=  - \f{A E - C D}{C^2 - B E} = - \f{d\ln T}{d\ln \tau}, \nn \\
\kappa_2(T,\mu) \!\!&=&\!\! \f{1}{3} \left( \f{\p{\cal P}^{\rm eq} }{ \p {\cal B}^{\rm eq}} \right)_{{\cal E}^{\rm eq}} 
=   \f{1}{3} \f{\p ({\cal P}^{\rm eq},{\cal E}^{\rm eq}) }{\p ({\cal B}^{\rm eq}, {\cal E}^{\rm eq} ) } 
=  - \f{D B - A C}{\tau (C^2 - B E)} = -  \tau T \f{d}{d\tau} \left(\f{\mu}{T} \right) .
\end{eqnarray} 

\section{Shear viscosity}
\label{s:Shear}

The shear viscosity can be obtained from the formula  $\eta = \tau ({\cal P}_T - {\cal P}_L)_{\rm NS}/2$, 
which in close-to-equilibrium situations leads to the expression
\begin{eqnarray}
\eta &=& \f{\tau}{2} \left( {\cal P}_T - {\cal P}_L \right)_{\rm NS}  \nn \\
&=&  \f{\tau}{2}   \int \f{dw d^2 \pT}{v}   \left[  \left( \f{\pT^2}{2} - \f{w^2}{\tau^2}   \right) 
\left[ k_\Q (f_{\Qp, \eq} + \delta f_\Qp +f_{\Qm, \eq} + \delta f_\Qm) + k_\G (f_{\G, \eq} + \delta f_\G) \right] \right] \nn \\
&=& 
 \f{\tau}{2}   \int \f{dw d^2 \pT}{v}   \left[  \left( \f{\pT^2}{2} - \f{w^2}{\tau^2}   \right) 
 \left[ k_\Q (\delta f_\Qp + \delta f_{\Qm} ) + k_\G  \delta f_\G \right] \right].
\label{shear1}
\end{eqnarray}
Here we used the property that the equilibrium distributions are isotropic and do not contribute to the integral \rfn{shear1}.
Using Eqs.~\rfn{df1q} and \rfn{df1g} we find
\begin{eqnarray}
\eta &=& 
- \f{\teq}{2}   \int \f{dw d^2 \pT}{v}   \left[  \left( \f{\pT^2}{2} - \f{w^2}{\tau}   \right) 
\f{w^2}{v \tau T}\right] f_{3S} ,
\label{shear2}
\end{eqnarray}
where the terms containing derivatives of $T$ and $\mu$ dropped out again due to symmetry reasons. 
Equation \rfn{shear2} can be rewritten as
\begin{eqnarray}
\eta &=& 
- \f{\teq}{2}   \int \f{d^3p}{E_p}   \left[  \left( p_x^2 - p_z^2   \right) 
 \f{p_z^2}{E_p T}\right] f_{3S} \label{shear3} \\
&=&  -\f{\teq}{2 T}   \int\limits \f{2\pi dp\, p^6 }{E_p^2}  \int\limits_0^\pi \sin\theta d\theta \left[  \left( \f{\sin^2\theta}{2} - \cos^2\theta   \right) 
 \cos^2\theta  \right] f_{3S}. 
\end{eqnarray}
The integral over the angle $\theta$ gives $-4/15$, hence the final result is
\begin{eqnarray}
\eta &=&  \f{4 \pi \teq}{15 T}   \int \f{dp\,p^6 }{E_p^2}   f_{3S}, 
\label{shear4}
\end{eqnarray}
which leads to Eqs.~\rfn{visceta}, \rfn{etaQ} and \rfn{etaG}.

For massless quarks, \rf{etaQ} gives $\eta_\Q = 7 g_\Q \pi^2 T^4 \teq/ 450$ and $\eta_\Q = 8 g_\Q T^4 \teq/(5 \pi^2)$ for Fermi-Dirac and Boltzmann statistics, respectively. The corresponding values of pressure are: $P^{\Q, \rm eq} = 7 g_\Q \pi^2 T^4/360$ and $P^{\Q, \rm eq} = 2 g_\Q T^4 /\pi^2$, hence, for the two statistics we find $\eta_\Q = 4 P^{\Q, \rm eq} /5$. In a similar way, from \rfn{etaG} we find for massless gluons: $\eta_\G = 2 g_\G \pi^2 T^4 \teq/225$ and $\eta_\G = 4 g_\G T^4 \teq/ (5 \pi^2)$ for Bose-Einstein and Boltzmann statistics. The corresponding pressures are: $P^{\G, \rm eq} = g_\G \pi^2 T^4/90$ and $P^{\G, \rm eq} = g_\G T^4/\pi^2$, which gives again the relation $\eta_\G = 4 P^{\G, \rm eq} /5$.

\section{Bulk viscosity}
\label{s:Bulk}

The bulk pressure is the difference between the average exact pressure, $({\cal P}_L + 2 {\cal P}_T)/3$, 
in the system and the reference equilibrium pressure, ${\cal P}^{\rm eq}$.
Close to local equilibrium, it can be defined by the following formula
\begin{eqnarray}
\Pi_{\rm NS} &=& \f{1}{3} \left( {\cal P}_L + 2 {\cal P}_T - 3 {\cal P}^{\rm eq} \right)_{\rm NS}   \nn \\
&=&  \f{1}{3}   \int \f{dw d^2 \pT}{v}   \left[  \left( 
\f{w^2}{\tau^2} + \pT^2 \right) 
\left[ k_\Q (f_{\Qp, \eq} + \delta f_\Qp +f_{\Qm, \eq} + \delta f_\Qm) + k_\G (f_{\G, \eq} + \delta f_\G) \right] \right.   \nn \\
&& 
\left.  \hspace{3cm}  - 3  \, \f{w^2}{\tau^2} \,  \left[ k_\Q (f_{\Qp, \eq} + f_{\Qm, \eq} + \delta f_\Qm) + k_\G  f_{\G, \eq}  \right] 
\right] \nn \\
&=& 
 \f{1}{3}   \int \f{dw d^2 \pT}{v}   \left[  \left( 
\f{w^2}{\tau^2} + \pT^2 \right)  \left[ k_\Q (\delta f_\Qp + \delta f_{\Qm} ) + k_\G  \delta f_\G \right] \right].
\label{bulk1}
\end{eqnarray}
To get the last line in \rfn{bulk1}, we have used the fact that equilibrium distributions are isotropic. It is interesting to notice that \rfn{bulk1}
can be also written as
\begin{eqnarray}
\Pi_{\rm NS}  &=& 
 \f{1}{3}   \int \f{dw d^2 \pT}{v}   \left[  \left( 
\f{w^2}{\tau^2} + \pT^2  + m^2 - m^2 \right)  \left[ k_\Q (\delta f_\Qp + \delta f_{\Qm} ) + k_\G  \delta f_\G \right] \right] \nn \\
&=& 
- \f{m^2}{3}   \int \f{dw d^2 \pT}{v}  \left[ k_\Q (\delta f_\Qp + \delta f_{\Qm} ) \right] ,
\label{bulk12}
\end{eqnarray}
where we used the Landau matching condition \rfn{LMdq} and the fact that gluons are massless.

Using the notation introduced above we find
\begin{eqnarray}
\Pi_{\rm NS}  &=& - \f{\teq}{3}   \int \f{dw d^2 \pT}{v} \left( \f{w^2}{\tau^2} + \pT^2 \right)  \left[ f_{3S} \left(\f{w^2}{v \tau^2 T} + \f{v}{\tau^2 T} \f{d\ln T}{d\ln \tau} \right) \right]
\nn \\
& & - \f{\teq}{3}   \int \f{dw d^2 \pT}{v} \left( \f{w^2}{\tau^2} + \pT^2 \right)  f_{2D}  \f{d}{d\tau} \left(\f{\mu}{T} \right) \nn \\
&=& - \f{\teq}{3}   \int \f{dw d^2 \pT}{v} \left( \f{w^2}{\tau^2} + \pT^2 \right)  \left[ f_{3S} \left(\f{w^2}{v \tau^2 T} - \f{v}{\tau^2 T} \, \kappa_1 \right) \right]
\nn \\
& & + \f{\teq}{3 \tau T}   \int \f{dw d^2 \pT}{v} \left( \f{w^2}{\tau^2} + \pT^2 \right)  f_{2D} \, \kappa_2. 
\label{bulk2}
\end{eqnarray}
Due to boost-invariance, the integral above can be done in the plane $z=0$, where $w = \pL t$, $v = E_p t$. Since $f_{3S}$ and $f_{2D}$ are isotropic, we obtain
\begin{eqnarray}
\Pi_{\rm NS}  &=& - \f{\teq}{3 \tau T}   \int d^3p \, p^2  \left[ f_{3S} \left(\f{p^2}{3 E^2_p} - \, \kappa_1 \right) \right]
 + \f{\teq}{3 \tau T}   \int \f{d^3p}{E_p} p^2  f_{2D} \, \kappa_2. 
\label{bulk3}
\end{eqnarray}
For the Bjorken flow we have $\p_\mu U^\mu = 1/\tau$, thus the Navier--Stokes relation $\Pi_{\rm NS}  = -\zeta \p_\mu U^\mu$ allows us to identify the bulk pressure as
\begin{eqnarray}
\zeta &=&  \f{\teq}{3 T}   \int d^3p \, p^2  \left[ f_{3S} \left(\f{p^2}{3 E^2_p} - \, \kappa_1 \right) \right]
 - \f{\teq}{3 T}   \int \f{d^3p}{E_p} p^2  f_{2D} \, \kappa_2. 
\label{bulk4}
\end{eqnarray}
Similarly, starting from \rfn{bulk12} we find
\begin{eqnarray}
\zeta &=&  \f{\teq m^2}{3 T}   \int d^3p   \left[ f_{2S}  \left(\kappa_1 - \f{p^2}{3 E^2_p} \right) \right]
 + \f{\teq m^2}{3 T}   \int \f{d^3p}{E_p}  f_{2D} \, \kappa_2,
\label{bulk5}
\end{eqnarray}
which leads to \rfn{zeta}.

\bibliographystyle{unsrt}
\bibliography{hydro_review}


\cleardoublepage
\thispagestyle{plain}

\begin{flushright} 
Kielce, dnia 1.09.2018
\end{flushright}
\begin{flushleft} 
......................................................\newline
\hspace{2in}/Imię i nazwisko doktoranta/\\[0.6cm]
......................................................\newline
/Numer albumu/\\[0.6cm]
......................................................\newline
\hspace{2in}/Kierunek/\\[0.6cm]
......................................................\newline
/Rodzaj studiów, forma studiów/\\[0.6cm]
......................................................\newline
/Data urodzenia/
\end{flushleft}
\vspace{0.5cm}
\begin{center}
\textsc{\Large Oświadczenie}\\[0.3cm]
\end{center}
\vspace{1cm}
Przedkładając pracę doktorską pod tytułem
\newline
\newline
\textit{Exact solutions of the relativistic Boltzmann equation in the relaxation time approximation}
\newline
\newline
oświadczam, że pracę napisałam samodzielnie, praca nie stanowi istotnego fragmentu lub innych elementów cudzego utworu, nie narusza żadnych innych istniejących praw autorskich,  wykorzystane w pracy materiały źródłowe zastosowane zostały z zachowaniem zasad prawa cytatu,
przedstawiona praca w całości ani też w części nie była wcześniej podstawą do ubiegania się o nadanie stopnia
naukowego doktora, zaś wersja elektroniczna (na nośniku elektronicznym i/lub w systemie Wirtualna Uczelnia) pracy jest
tożsama z wersją drukowaną. \newline

Wyrażam zgodę na udostępnianie mojej pracy doktorskiej przez Uniwersytet Jana
Kochanowskiego w Kielcach dla celów naukowych i dydaktycznych.\newline

Nie wyrażam zgody na upowszechnianie mojej daty urodzenia przez Uniwersytet
Jana Kochanowskiego w Kielcach, w celu tworzenia autorskich rekordów wzorcowych w katalogach
bibliotecznych i bazach danych.

Prawdziwość powyższego oświadczenia potwierdzam własnoręcznym podpisem.
\vspace{2cm}

\begin{flushright}
.....................................\\
/Podpis autora pracy/
\end{flushright}

\end{document}